\documentclass[a4paper,fleqn]{cas-dc}

\usepackage[numbers]{natbib}
\usepackage{stfloats} 
\usepackage{algpseudocode}
\usepackage{algorithm}
\usepackage[caption=false,font=normalsize,labelfont=sf,textfont=sf]{subfig} 

\def\tsc#1{\csdef{#1}{\textsc{\lowercase{#1}}\xspace}}
\tsc{WGM}
\tsc{QE}
\tsc{EP}
\tsc{PMS}
\tsc{BEC}
\tsc{DE}

\begin{document}
\let\WriteBookmarks\relax
\def\floatpagepagefraction{1}
\def\textpagefraction{.001}
\shorttitle{HSI Denoising via SSTV Regularized Nonconvex Local Low-Rank Method}
\shortauthors{Haijin Zeng et~al.}

\title [mode = title]{Hyperspectral Image Denoising via Global Spatial-Spectral Total Variation Regularized Nonconvex Local Low-Rank Tensor Approximation}


\author[1]{Haijin Zeng}[style=chinese]

\author[1]{Xiaozhen Xie}[style=chinese]
\address[1]{College of Science, Northwest A\&F University, Yangling 712100, China, Email: xiexzh@nwafu.edu.cn}

\author[2]{Jifeng Ning}[style=chinese]
\address[2]{College of Information Engineering, Northwest A\&F University, Yangling 712100, China}

%

%

\begin{abstract}
Hyperspectral image (HSI) denoising aims to restore clean HSI from the noise-contaminated one.
Noise contamination can often be caused during data acquisition and conversion.
In this paper, we propose a novel spatial–spectral total variation (SSTV) regularized nonconvex local low-rank (LR) tensor approximation method to remove mixed noise in HSIs.
From one aspect, the clean HSI data have its underlying local LR tensor property, even though the real HSI data may not be globally low-rank due to out-liers and non-Gaussian noise.
According to this fact, we propose a novel tensor $L_{\gamma}$-norm to formulate the local LR prior.
From another aspect, HSIs are assumed to be piecewisely smooth in the global spatial and spectral domains.
Instead of traditional bandwise total variation, we use the SSTV regularization to simultaneously consider global spatial structure and spectral correlation of neighboring bands.
Results on simulated and real HSI datasets indicate that the use of local LR tensor penalty and global SSTV can boost the preserving of local details and overall structural information in HSIs.
\end{abstract}



\maketitle

\section{Introduction}

Hyperspectral images (HSIs) can provide spectral information of hundreds of continuous bands in the same scene.
Therefore, they are widely used in many fields, such as environmental research, agriculture, military and geography \cite{HSI_application_1,HSI_application_2}.
In recent years, HSIs have attracted great research interest in the field of remote sensing.
However, due to the limitations of observation conditions and sensors, the observed HSI is usually contaminated by a variety of noises, such as Gaussian noise, stripes, deadlines, impulse noise and their hybrids.
These noises adversely affect the image quality of HSIs, the subsequent processing and applications,
e.g., feature classification \cite{HSI_class_1}, target detection \cite{HSI_detection_1}, unmixing \cite{HSI_unmixing_1} and so on.
Therefore, as a pre-processing step, HSI denoising is an important research topic.

The high-dimensional HSI is composed of hundreds of single grey-scale images.
Therefore, a natural way for HSI denoising is to use the 1-D or 2-D denoising model to remove the noise pixel by pixel or band by band.
For example, the classical image denoising algorithm based on an enhanced sparse representation in transform domain \cite{BM3D},
the image denoising model via sparse and redundant representations over learned dictionaries \cite{2_D_dictionary},
and the nonlocal image restoration method with bilateral variance estimation \cite{2_D_nonlocal}.
However, these 1-D or 2-D denoising methods can only explore the structural characteristics of each pixel or each band individually,
they ignore the high correlations among all the spectral bands \cite{LRTDTV, LLRSSTV}.
Therefore, the quality of their restored images is relatively low.

In order to explore the spectral correlations that 1-D or 2-D methods ignored,
a number of studies have been conducted in the literature.
For example, a noise reduction algorithm is introduced and applied to HSI denoing \cite{denoising_spectral_correlation_Othman}.
That algorithm resorts to the spectral derivative domain where the noise level is elevated, and the dissimilarity of the signal regularity in the spatial and the spectral dimensions of HSI.
Zhong et al. \cite{denoising_spectral_correlation_random_field} propose a denoising model in the domain of imaging spectroscopy by conditional random fields,
which can simultaneously model and use the spatial and spectral dependencies in a unified probabilistic framework.
Majumdar et al. \cite{denoising_spatiospectral_correlation} exploit the spatiospectral correlation to sparsify the HSI datacube for reducing impulse noise in corrupted HSIs.

Furthermore, low-rank technique is a powerful tool to describe the spectral correlations presented in HSIs.
Based on low-rank technique, a robust principle component analysis (RPCA) model \cite{RPCA} decomposes the observed data into a low-rank matrix representing data information and a sparse matrix representing noise information.
Motivated by the idea of RPCA model, numerous low-rank based approaches have been proposed for HSI restoration.
Zhang et al. \cite{LRMR} propose a HSI restoration method based on low-rank matrix recovery,
which explored the low-rank property by lexicographically ordering a patch of the HSI into a 2-D matrix.
Consider that the noise intensity in different frequency bands in the HSI is usually different,
He et al. propose a noise-adjusted iterative low-rank matrix approximation (NAILRMA) method for HSI denoising.
To effectively explore the local low-rank (LR) property of HSIs and reduce the dependence on the noise independent and identical distribution assumption,
a variety of patch based RPCA models \cite{LRTV,LLRSSTV,local_zeng_HSI} with outstanding recovery performance is proposed to denoise HSIs patch by patch.


Although the above models and their extended models have been successfully used in various applications,
they need to convert 3-D HSI into 2-D matrices.
This strategy will introduce loss of useful multi-order structure information.
To alleviate the above problem, a tensor nuclear norm (TNN) induced by the tensor Singular Value Decomposition (t-SVD) is embedded into the RPCA model,
and a tensor robust principal component (TRPCA) model is proposed in \cite{TRPCA, zhang2016exact}.
In this model, the tensor nuclear norm is used to represent low-rank priors of multi-order data.
Furthermore, TNN based methods have reported success on low-rank tensor completion and HSI denoising.
However, the t-SVD used to induce TNN is defined based on tensor-tensor product (t-prod).
The structure of t-prod shows that it is equivalent to the matrix-matrix product in the Fourier transform domain.
Therefore, TNN is essentially the matrix nuclear norm of the block diagonal unfolding of the Fourier transformed tensor \cite{PSTNN}.
Unfortunately, since the nuclear norm may not be a perfect approximation to the rank function \cite{nonconconvex_TNN, Non_LRMA},
the TNN based models may obtain suboptimal performance in real applications.
Specifically, compared to the rank function in which all the nonzero singular values have equal contributions,
the nuclear norm treats the singular values differently by adding them together.
Moreover, the theoretical requirements (e.g., incoherence property) of the nuclear norm heuristic are usually very hard to satisfy in practical scenarios \cite{nonconvex_power, RPCA}.

Recently, a number of studies \cite{nonconvex_optimal, nonconvex_log_determinant}, both practically and theoretically,
have shown that the nonconvex approximation of rank function can provide better estimation accuracy and variable selection consistency than the nuclear norm.
Motivated by such facts, several nonconvex penalties have been proposed and studied as alternatives to nuclear norm.
A few notable examples are the minimax concave penalty \cite{minimax_concave_penalty},
$L_p$ for $p\in (0, 1)$ \cite{nonconvex_Lp, nonconvex_L12},
log-sum penalty \cite{nonconvex_log_sum},
log-determinant penalty \cite{nonconvex_log_determinant},
truncated nuclear norm \cite{nonconconvex_TNN},
$L_{1/2}$ norm \cite{ratio_of_L1_L2}, the weighted schatten $p$-norm \cite{HSI_Sp} and $\gamma$ norm \cite{Non_LRMA}.

TV regularization is another efficient tool in image processing field.
Due to its good performance in preserving the spatial piecewise smoothness and edge structures of images, i.e., spatial sparsity,
it is widely used in the denoising task \cite{osher_iterative_2005_4_13,iordache_sparse_2011_4_14},
Magnetic Resonance processing \cite{MRI},
image superresolution \cite{zhang_super-resolution_2012_4_15},
reconstruction \cite{needell2013stable_4_16} and so on.
At first, TV regularization is introduced to denoise gray-level images \cite{rudin_nonlinear_1992_8_56_24} and colour images \cite{li2014denoising_4_22}.
Compared to gray or colour images, HSIs own hundreds of spectral bands.
The distributions and variances of the noises in each band are even different.
Then, for HSI restoration problems, LRTV model \cite{LRTV} uses TV regularization to preserve the spatial sparsity bands by bands.
Lou et al. \cite{L1-2TV} propose the $L_{1-2}$TV regularization to more approximately represent the spatial sparsity in gray-level images.
Xie et al \cite{L12TV_HSI_denoising} substitute the classical TV for the $L_{1-2}$TV in LRTV model and reach better restoration results.
Recently, Chang et al. \cite{chang_anisotropic_2015_8_5} extend the classical TV to the 3-D anisotropic spatial-spectral total variation (SSTV)
which regularizes not only the spatial sparsity but also the spectral sparsity.
Notably, although the TV-based method excels at single-type noise removal,
they cannot effectively remove mixed noise in HSIs, especially in some heavy noise situations \cite{LRTDTV}.

In this paper, we propose a global SSTV regularized nonconvex local low-rank tensor approximation (LLxRGTV) model for HSI denoising.
The HSI data are first divided into overlapping patches.
Then, from one aspect, the clean HSI data have its underlying local LR tensor property,
even though the real HSI data may not be globally low-rank due to outliers and non-Gaussian noise \cite{HSI_using_tucker}.
According to this fact, we proposed a novel tensor $L_{\gamma}$-norm to formulate the local LR of HSIs.
From another aspect, HSIs are assumed to be piecewisely smooth in the global spatial domain.
The TV regularization is effective in preserving the spatial piecewise smoothness and removing Gaussian noise.
These facts inspire the integration of the local tensor LR with TV regularization.
To address the limitations of bandwise TV, we use the SSTV regularization to simultaneously consider global spatial structure and spectral correlation of neighboring bands.

The main contributions of this paper are summarized as follows.
\begin{itemize}
\item A nonconvex LR tensor approximation is proposed and be used to formulate the local LR prior of HSI.
\item Global SSTV regularization is incorporated into the nonconvex local LR model.
      The nonconvex local LR is used to separate the clean local HSI from the sparse noise,
      and the global SSTV regularization is utilized to remove Gaussian noise and to simultaneously consider spatial structure and spectral correlation.
\item An ADMM-based algorithm is designed to efficiently solve the proposed model.
      Experimental results demonstrate that the proposed method clearly improves the denoising results in terms of both quantitative evaluation and visual inspection,
      as compared to the state-of-the-art TV regularized and LR based methods.
\end{itemize}

This paper is organized as follows:
The related works for the denoising task is described in Section \ref{related work}.
Section \ref{key model} gives the proposed model and Section \ref{opti_procedure} lists the optimization procedure of the proposed model.
Section \ref{results} includes experimental results and discussions. Finally, Section \ref{conclusion} concludes the paper.

\section{Notations and problem formulation}
\label{related work}

\subsection{Notations}

The Discrete Fourier Transformation (DFT) plays a core role in t-prod.
We give some related background knowledge and notations here.
The DFT on $\boldsymbol{v} \in \mathbb{R}^{n}$ denoted as $\widehat{\boldsymbol{v}},$ is given by
\begin{equation}
\label{equation:computefft}
  \widehat{\boldsymbol{v}}=\boldsymbol{F}_{n} \boldsymbol{v} \in \mathbb{C}^{n},
\end{equation}
where $\boldsymbol{F}_{n}$ is the DFT matrix defined as
\begin{equation}
  \boldsymbol{F}_{n}=\left[\begin{array}{ccccc}1 & 1 & 1 & \cdots & 1 \\ 1 & \omega & \omega^{2} & \cdots & \omega^{n-1} \\ \vdots & \vdots & \vdots & \ddots & \vdots \\ 1 & \omega^{n-1} & \omega^{2(n-1)} & \cdots & \omega^{(n-1)(n-1)}\end{array}\right] \in \mathbb{C}^{n \times n},
\end{equation}
where $\omega=\mathrm{e}^{-\frac{2 \pi i}{n}}$ is a primitive $n$-th root of unity in which $i=\sqrt{-1} .$

Computing $\widehat{\boldsymbol{v}}$ by using (\ref{equation:computefft}) costs $O\left(n^{2}\right)$.
A more widely used method is the Fast Fourier Transform (FFT) whichcosts $O(n \log n)$.
By using the Matlab command fft, we have $\widehat{\boldsymbol{v}}=$ fft $(\boldsymbol{v}) .$
For $\mathcal{L} \in \mathbb{R}^{m \times n \times p},$ we denote $\widehat{\mathcal{L}} \in \mathbb{C}^{m \times n \times p}$ as the result of FFT on $\mathcal{L}$ along the 3-rd dimension,
i.e., performing the FFT on all the tubes of $\mathcal{L} .$
Then, we have
\begin{equation}
\widehat{\mathcal{L}}=\operatorname{fft}(\mathcal{L},[], 3).
\end{equation}
Similarly, we can compute $\mathcal{L}$ from $\widehat{\mathcal{L}}$ by using the inverse FFT, i.e.,
\begin{equation}
\mathcal{L}=\operatorname{ifft}(\widehat{\mathcal{L}},[], 3).
\end{equation}

Let $\overline{\mathcal{L}}$ denote the block-diagonal matrix of the tensor $\mathcal{L}$ in the Fourier domain, i.e.
\begin{equation}
  \begin{array}{l}
\overline{\mathcal{L}} \triangleq \text { blockdiag }(\widehat{\mathcal{L}}) \\
\triangleq\left[\begin{array}{cccc}
\widehat{\mathcal{L}}^{(1)} & & & \\
& \widehat{\mathcal{L}}^{(2)} & & \\
& & \ddots & \\
& & & \widehat{\mathcal{L}}^{\left(p\right)}
\end{array}\right] \in \mathbb{C}^{m p \times n p},
\end{array}
\end{equation}
where $\widehat{\mathcal{L}}^{(i)}$ denotes the $i$-th frontal slices of $\widehat{\mathcal{L}}$, $i=1,2,\cdots,p$.

\subsection{Problem formulation}

The observed HSI is corrupted by mixed noise which typically consists of Gaussian noise, stripes, impulse noise, deadlines and so on \cite{LRMR,NAILRMA}.
Let 3rd-order tensor $\mathcal{O} \in \mathbb{R}^{m \times n \times p}$ denote the observed HSI,
where the spatial information lies in the first two dimensions and the spectral information lies in the third dimension.
Then the degradation model of the HSI can be formulated as
\begin{equation}
\mathcal{O}=\mathcal{L}+\mathcal{S}+\mathcal{N},
\label{eq:framework}
\end{equation}
where $\mathcal{O}, \mathcal{L}, \mathcal{S}, \mathcal{N} \in \mathbb{R}^{m \times n \times p}$;
$\mathcal{O}$ denotes the observed HSI;
$\mathcal{L}$ represents the clean HSI data;
$\mathcal{S}$ is the sparse noise which consists of impulse noise, stripes, deadline and so on;
$\mathcal{N}$ is the Gaussian noise;
$m \times n$ is the spatial size of the HSI, and $p$ is the number of spectral bands.

Under the framework of degradation model (\ref{eq:framework}),
HSI denoising is a process of separating the mixed noise $\mathcal{S}, \mathcal{N}$ from the observed HSI $\mathcal{O}$, and restoring the clear HSI $\mathcal{L}$.
In mathematical theory, this is a serious ill-posed problem.
The regularization method is an effectively and widely used method for solving such inverse problems.
It establishes the following regularized restoration framework by adding prior information of the unknown clear HSI and mixed noise, i.e.,
\begin{equation}
\arg \min _{\mathcal{L}} \mathrm{J}(\mathcal{S}, \mathcal{N})+\tau \mathrm{R}(\mathcal{L}),
\label{regularframework}
\end{equation}
where $\mathrm{J}(\mathcal{S}, \mathcal{N})$ is a regular term to describe the distribution of different noises;
$\mathrm{R}(\mathcal{L})$ is a regular term to represent the prior information of unknown clear HSIs;
$\tau$ is a non-negative regularization parameter used to balance two regular terms.
In restoration framework (\ref{regularframework}), both prior information and the formulations of regular terms are important,
which determine the accuracy of the restoration results.
Therefore, the research on the HSI restoration mainly focuses on the exploration of prior information and the improvement of regularization formulations.

\begin{figure*}
	\centering
	\includegraphics[width=0.95\linewidth]{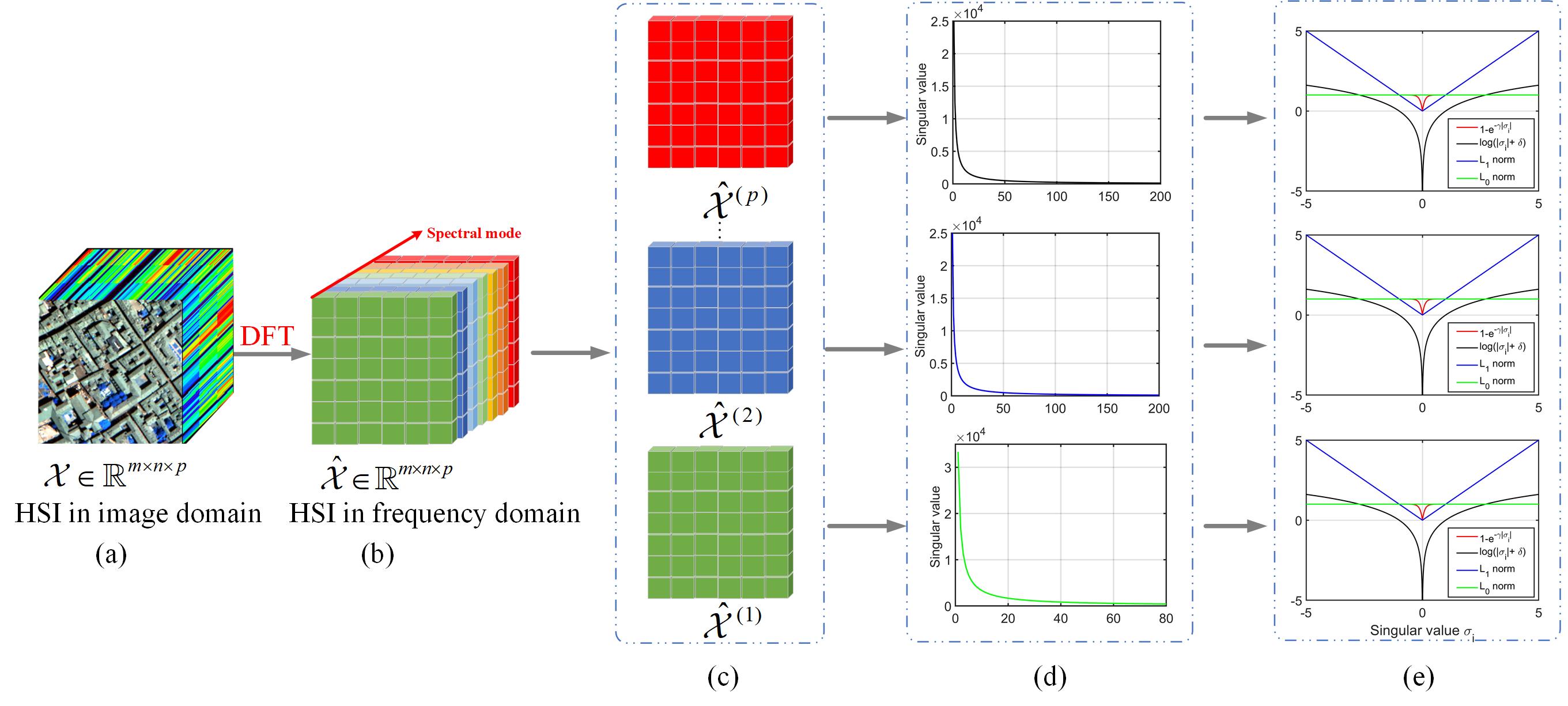}
	\caption{Flowchart of the proposed low-rank HSI approximation. (a) HSI in image domain, denoted as $\mathcal{X}$ . (b) HSI in the frequency domain by
		performing the Discrete Fourier Transformation (DFT) along spectral mode, denoted as $\widehat{\mathcal{X}}$. (c) All frontal slices of $\widehat{\mathcal{X}}$ (along the spectral mode). (d) Singular value curves of frontal slices in $\widehat{\mathcal{X}}$. (e) Different approximations of $L_0$ norm (true rank). }
	\label{figure: approximaterank}
\end{figure*}

\section{HSI denoising via global SSTV regularized local LR tensor recovery}
\label{key model}
\subsection{Low-rank approximation for HSI denoising}

Based on framework (\ref{regularframework}), many classic HSI denoising methods have been proposed.
Among them, RPCA-based methods are widely used.
The RPCA model \cite{RPCA_convex} uses the rank function of the matrix to describe the LR prior of the image,
and uses the $L_0$ norm to describe the sparsity of the noise,
i.e., $\mathrm{J}(\mathcal{S}$, $\mathcal{N})= \lambda\|S\|_{0}$,
$\mathrm{R}(\mathcal{L})=\operatorname{rank}(L)$.
It can be formulated as
\begin{equation}
\label{equation:matrix_rank}
\begin{array}{rl}
\displaystyle \arg \min _{L} & \operatorname{rank}(L) + \lambda\|S\|_{0}\\
s.t. & O=L+ S,
\end{array}
\end{equation}
where $O, L, S \in \mathbb{R}^{mn \times p}$ are the Casorati matrices of $\mathcal{O}$, $\mathcal{L}$ and $\mathcal{S}$, respectively.
The optimization problem in (\ref{equation:matrix_rank}) is a nonconvex optimization problem,
since the function $\operatorname{rank}(L)$ is nonconvex.
One common approach is to use the matrix nuclear norm $\|\cdot\|_{*}$ to approximate the $\operatorname{rank}(L)$.
The advantage of the nuclear norm is that it is the tightest convex envelop for the rank of matrices.
This leads to the following convex optimization problem \cite{guaranteed_NN, power_NN, SVT_NN}:
\begin{equation}
\begin{array}{rl}
\displaystyle \arg \min_{L} & \|L\|_{*} +\lambda\|S\|_{1}\\
s.t. & O=L+S+N,
\end{array}
\end{equation}
where $\|L\|_{*}=\sum_{i} \sigma_{i}(L)$ and $\sigma_{i}(L)$ is the $i$-th singular value of the matrix $L$;
$\|S\|_{1}=\sum_{i, j}|S_{i, j}|$; $N$ is an additional noise term representing independent and identically distributed Gaussian noise.

Although the above model and its extended models have been successfully used in various applications, they need to convert 3-D HSI into 2-D matrices.
This strategy will introduce loss of useful multiorder structure information.
Recent years, many studies \cite{liu_tensor_2012_lrtdtv_18,yuan_tensor_2016_lrtdtv_19,cao_total_2016_lrtdtv_20,anandkumar_tensor_2016_lrtdtv_21}
have proven that denoising methods directly modeling tensors can achieve better results than the ones modeling the tensor’ matriczation.
The tensor is the generalization of the matrix concept.
We generalize the denoising algorithm for the matrix (i.e., 2-order tensor) case to 3-order tensors by solving the following optimization problem:
\begin{equation}
\label{equation:tensor_rank}
\begin{array}{rl}
\displaystyle \arg \min _{\mathcal{L}} & \operatorname{rank}(\mathcal{L}) + \lambda\|\mathcal{S}\|_0\\
s.t. & \mathcal{O}=\mathcal{L}+\mathcal{S}+\mathcal{N},
\end{array}
\end{equation}
where $\mathcal{O}, \mathcal{L}, \mathcal{S}, \mathcal{N}$ are 3-order tensors.
The first issue is how to define or approximate $\operatorname{rank}(\mathcal{L})$ in the case of tensors,
and there is not much work on this issue.
In these rare works, the TNN based on t-SVD is a classic method, in which the sum of singular values of all the frontal slices of $\widehat{\mathcal{L}}$ is used to approximate $\operatorname{rank}(\mathcal{L})$.
The TNN based optimization problem can be formulated as
\begin{equation}
\begin{array}{rl}
\displaystyle \arg \min _{\mathcal{L}} & \left\|\mathcal{L}\right\|_{*} + \lambda\|\mathcal{S}\|_1\\
s.t. & \mathcal{O}=\mathcal{L}+\mathcal{S}+\mathcal{N}.
\end{array}
\end{equation}

Although TNN based method reported success on low-rank completion and HSI denoising.
TNN is induced by t-SVD which is defined based on tensor-tensor product (t-prod).
The structure of t-prod shows that it is equivalent to the matrix-matrix product in the Fourier transform domain.
Therefore, TNN is essentially the matrix nuclear norm of the block diagonal unfolding of the Fourier transformed tensor \cite{PSTNN}.
However, in the case of matrices, minimizing the nuclear norm would cause some inevitable deviations \cite{drawback1_TNN, drawback2_TNN}.
For example, the variance of the recovered data would be smaller than original data when equally shrinking every singular value.
Similarly, the estimated results may be lower rank than original data.
To alleviate this issue, we propose the nonconvex tensor $L_{\gamma}$ norm of $\mathcal{L} \in \mathbb{R}^{m \times n \times p}$ to approximate $\operatorname{rank}(\mathcal{L})$ as follows:
\begin{equation}
\label{LxTNN}
  \left\|\mathcal{L}\right\|_{\gamma}=\left\|\overline{\mathcal{L}}\right\|_{\gamma}=\left\|\text{ blockdiag }(\widehat{\mathcal{L}})\right\|_{\gamma}=\frac{1}{p} \sum_{i=1}^{p}\left\|\widehat{\mathcal{L}}^{(i)}\right\|_{\gamma},
\end{equation}
where $$\left\|\widehat{\mathcal{L}}^{(i)}\right\|_{\gamma}=\sum_{i=1}^{\min \{m, n\}}\left(1-e^{- \gamma \left|\sigma_{i}( \widehat{\mathcal{L}}^{(i)})\right|}\right).$$

To better understand the novel $L_{\gamma}$ norm, we plot the function curves
corresponding to $L_{\gamma}$ norm of frontal slice $\widehat{\mathcal{L}}^{(i)}$
in comparison with the $L_1$ norm, the $L_0$ norm and log-sum function \cite{nonconvex_log_sum} in Fig. \ref{figure: approximaterank}.
It can be seen from the fifth column of Fig. \ref{figure: approximaterank} that when the singular value is greater than 1, the blue nuclear norm deviates significantly from 1, indicating that it excessively reduces the rank component.
In contrast, the red $L_{\gamma}$ norm matches well with the $L_0$ norm, which means that the it is closer to the green $L_0$ norm than the nuclear norm.
Under definition (\ref{LxTNN}), the optimization in (\ref{equation:tensor_rank}) can be written as:
\begin{equation}
\label{equation:LxTNN_arg}
\begin{array}{rl}
\displaystyle \arg \min _{\mathcal{L}} & \left\|\mathcal{L}\right\|_{\gamma} + \lambda\|\mathcal{S}\|_1\\
s.t. & \mathcal{O}=\mathcal{L}+\mathcal{S}+\mathcal{N}.
\end{array}
\end{equation}

\subsection{The proposed local LR model}

\begin{figure}
	\centering
	\includegraphics[width=0.7\linewidth]{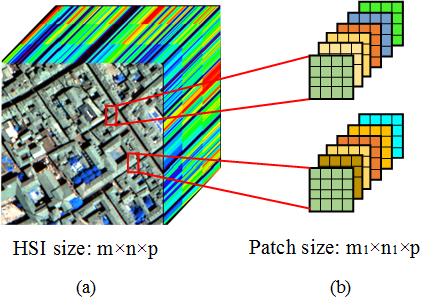}
	\caption{Formulation of the local low-rank patches from an HSI.}
	\label{local_low_rank}
\end{figure}

Although (\ref{equation:LxTNN_arg}) can be directly used to remove the mixed noise in HSIs,
as traditional LR based models are proposed in literatures, it needs to assume that the underlying HSI cube has the LR tensor property.
However, in reality, the observed HSIs may not be globally low-rank due to outliers and non-Gaussian noise \cite{HSI_using_tucker}.
Fortunately, pixels from the same local area are more likely to belong to the same material, and the spectral signatures of the same material are more likely to be the same.
These facts mean that the clean HSI data have its underlying local LR tensor property, even though the entire HSIs may not be low-rank \cite{HSI_using_tucker}.
Therefore, to effectively explore the local LR structure of underlying HSIs and reduce the impact of noise and outliers on the LR hypothesis, we denoise HSIs patch by patch.
Firstly, we define an operator $\mathrm{P}_{i, j}:\mathcal{L} \rightarrow \mathcal{L}_{i, j}$.
This binary operator $\mathrm{P}_{i, j}$ is used to extract $m_1 \times n_1 \times p$ patches from HSI data $\mathcal{L} \in \mathbb{R}^{m \times n \times p}$, i.e., $\mathcal{L}_{i, j}=\mathrm{P}_{i, j}(\mathcal{L})$,
where the spatial size of $m_1 \times n_1$ is centralized at pixel $(i, j)$ of HSI data, $(i,j)\in[1,m-m_1+1]\times[1,n-n_1+1]$.
$\mathrm{P}_{i, j}^{\text{T}}$ is the inverse of $\mathrm{P}_{i, j}$. The process is shown in Fig. \ref{local_low_rank}.

Based on binary operator $\mathrm{P}_{i, j}$, we first use the proposed tensor $L_{\gamma}$ norm (\ref{LxTNN}) to represent the local LR property of HSI data,
i.e.,  performing (\ref{equation:LxTNN_arg}) on all the patches of $\mathcal{L}$.
In addition,
to further model Gaussian noise and enhance the performance of the proposed model in some heavy Gaussian noise situations,
we introduce the Frobenius norm into the local version of model (\ref{equation:LxTNN_arg}) to describe the Gaussian noise in patch $\mathcal{N}_{i,j}$.
Then, local patch-based (\ref{equation:LxTNN_arg}) is formulated as follows:
\begin{equation}
\label{eq:LTRPCA}
\begin{split}
\displaystyle \arg \min_{\mathcal{L}_{i,j}, \mathcal{S}_{i,j}, \mathcal{N}_{i,j}} &~  \sum_{i,j} \left\|\mathcal{L}_{i,j}\right\|_{\gamma}+\lambda\left\|\mathcal{S}_{i, j}\right\|_{1}+\beta\left\|\mathcal{N}_{i, j}\right\|_{\mathrm{F}}^2 \\
s.t.&~ \mathcal{O}_{i, j}=\mathcal{L}_{i, j}+\mathcal{S}_{i, j}+\mathcal{N}_{i, j}.
\end{split}
\end{equation}

\subsection{Global SSTV regularized local LR model}

The proposed denoising model (\ref{eq:LTRPCA}) is a local LR model,
which only exploits the local LR tensor property of HSIs and ignores the global geometrical structure in spectral and spatial domains.
Considering that TV-based methods can effectively remove the Gaussian noise and preserve the spatial piecewise smoothness of HSIs,
we are inspired to incorporate TV regularization into the proposed local LR model.

For HSI $\mathcal{L}\in\mathbb{R}^{m \times n \times p}$, a bandwise TV is commonly used as follows:
\begin{equation}
\|\mathcal{L}\|_{\mathrm{HTV}}=\sum_{j=1}^{p}\left\|\mathcal{L}^{(j)}\right\|_{\mathrm{TV}}.
\end{equation}
However, it only studies the spatial sparsity and ignores the spectral sparsity.
To address this limitation, an additional 1-D finite difference operation is applied in the spectral dimension in the SSTV model \cite{LRTDTV}, i.e.,
\begin{equation}
\label{equation:the SSTV}
\begin{aligned}
\|\mathcal{L}\|_{\text {SSTV }}:=& \sum_{i, j, k} w_{1}\left|l_{i, j, k}-l_{i, j, k-1}\right|+w_{2}\left|l_{i, j, k}-l_{i, j-1, k}\right| \\ &+w_{3}\left|l_{i, j, k}-l_{i-1, j, k}\right|,
\end{aligned}
\end{equation}
where the $w_{i}(i=1,2,3)$ in (\ref{equation:the SSTV}) is the weight along the $i$-th mode of $\mathcal{L}$ that controls its regularization strength.

Finally, by combining the local LR and spectral-spatial sparse properties in both spatial and spectral domains,
we propose the following LLxRGTV model
\begin{equation}
\label{eq:TLRL1-2LSSTV}
\begin{split}
\displaystyle\arg\min_{\mathcal{L},\mathcal{S}} &~ \sum_{i, j}\left(\left\|\mathcal{L}_{i, j}\right\|_{\gamma}+\lambda\left\|\mathcal{S}_{i, j}\right\|_{1}+\beta\left\|\mathcal{N}_{i, j}\right\|_{\mathrm{F}}^2 \right) \\
&~ +\tau\|\mathcal{L}\|_ \mathrm{SSTV}  \\
s.t.&~\mathcal{O}_{i, j}=\mathcal{L}_{i, j}+\mathcal{S}_{i, j}+\mathcal{N}_{i, j}.
\end{split}
\end{equation}
In the next section, we design algorithms to efficiently solve the proposed model. In addition, the convergence and parameter analysis of the proposed model will be discussed in detail in Section \ref{discussion}.

\section{Optimization procedure}
\label{opti_procedure}
In this sention, we use the alternating direction method of multipliers (ADMM) to solve the proposed model model (\ref{eq:TLRL1-2LSSTV}).
Auxiliary variables $\mathcal{J}, \mathcal{X} \in \mathbb{R}^{m \times n \times p}$ are first introduced,
and the proposed model (\ref{eq:TLRL1-2LSSTV}) can be rewritten as following:
\begin{equation}
\nonumber
\begin{split}
\arg\min_{\mathcal{L}, \mathcal{S}, \mathcal{J}, \mathcal{X}} & \sum_{i, j}\left( \left\|\mathcal{L}_{i, j}\right\|_{\gamma}+\lambda\left\|\mathcal{S}_{i, j}\right\|_{1}+\beta\left\|\mathcal{N}_{i, j}\right\|_{\mathrm{F}}^2\right)\\
&~ +\tau\|\mathcal{X}\|_ \mathrm{SSTV}\\
s.t.&~\mathcal{J}=\mathcal{X}, \mathcal{L}_{i, j}=\mathcal{J}_{i, j}, \mathcal{U}=\textbf{D}\mathcal{X},\\
&~\mathcal{O}_{i, j}=\mathcal{L}_{i, j}+\mathcal{S}_{i, j}+\mathcal{N}_{i, j},
\end{split}
\end{equation}
where $\mathbf{D}(\cdot)=\left[w_{1} \times \mathbf{D}_{1}(\cdot) ; w_{2} \times \mathbf{D}_{2}(\cdot) ; w_{3} \times \mathbf{D}_{3}(\cdot)\right]$ is the so-called weighted three-dimensional difference operator
and $\mathbf{D}_{1}, \mathbf{D}_{2}, \mathbf{D}_{3}$ are the first-order difference operators with respect to three different directions of a HSI cube.
By the augmented Lagrangian multiplier (ALM) method, the above optimization model can be rewritten as:
\begin{equation}
\label{eq:ALM-opt}
\begin{aligned}
&\arg\min_{\mathcal{L}, \mathcal{S}, \mathcal{J}, \mathcal{X}, \mathcal{N}}  \ell(\mathcal{L}, \mathcal{S}, \mathcal{J}, \mathcal{X}, \mathcal{N}) \\
&=\arg\min_{\mathcal{L}, \mathcal{S}, \mathcal{J}, \mathcal{X}, \mathcal{N}} \sum_{i,j} \left( \left\|\mathcal{L}_{i,j}\right\|_{\gamma} +\lambda\left\|\mathcal{S}_{i, j}\right\|_{1}+\beta\left\|\mathcal{N}_{i, j}\right\|_{\mathrm{F}}^2 \right. \\
&+<\Lambda_{i, j}^{\mathcal{O}}, \mathcal{O}_{i, j}-\mathcal{L}_{i, j}-\mathcal{S}_{i, j}-\mathcal{N}_{i, j}>  +\frac{\mu}{2}\left\|\mathcal{L}_{i, j}-\mathcal{J}_{i, j}\right\|_{\mathrm{F}}^{2}\\
&\left. +\frac{\mu}{2} \| \mathcal{O}_{i, j}-\mathcal{L}_{i, j}-\mathcal{S}_{i, j}-\mathcal{N}_{i, j} \|_{\mathrm{F}}^{2}+<\Lambda_{i, j}^{\mathcal{L}}, \mathcal{L}_{i, j}-\mathcal{J}_{i, j}> \right)\\
&+<\Lambda, \mathcal{U}-\textbf{D}\mathcal{X}>+\frac{\mu}{2}\|\mathcal{U}-\textbf{D}\mathcal{X}\|_{\mathrm{F}}^{2} +\tau\|\mathcal{U}\|_1 \\
&+<\Lambda_{\mathcal{X}}, \mathcal{J}-\mathcal{X}>+\frac{\mu}{2}\|\mathcal{J}-\mathcal{X}\|_{\mathrm{F}}^{2},
\end{aligned}
\end{equation}
where $\mu$ is the penalty parameter; $\Lambda, \Lambda_{i, j}^{\mathcal{O}}, \Lambda_{i, j}^{\mathcal{L}}$ and $\Lambda_{\mathcal{X}}$ are the Lagrangian multipliers.
The above minimization (\ref{eq:ALM-opt}) can be solved by the ADMM method.
In the $k$-th iteration, its solution in the next iteration can be transformed into the following two subproblems:
\begin{equation}
\label{eq:ADMM1}
\left(\mathcal{L}^{k+1}, \mathcal {S}^{k+1}, \mathcal {N}^{k+1}\right) =\arg \min_{\mathcal{L}, \mathcal{S}, \mathcal{N}} \ell\left(\mathcal{L}, \mathcal {S}, \mathcal{N}, \mathcal {J}^{k}\right),
\end{equation}
\begin{equation}
\label{eq:ADMM2}
\left(\mathcal{J}^{k+1}, \mathcal{X}^{k+1}, \mathcal{U}^{k+1}\right) =\arg \min_{\mathcal{J}, \mathcal{X}, \mathcal{U}} \ell\left(\mathcal{L}^{k+1}, \mathcal {J}, \mathcal{X}, \mathcal{U}\right).
\end{equation}
The subproblem (\ref{eq:ADMM1}) can be regraded as the local LR model defined in (\ref{eq:LTRPCA});
Meanwhile, the subproblem (\ref{eq:ADMM2}) is the global SSTV regularization problem with respect to the denoised patches in subproblem (\ref{eq:ADMM1}).

\subsection{Local LR optimization for $(\mathcal{L}, \mathcal{S}, \mathcal{N})$}
With the other variables fixed, the subproblem (\ref{eq:ADMM1}) for $(\mathcal{L}, \mathcal{S}, \mathcal{N})$ can be reformulated as
\begin{equation}
\begin{split}
&\arg \min_{\mathcal{L}, \mathcal{S}, \mathcal{N}} \ell\left(\mathcal{L}, \mathcal {S}, \mathcal{N}, \mathcal {J}^{k}\right) \\
& =\arg \min_{\mathcal{L}_{i,j}, \mathcal{S}_{i,j}, \mathcal{N}_{i,j}} \sum_{i, j}\left\|\mathcal{L}_{i, j}\right\|_{\gamma}+\lambda\left\|\mathcal{S}_{i, j}\right\|_{1} + \beta\left\|\mathcal{N}_{i, j}\right\|_{1}\\
&+<\Lambda_{i, j}^{\mathcal{O}}, \mathcal{O}_{i, j}-\mathcal{L}_{i, j}-\mathcal{S}_{i, j}> + \frac{\mu}{2} \| \mathcal{O}_{i, j}-\mathcal{L}_{i, j}-\mathcal{S}_{i, j} \|_{\mathrm{F}}^{2} \\
&+<\Lambda_{i, j}^{\mathcal{L}}, \mathcal{L}_{i, j}-\mathcal{J}_{i, j}>+\frac{\mu}{2}\left\|\mathcal{L}_{i, j}-\mathcal{J}_{i, j}\right\|_{\mathrm{F}}^{2}.
\end{split}
\end{equation}
For the above optimization problem, we solve each patch separately and accumulate a weighted sum of $\left(\mathcal{L}_{i, j}, \mathcal{S}_{i, j}, \mathcal{N}_{i, j}\right)$ to reconstruct $(\mathcal{L}, \mathcal{S}, \mathcal{N})$.
With respect to each $\left(\mathcal{L}_{i, j}, \mathcal{S}_{i, j}, \mathcal{N}_{i, j}\right)$, we have the following patchwise optimization problem
\begin{equation}
\label{eq:LSmodel}
\begin{split}
&\arg \min_{\mathcal{L}_{i,j}, \mathcal{S}_{i,j}, \mathcal{N}_{i,j}} \left\|\mathcal{L}_{i, j}\right\|_{\gamma}+\lambda\left\|\mathcal{S}_{i, j}\right\|_{1}+\beta\left\|\mathcal{N}_{i, j}\right\| \\
&+<\Lambda_{i, j}^{\mathcal{O}}, \mathcal{O}_{i, j}-\mathcal{L}_{i, j}-\mathcal{S}_{i, j}-\mathcal{N}_{i, j}> +\frac{\mu}{2}\left\|\mathcal{L}_{i, j}-\mathcal{J}_{i, j}\right\|_{\mathrm{F}}^{2} \\
&+ \frac{\mu}{2} \| \mathcal{O}_{i, j}-\mathcal{L}_{i, j}-\mathcal{S}_{i, j}-\mathcal{N}_{i, j} \|_{\mathrm{F}}^{2} +<\Lambda_{i, j}^{\mathcal{L}}, \mathcal{L}_{i, j}-\mathcal{J}_{i, j}>.
\end{split}
\end{equation}
We alternately update the three variables $\mathcal{L}_{i, j}$, $\mathcal{S}_{i, j}$ and $\mathcal{N}_{i, j}$.
Then, the optimization problem (\ref{eq:LSmodel}) can be separated into three simpler minimization subproblems.

\subsubsection{Update $\mathcal{L}_{i, j}$}
With the other variables fixed, the minimization subproblem for $\mathcal{L}_{i, j}$ can be deduced from (\ref{eq:LSmodel}) as follows
\begin{equation}
\label{eq:Lmodel}
\begin{split}
\mathcal{L}_{i, j}^{k+1}= & \arg \min_{ \mathcal{L}_{i, j} } \left\|\mathcal{L}_{i, j}\right\|_{\gamma}+2\times\frac{\mu}{2}\| \mathcal{L}_{i,j}-\frac{1}{2}\left(\mathcal{O}_{i,j}\right. \\
&+\mathcal{J}_{i, j}-\mathcal{S}_{i, j}-\mathcal{N}_{i, j}+\left(\Lambda_{i, j}^{\mathcal{O}}+\Lambda_{i, j}^{\mathcal{L}}\right) / \mu)\|_{\mathrm{F}}^{2}.
\end{split}
\end{equation}
For the sake of simplicity, we denote the iteration of $\mathcal{L}_{i, j}$ in (\ref{eq:Lmodel}) as
\begin{equation}
\label{eq:Lmodel-simpl}
\mathcal{L}_{i, j}^{k+1}=\arg \min_{ \mathcal{L}_{i, j} } \left\|\mathcal{L}_{i, j}\right\|_{\gamma}+2\times\frac{\mu}{2}\left\|\mathcal{L}_{i, j}-\mathcal{M}_{i, j}\right\|_{\mathrm{F}}^{2},
\end{equation}
where
$$\mathcal{M}_{i, j}=\frac{1}{2}\left(\mathcal{O}_{i, j}+\mathcal{J}_{i, j}-\mathcal{S}_{i, j}-\mathcal{N}_{i, j}+\left(\Lambda_{i, j}^{\mathcal{O}}+\Lambda_{i, j}^{\mathcal{L}}\right) / \mu\right).$$

Solving optimization subproblem (\ref{eq:Lmodel-simpl}) is equivalent to solving the following tensor recovery problem in frequency domain
\begin{equation}
\label{eq:Lmodel-fft}
\widehat{\mathcal{L}}_{i, j}^{k+1}=\arg \min_{ \widehat{\mathcal{L}}_{i, j} } \frac{1}{p}\sum_{q=1}^{p}\left\|\widehat{\mathcal{L}}_{i, j}^{(q)}\right\|_{\gamma}+2\times\frac{\mu}{2}\left\|\widehat{\mathcal{L}}_{i, j}^{(q)}-\widehat{\mathcal{M}}_{i, j}^{(q)}\right\|_{\mathrm{F}}^{2},
\end{equation}
where $\widehat{\mathcal{L}}_{i, j} = \operatorname{fft}\left(\mathcal{L}_{i, j},[~], 3\right)$;
$\widehat{\mathcal{M}}_{i, j} = \operatorname{fft}\left(\mathcal{M}_{i, j},[~], 3\right)$;
$\widehat{\mathcal{L}}^{(q)}_{i, j}$ and $\widehat{\mathcal{M}}^{(q)}_{i, j}$ denotes the $q$-th frontal slice of $\widehat{\mathcal{L}}_{i, j}$ and $\widehat{\mathcal{M}}_{i, j}$, respectively.
It can be broken up to $p$ independent minimization subproblems:
\begin{equation}
\label{eq:Lmodel-fft-slice}
\widehat{\mathcal{L}}_{i, j}^{k+1,(q)}=\arg \min_{\widehat{\mathcal{L}}_{i, j}^{(q)}} \left\|\widehat{\mathcal{L}}_{i, j}^{(q)}\right\|_{\gamma}+
2\times\frac{\mu}{2}\left\|\widehat{\mathcal{L}}_{i, j}^{(q)}-\widehat{\mathcal{M}}_{i,j}^{(q)}\right\|_{\mathrm{F}}^{2}.
\end{equation}

Let $\sigma_{1}^{k} \geq \sigma_{2}^{k} \geq \cdots \geq \sigma_{s}^{k}$ represent the singular values of
$\widehat{\mathcal{L}}_{i, j}^{k,(q)}$ with $s=\min \left\{m_{1}n_{1}, p\right\}$,
$\phi\left(x\right)=1-e^{- \gamma |x|}$ and $\nabla \phi\left(x\right)$ denote the gradient of $\phi$ at point $x$,
$f(\widehat{\mathcal{L}}_{i,j}^{(q)})=(1 / 2)\|\widehat{\mathcal{L}}_{i,j}^{(q)}-\widehat{\mathcal{M}}_{i,j}^{(q)}\|_{\text{F}}^{2}$.
It is easy to prove that the gradient of $f(\widehat{\mathcal{L}}_{i,j}^{(q)})$ is Lipschitz continuous by setting the Lipschitz constant being $1$.
As stated in \cite{Non_LRMA},
considering the nonascending order of singular values and according to the antimonotone property of gradient of our nonconvex function,
we have
\begin{equation}
\begin{aligned}
& 0 \leq \nabla \phi\left(\sigma_{1}^{k}\right) \leq \nabla \phi\left(\sigma_{2}^{k}\right) \leq \cdots \leq \nabla \phi\left(\sigma_{s}^{k}\right), \\
& \phi\left(\sigma_{n}(\widehat{\mathcal{L}}_{i,j}^{(q)})\right) \leq \phi\left(\sigma_{n}^{k}\right)+\nabla \phi\left(\sigma_{n}^{k}\right)\left(\sigma_{n}(\widehat{\mathcal{L}}_{i,j}^{(q)})-\sigma_{n}^{k}\right),
\end{aligned}
\label{equ:gradient}	
\end{equation}
where $n=1,2,\cdots,s$.

Based on (\ref{equ:gradient}), (\ref{eq:Lmodel-fft-slice}) can be converted into following relaxation problem:
\begin{equation}
\begin{split}
& \arg \min _{\widehat{\mathcal{L}}_{i,j}^{(q)}} \frac{1}{2\mu} \sum_{n=1}^{s} \phi\left(\sigma_{n}^{k}\right)+\nabla \phi\left(\sigma_{n}^{k}\right)\left(\sigma_{n}(\widehat{\mathcal{L}}_{i,j}^{(q)})-\sigma_{n}^{k}\right) \\ &~~~~~~~~~~~~~~~~~~~~~~~~+f(\widehat{\mathcal{L}}_{i,j}^{(q)}) \\
= &  \arg \min _{\widehat{\mathcal{L}}_{i,j}^{(q)}} \frac{1}{2\mu} \sum_{n=1}^{s} \nabla \phi\left(\sigma_{n}^{k}\right) \sigma_{n}(\widehat{\mathcal{L}}_{i,j}^{(q)})+\frac{1}{2}\left\|\widehat{\mathcal{L}}_{i,j}^{(q)}-\widehat{\mathcal{M}}_{i,j}^{(q)}\right\|_{\text{F}}^{2}	
\end{split}
\label{equ:relaxation}
\end{equation}
Then, following \cite{lu_nonconvex, Non_LRMA}, the subproblem (\ref{equ:relaxation}) can be efficiently solved by generalized weight singular value thresholding \cite{WSVT},
as shown in Lemma 1.

\textbf{Lemma 1}:
For any $1/(2\mu)>0$, the given matrix $\widehat{\mathcal{M}}_{i,j}^{(q)}$
and $0 \leq \nabla \phi\left(\sigma_{1}^{k}\right) \leq \nabla \phi\left(\sigma_{2}^{k}\right) \leq \cdots \leq \nabla \phi\left(\sigma_{s}^{k}\right)$,
a globally optimal solution to problem (\ref{equ:relaxation}) is given as follows:
\begin{equation}
\widehat{\mathcal{L}}_{i,j}^{k+1,(q)} = U \operatorname{S}_{\frac{\nabla \phi}{2\mu}}(\Sigma) V^{T}, ~q=1,2,\cdots,p,
\end{equation}
where $\widehat{\mathcal{M}}_{i,j}^{(q)}=U \Sigma V^{T}$ is the SVD of $\widehat{\mathcal{M}}_{i,j}^{(q)}$, and
\begin{equation}
 \operatorname{S}_{\frac{\nabla \phi}{2\mu}}(\Sigma)=
\operatorname{Diag}\left\{\max \left(\Sigma_{n n}-\frac{\nabla \phi(\sigma_{n}^{k})}{2\mu}, 0\right)\right\}.
\end{equation}

Then, the $(k+1)$-th updating of $\mathcal{L}^{k+1}_{i, j}$ can be obtained via inverse Fourier transform
\begin{equation}
\label{eq:Lmodel-solution}
\mathcal{L}_{i, j}^{k+1} =\operatorname{ifft}\left(\widehat{\mathcal{L}}_{i, j}^{k+1},[~], 3\right).
\end{equation}

\subsubsection{Update $\mathcal{S}_{i, j}$}
With the other variables fixed, the minimization subproblem for $\mathcal{S}_{i, j}$ can be deduced from (\ref{eq:LSmodel}) as follows:
\begin{equation}
\begin{split}
\mathcal{S}_{i, j}^{k+1} = & \arg\min_{\mathcal{S}_{i, j}} \lambda\left\|\mathcal{S}_{i, j}\right\|_{1} \\
&+ \frac{\mu}{2} \|\mathcal{S}_{i, j}- ( \mathcal{O}_{i, j} - \mathcal{L}_{i, j}^{k+1}- \mathcal{N}_{i, j}^{k+1}+\Lambda_{i, j}^{\mathcal{O}} /\mu )\|_{\mathrm{F}}^{2}.
\end{split}
\end{equation}
Its solution can be directly obtained by the soft threshold 
\begin{equation}
\label{eq:Smodel-solution}
\mathcal{S}_{i, j}^{k+1} = \operatorname{Soft}_{\lambda / \mu} \left( \mathcal{O}_{i, j}-\mathcal{L}_{i, j}^{k+1}-\mathcal{N}_{i, j}^{k+1}+\Lambda_{i, j}^{\mathcal{O}} / \mu\right),
\end{equation}
where
\begin{equation}
\nonumber
\operatorname{Soft}_{\delta}(x)=
\left\{
\begin{array}{ll}
x-\delta, &\text{ if } x>\delta, \\
x+\delta, &\text { if } x<-\delta, \\
0,        &\text { otherwise. }
\end{array}
\right.
\end{equation}

\subsubsection{Update $\mathcal{N}_{i, j}$}
With the other variables fixed, the subproblem for $\mathcal{N}_{i,j}$ from (\ref{eq:LSmodel}) can be reformulated as
\begin{equation}
\begin{array}{l}
\displaystyle \arg \min_{\mathcal{N}_{i,j}} \beta\|\mathcal{N}_{i,j}\|_{\mathrm{F}}^{2}+\left\langle\Lambda_{i, j}^{\mathcal{O}}, \mathcal{O}_{i,j}-\mathcal{L}^{k+1}_{i,j}-\mathcal{S}^{k+1}-\mathcal{N}_{i,j}\right\rangle \\
~~~~~~~~~~~~~+\frac{\mu}{2}\left\|\mathcal{O}_{i,j}-\mathcal{L}^{k+1}_{i,j}-\mathcal{S}^{k+1}_{i,j}-\mathcal{N}_{i,j}\right\|_{\mathrm{F}}^{2}\\	=\underset{\mathcal{N}_{i,j}}{\operatorname{argmin}}\left(\beta+\frac{\mu}{2}\right)\left\|\mathcal{N}_{i,j}-\frac{\mu\left(\mathcal{O}_{i,j}-\mathcal{L}^{k+1}_{i,j}-\mathcal{S}^{k+1}_{i,j}\right)+\Lambda_{i, j}^{\mathcal{O}}}{\mu+2 \beta}\right\|_{\mathrm{F}}^{2}.
	\end{array}
\end{equation}
Its closed-form solution can be easily obtained as follows:
\begin{equation}
\label{equation:solution of N}
	\mathcal{N}_{i,j}^{k+1}=\frac{\mu\left(\mathcal{O}_{i,j}-\mathcal{L}^{k+1}_{i,j}-\mathcal{S}^{k+1}_{i,j}\right)+\Lambda_{i, j}^{\mathcal{O}}}{\mu+2 \beta}.
\end{equation}

\subsection{Global SSTV regularization problem for $(\mathcal{J}, \mathcal{X}, \mathcal{U})$}
With the other variables fixed, the subproblem (\ref{eq:ADMM2}) for $(\mathcal{J}, \mathcal{X}, \mathcal{U})$ can be reformulated as
\begin{equation}
\label{eq:JXmodel}
\begin{split}
& \arg \min_{\mathcal{J}, \mathcal{X}, \mathcal{U}}  \ell\left(\mathcal{L}^{k+1}, \mathcal {J}, \mathcal{X}, \mathcal{U} \right) \\
& = \arg \min_{\mathcal{J}, \mathcal{X}, \mathcal{U}} \sum_{i,j} \left( <\Lambda_{i, j}^{\mathcal{L}}, \mathcal{L}_{i, j}-\mathcal{J}_{i, j}>  \right. \left. +\frac{\mu}{2}\left\|\mathcal{L}_{i, j}-\mathcal{J}_{i, j}\right\|_{\mathrm{F}}^{2} \right) \\
& +<\Lambda_{\mathcal{X}}, \mathcal{J}-\mathcal{X}>+\frac{\mu}{2}\|\mathcal{J}-\mathcal{X}\|_{\mathrm{F}}^{2}+<\Lambda, \mathcal{U}-\textbf{D}\mathcal{X}>\\
& +\frac{\mu}{2}\|\mathcal{U}-\textbf{D}\mathcal{X}\|_{\mathrm{F}}^{2}+\tau\|\mathcal{U}\|_1.
\end{split}
\end{equation}
Similarly, we alternately update the three variables $\mathcal{J}$, $\mathcal{X}$ and $\mathcal{U}$,
then the optimization problem (\ref{eq:JXmodel}) can be separated into three simpler minimization subproblems.

\subsubsection{Update $\mathcal{J}$}
With  the other variables fixed, the minimization subproblem for $\mathcal{J}$ can be deduced from (\ref{eq:JXmodel}) as follows:
\begin{equation}
\begin{split}
\arg \min _{\mathcal{J}} &\frac{\mu}{2}\left\|\mathcal{J}-\mathcal{X}+\Lambda_{\mathcal{X}} / \mu\right\|_{2}^{2} \\
&+\sum_{i, j}\left(\frac{\mu}{2}\left\|\mathcal{L}_{i, j}-\mathcal{J}_{i, j}+\Lambda_{i, j}^{\mathcal{L}} / \mu\right\|_{F}^{2}\right).
\end{split}
\end{equation}
It is convex and has the following closed-form solution
\begin{equation}
\label{eq:Jmodel-solution}
\begin{split}
\mathcal{J}= & \left(\mathcal{X}-\Lambda_{\mathcal{X}}/\mu + \sum_{i, j} \operatorname{P}_{i, j}^{\text{T}}\left(\mathcal{L}_{i, j}+\Lambda_{i, j}^{\mathcal{L}} / \mu\right)\right)\\
& ./\left(\mathbf{1}+\sum_{i, j} \operatorname{P}_{i, j}^{\text{T}} \operatorname{P}_{i, j}\right),
\end{split}
\end{equation}
where $\mathbf{1}$ stands for an all-one tensor of size $m\times n \times p$.

\subsubsection{Update $\mathcal{X}$}
With the other variables fixed, the minimization subproblem for $\mathcal{X}$ can be deduced from (\ref{eq:JXmodel}) as follows:
\begin{equation}
\arg\min_{\mathcal{X}} \frac{\mu}{2}\|\mathcal{U}-\mathbf{D} \mathcal{X}+\Lambda / \mu\|_{2}^{2}+\frac{\mu}{2}\left\|\mathcal{J}-\mathcal{X}+\Lambda_{\mathcal{X}} / \mu\right\|_{2}^{2}.
\end{equation}
It can be solved by considering the following normal equation:
\begin{equation}
\left(\mathbf{D}^{\text{T}} \mathbf{D}+\mathbf{1}\right) \mathcal{X}=\mathbf{D}^{\text{T}}(\mathcal{U}+\Lambda / \mu)+\left(\mathcal{J}+\Lambda_{\mathcal{X}} / \mu\right)
\end{equation}
which can be efficiently solved by the fast Fourier transform method:
\begin{equation}
\label{eq:Xmodel-solution}
\mathcal{X}=\mathcal{F}^{-1}\left[\frac{\mathcal{F}\left(\left(\mathcal{J}+\Lambda_{\mathcal{X}} / \mu\right)+\mathbf{D}^{\text{T}}(\mathcal{U}+\Lambda / \mu)\right)}{1+\sum_{i=1}^{3}\left(\mathcal{F}\left(w_{i} \mathbf{D}_{i}\right)\right)^{2}}\right]
\end{equation}
where $\mathcal{F}(\cdot)$ denotes the FFT, and $\mathcal{F}^{-1}$ is the inverse FFT;
$\mathbf{D}^{\text{T}}$ represents the adjoint operator of $\mathbf{D}$.

\subsubsection{Update $\mathcal{U}$}
With the other variables fixed, the minimization subproblem for $\mathcal{U}$ can be deduced from (\ref{eq:JXmodel}) as follows:
\begin{equation}
\begin{array}{l}
\displaystyle \arg\min_{\mathcal{U}} \tau\|\mathcal{U}\|_{1}+\langle \Lambda, \mathcal{U}-\mathbf{D} \mathcal{X}\rangle+\frac{\mu}{2}\|\mathcal{U}-\mathbf{D} \mathcal{X}\|_{2}^{2} \\
\displaystyle \arg\min_{\mathcal{U}} \tau\|\mathcal{U}\|_{1}+\frac{\mu}{2}\|\mathcal{U}-\mathbf{D} \mathcal{X}+\Lambda / \mu\|_{2}^{2},
\end{array}
\label{equ:U_function}
\end{equation}
where $\Lambda=\left[\Lambda_{1}, \Lambda_{2}, \Lambda_{3}\right]$ and $\mathcal{U}=\left[\mathcal{U}_{1}, \mathcal{U}_{2}, \mathcal{U}_{3}\right]$.
Likewise, the optimization (\ref{equ:U_function}) can be solved by the  soft threshold operator defined in (\ref{eq:Smodel-solution}) as follows:
\begin{equation}
\label{eq:Umodel-solution}
\mathcal{U}_{i}=\operatorname{Soft}_{\frac{\tau}{\mu}} \left(w_{i} \mathbf{D}_{i} \mathcal{X}-\Lambda_{1} / \mu\right), i=1,2,3.
\end{equation}

\subsection{Updating the Lagrangian parameters}
Finally, the Lagrangian parameters can be updated as follows:
\begin{equation}
\label{eq:Lagparas}
\left\{
\begin{split}
& \Lambda_{i, j}^{\mathcal{O}}=\Lambda_{i, j}^{\mathcal{O}}+\mu\left(\mathcal{O}_{i, j}-\mathcal{L}_{i, j}-\mathcal{S}_{i, j}-\mathcal{N}_{i, j}\right),\\
& \Lambda_{i, j}^{\mathcal{L}}=\Lambda_{i, j}^{\mathcal{L}}+\mu\left(\mathcal{L}_{i, j}-\mathcal{J}_{i, j}\right), \\
& \Lambda_{\mathcal{X}}=\Lambda_{\mathcal{X}}+\mu(\mathcal{J}-\mathcal{X}),\\
& \Lambda = \Lambda + \mu(\mathcal{U}-\mathbf{D}\mathcal{X}).
\end{split}
\right.
\end{equation}

\renewcommand{\algorithmicrequire}{\textbf{Input:}} 
\renewcommand{\algorithmicensure}{\textbf{Output:}} 
\begin{algorithm}
	\caption{HSI denoising via LLxRGTV model.} \label{algorightm-1}
	\begin{algorithmic}[1]
		\Require
		$m \times n \times p$ observed HSI $ \mathcal{O}$, patch size $m_1 \times n_1 \times p$, stopping criterion $\varepsilon$, regularization parameters $\lambda$, $\tau$, $\beta$.
		\Ensure
		Denoised HSI $\mathcal{L}$;
		\State Initialize: $\mathcal{L}=\mathcal{X}=\mathcal{S}=\mathcal{J}=0,$
		$\Lambda_{i, j}^{\mathcal{O}}=0, \Lambda_{i, j}^{\mathcal{L}}=0, \Lambda_{\mathcal{X}}=0, \Lambda=0, \mu=10^{-2}, \mu_{\max }=10^{6}, \rho=1.5$
		and $k=0$;
		\State Update all patches $\left(\mathcal{L}_{i, j}, \mathcal{S}_{i, j}, \mathcal{N}_{i, j}\right)$  by (\ref{eq:Lmodel-solution}), (\ref{eq:Smodel-solution}) and (\ref{equation:solution of N}), respectively;
		\State Update $\left(\mathcal{J}, \mathcal{X}, \mathcal{U} \right)$ by (\ref{eq:Jmodel-solution}), (\ref{eq:Xmodel-solution}), (\ref{eq:Umodel-solution}), respectively;
		\State Update the Lagrangian multipliers by (\ref{eq:Lagparas});
		\State Update the penalty parameter by $\mu :=\min \left(\rho \mu, \mu_{\max }\right)$;
		\State Check the convergence condition
		$$
          \max \left\{
          \begin{array}{l}
          \|\mathcal{O}_{i, j}-\mathcal{L}_{i, j}^{k+1}-\mathcal{S}_{i, j}^{k+1} -\mathcal{N}_{i, j}^{k+1}\|_{\infty} \\
          \|\mathcal{L}_{i,j}^{k+1} - \mathcal{J}_{i,j}^{k+1} \|_{\infty} \\
          \|\mathcal{J}^{k+1}-\mathcal{X}^{k+1}\|_{\infty}
          \end{array}
          \right\}
          \leq \varepsilon.
        $$
	\end{algorithmic}
\end{algorithm}
Summarizing the optimization strategy of step-by-step iteration as above, the solution of the proposed LLxRGTV model can be obtained in Algorithm \ref{algorightm-1}.
In the proposed algorithm, the inputs include the observed HSI $\mathcal{O} \in \mathbb{R}^{m \times n \times p}$ ,
the stopping criteria $\epsilon$, and the regularized parameters $\tau$, $\lambda$, $\beta$.
Considering the fact that these parameters have certain proportional relationship,
we need to tune $\lambda$, $\beta$ and $\tau$ carefully.
More details and discussions would be presented in Section \ref{results}.
In addition for $\mu$, we first initialize it as $\mu=10^{-2}$ and then update it via $\mu=\min \left(\rho \mu, \mu_{\max }\right)$, which has been widely used in the ALM-based algorithms \cite{ALM_1,ALM_2}.

\begin{table*}
\label{tab:indian}
\begin{center}
	\caption{Quantitative evaluation of different methods in all noise cases of USGS Indian Pines}	
	\begin{tabular}{llllllllllllr}
		\hline \hline
		Noise Case & Level & Evaluation index & BM3D & LRTA & NAILRMA & LRMR & LLRSSTV & LLxRGTV  \\
		\hline \hline
		\multirow{5}{*}{Case 1} & & MPSNR & 28.676  & 29.031  & 24.295  & 33.757  & \underline{34.497}  & \textbf{37.465}\\
		& G=0.1 & MSSIM & 0.945  & 0.833  & 0.768  & 0.892  & \underline{0.886}  & \textbf{0.978}\\
		& P=0.2 & MFSIM & 0.942  & 0.857  & 0.797  & 0.898  & \underline{0.893}  & \textbf{0.974}\\
		& & ERGAS & 88.071  & 85.23  & 145.864  & 47.974  & \underline{44.081}  & \textbf{33.455}\\
		& & MSAD & 2.897  & 3.224  & 5.554  & 1.931  & \underline{1.775}  & \textbf{1.205}\\
		\hline
		\multirow{5}{*}{Case 2} & & MPSNR & 28.779 & 29.499 & 28.19 & 33.986 & \underline{35.642} & \textbf{38.785}\\
		& impulse & MSSIM & 0.946 & 0.863 & 0.841 & 0.893 & \underline{0.904} & \textbf{0.956}\\
		& +Gaussion & MFSIM & 0.944 & 0.883 & 0.857 & 0.9 & \underline{0.91} & \textbf{0.98}\\
		& & ERGAS & 88.129 & 81.718 & 101.567 & \underline{48.852} & 58.271 & \textbf{29.209}\\
		& & MSAD & 2.948 & 3.068 & 3.95 & \underline{2.017} & 2.504 & \textbf{1.056}\\
		\hline
		\multirow{5}{*}{Case 3} & & MPSNR & 29.277 & 30.279 & \underline{37.144} & 35.32 & 35.854 & \textbf{40.189}\\
		& & MSSIM & \underline{0.949} & 0.88 & 0.937 & 0.911 & 0.903 & \textbf{0.987}\\
		& Gaussian & MFSIM & \underline{0.945} & 0.897 & 0.936 & 0.911 & 0.908 & \textbf{0.984}\\
		& & ERGAS & 83.134 & 74.881 & \underline{34.075} & 42.107 & 44.591 & \textbf{24.118}\\
		& & MSAD & 2.764 & 2.832 & \underline{1.322} & 1.766 & 1.84 & \textbf{0.833}\\
		\hline
		\multirow{5}{*}{Case 4} & & MPSNR & 28.718 & 29.391 & 28.089 & 33.679 & \underline{35.258} & \textbf{38.331}\\
		& impulse & MSSIM & \underline{0.946} & 0.859 & 0.841 & 0.891 & 0.899 & \textbf{0.986}\\
		& +Gaussian & MFSIM & \underline{0.943} & 0.879 & 0.856 & 0.897 & 0.906 & \textbf{0.979}\\
		& +Stripes & ERGAS & 88.773 & 82.711 & 102.25 & \underline{50.465} & 57.613 & \textbf{30.355}\\
		& & MSAD & 2.976 & 3.11 & 3.996 & \underline{2.102} & 2.479 & \textbf{1.118}\\
		\hline
		\multirow{5}{*}{Case 5} & & MPSNR & 28.647 & 29.335 & 27.467 & 33.523 & \underline{34.854} & \textbf{38.293}\\
		& Gaussian & MSSIM & \underline{0.946} & 0.862 & 0.826 & 0.89 & 0.91 & \textbf{0.985}\\
		& +impulse & MFSIM & \underline{0.943} & 0.882 & 0.843 & 0.897 & 0.909 & \textbf{0.979}\\
		& +deadline & ERGAS & 89.197 & 82.905 & 111.244 & \underline{52.167} & 75.319 & \textbf{30.439}\\
		& & MSAD & 2.987 & 3.099 & 4.44 & \underline{2.191} & 3.249 & \textbf{1.141}\\
		\hline
		\multirow{5}{*}{Case 6} & & MPSNR & 28.573 & 29.219 & 27.412 & 33.207 & \underline{34.271} & \textbf{38.101}\\
		& Gaussian & MSSIM & \underline{0.945} & 0.859 & 0.831 & 0.886 & 0.9 & \textbf{0.986}\\
		& +impulse & MFSIM & \underline{0.943} & 0.879 & 0.847 & 0.893 & 0.901 & \textbf{0.978}\\
		& +deadline & ERGAS & 89.958 & 83.993 & 111.299 & \underline{54.884} & 79.462 & \textbf{31.351}\\
		& +stripe & MSAD & 3.034 & 3.156 & 4.45 & \underline{2.311} & 3.448 & \textbf{1.186}\\
		\hline \hline
	\end{tabular}
\end{center}
\end{table*}

\section{Experimental results and discussion}
\label{results}

In this section, to verify the effectiveness of our proposed model for HSI denoising, various experiments are performed on a set of challenging simulated and real HSI dataset.
For comparison, five different state-of-the-art HSI denoising methods are employed as the benchmark in the experiments, i.e.,
BM3D \cite{BM3D}, LRTA \cite{HSI_using_tucker}, NAILRMA\cite{NAILRMA}, LRMR \cite{LRMR} and LLRSSTV \cite{LLRSSTV}.
Since the BM3D method is only suitable to remove Gaussian noise,  we implement it on HSIs which are preprocessed by the RPCA restoration method.

Before performing the denoising model, the gray values in each HSI band are normalized to $[0, 1]$.
After the denoising, each band in HSIs is converted to the original gray level.
The parameter selection in the comparison models is consistent with the description in the original papers.
And in Section \ref{discussion}, the parameters in our proposed LLxRGTV model are discussed in detail.
To thoroughly evaluate the performance of different denoising methods, the visual comparison and quantitative comparison are adopted to give the quality assessments.
Especially for quantitative comparison, five quantitative picture quality indices (PQIs),
including the mean peak signal-to-noise ratio (MPSNR) \cite{PSNR}, mean structural similarity (MSSIM) \cite{SSIM}, mean feature similarity (MFSIM) \cite{FSIM},
erreur relative globale adimensionnelle de synth\`{e}se (EGRAS) \cite{EGRAS} and
the mean spectral angle distance (MSAD), are used to obtain the qualitative evaluations of the denoised HSIs.
PSNR and SSIM are two conventional PQIs in image processing and computer vision.
They evaluate the similarity between the target image and the reference image based on MSE and structural consistency, respectively.
FSIM emphasizes the perceived consistency with the reference image.
The higher the values of MPSNR, MFSIM, and MSSIM are, the smaller the values of EGRAS and MSAD are, the higher the quality of the denoised HSIs is and the better the denoising method is.
The MSAD is defined as
\begin{equation}
\mathrm{MSAD}=\frac{1}{m n} \sum_{i=1}^{m n} \frac{180}{\pi} \times \arccos \frac{\left(\mathcal{X}^{i}\right)^{T} \cdot\left(\hat{\mathcal{X}}^{i}\right)}{\left\|\mathcal{X}^{i}\right\| \cdot\left\|\hat{\mathcal{X}}^{i}\right\|},
\end{equation}
where $\mathcal{X}^{i}$ and $\hat{\mathcal{X}}^{i}$ denote the $i$ th spectral signatures of the noise-free and denoised HSIs, respectively.

\begin{figure}[!t]
	\centering
	\includegraphics[width=0.35\linewidth]{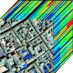}
	\includegraphics[width=0.35\linewidth]{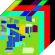}
	\caption{Datasets used in the simulated experiments. (a) Pavia City Centre dataset (R: 20, G: 50, B: 80). (b) USGS Indian Pines dataset  (R: 6, G: 88, B: 221).}
	\label{3dcubefig}
\end{figure}

\begin{figure*}[!t]
\begin{center}
	\includegraphics[width=0.5\linewidth]{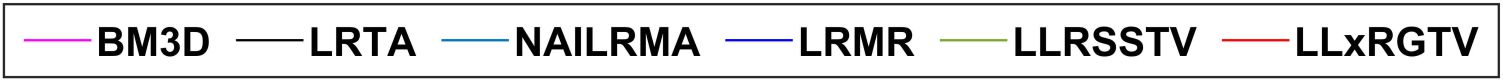}\\
	\subfloat[Noise Case 1]{\includegraphics[width=0.3\linewidth]{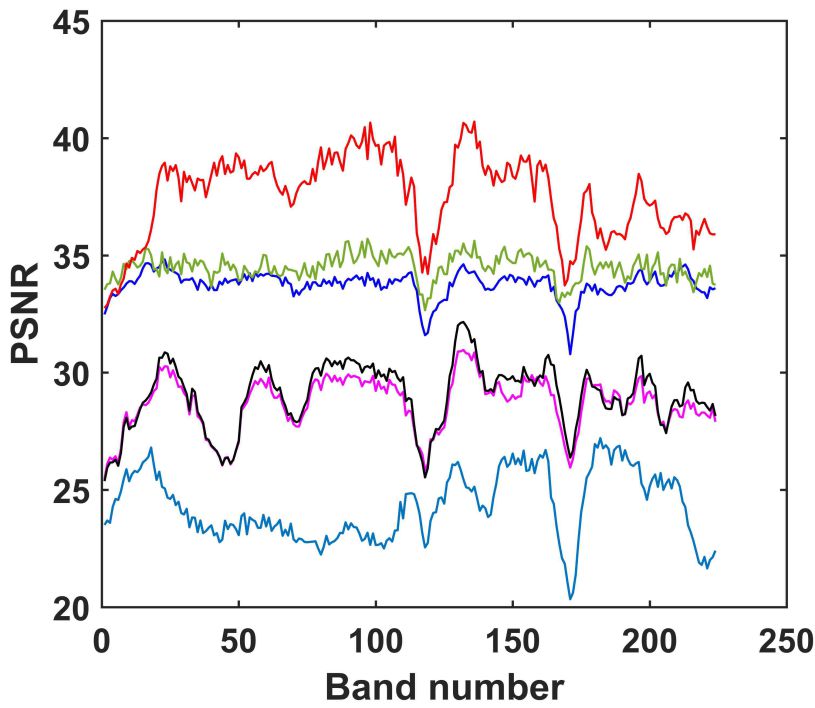}}%
	\hfil
	\subfloat[Noise Case 2]{\includegraphics[width=0.3\linewidth]{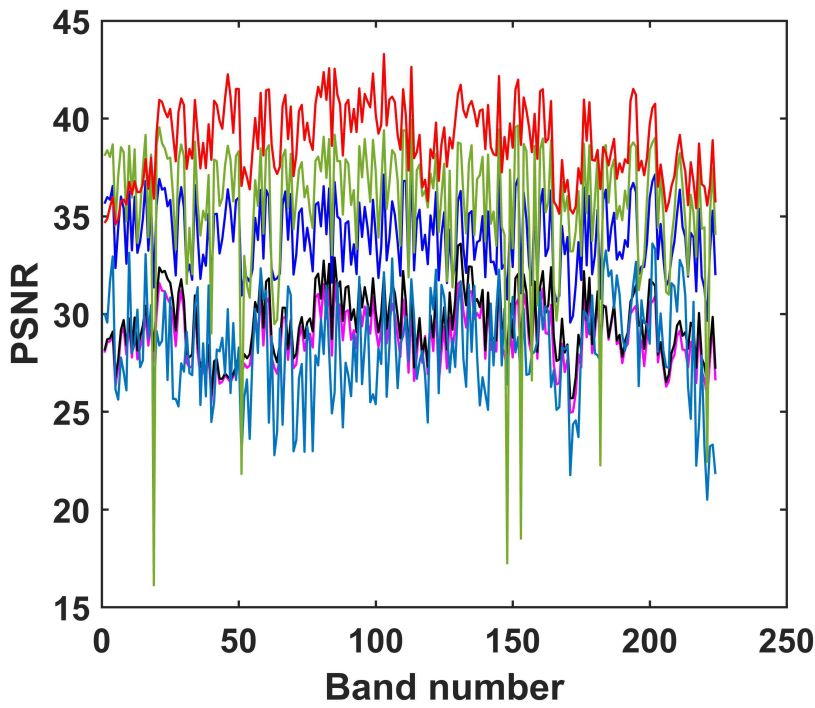}}%
	\hfil
	\subfloat[Noise Case 3]{\includegraphics[width=0.3\linewidth]{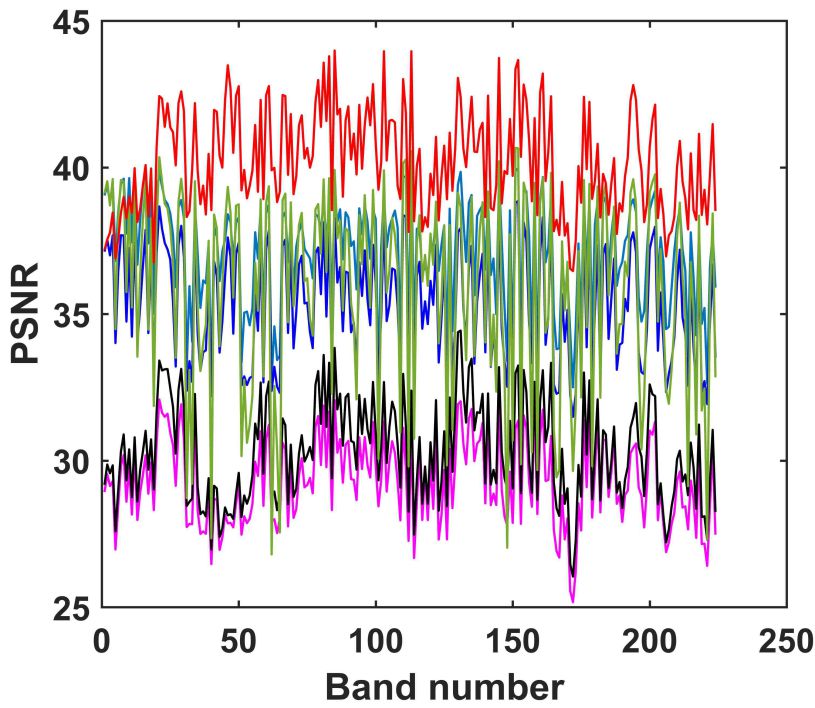}}%
	\hfil
	\subfloat[Noise Case 4]{\includegraphics[width=0.3\linewidth]{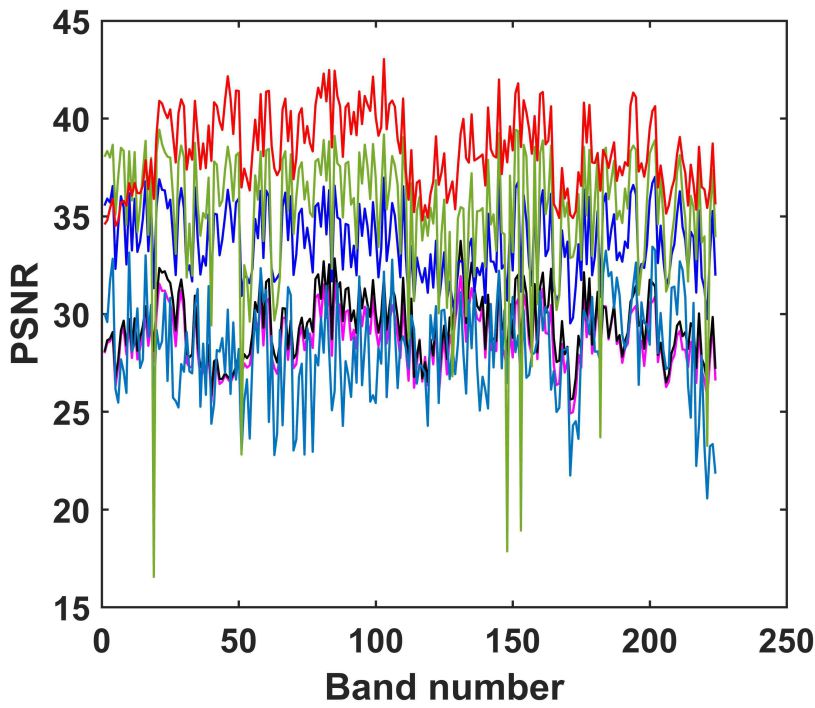}}%
	\hfil
	\subfloat[Noise Case 5]{\includegraphics[width=0.3\linewidth]{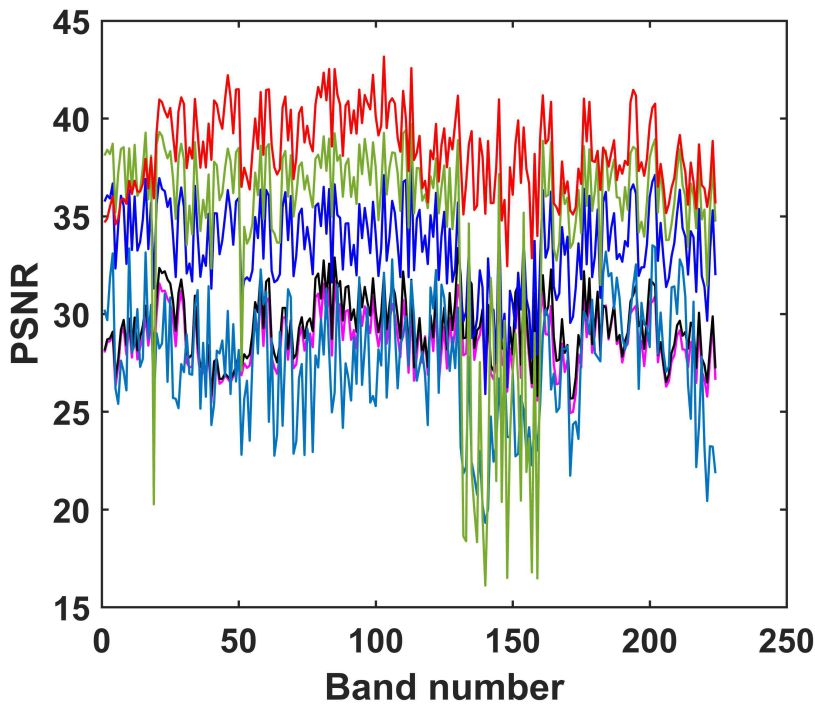}}%
	\hfil
	\subfloat[Noise Case 6]{\includegraphics[width=0.3\linewidth]{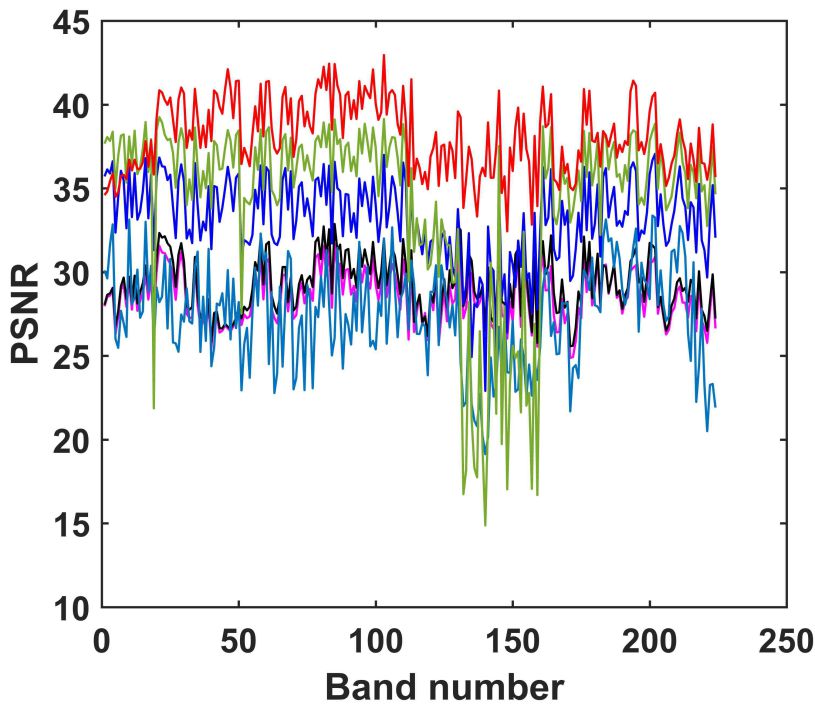}}%
\caption{Detailed PSNR values of different denoising methods in each band of USGS Indian Pines.}
	\label{fig:psnr_indian}
\end{center}
\end{figure*}

\begin{figure*}[!t]
	\centering
	\includegraphics[width=0.5\linewidth]{tuli_indian.jpg}\\
	\subfloat[Noise Case 1]{\includegraphics[width=0.3\linewidth]{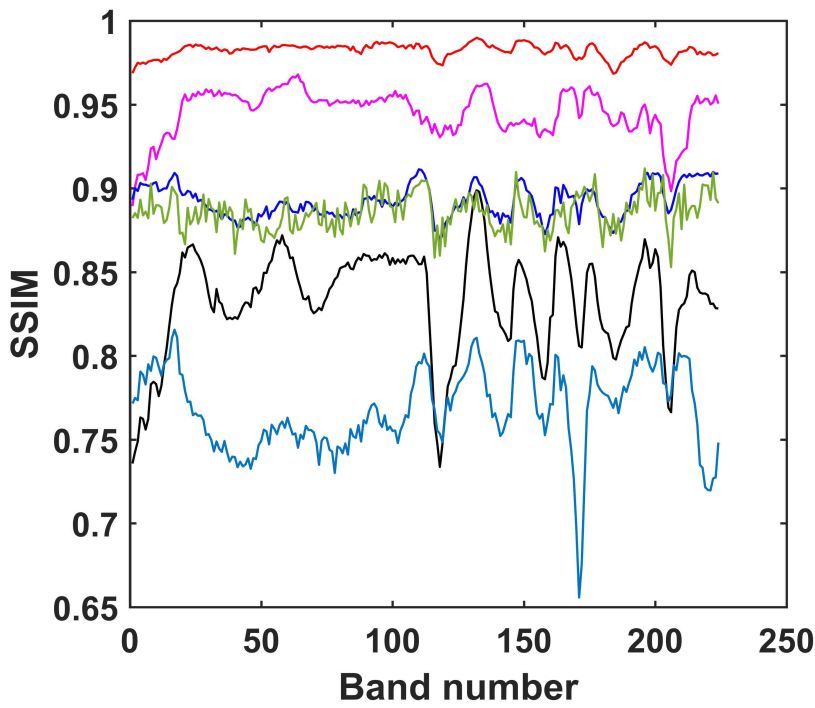}}%
	\hfil
	\subfloat[Noise Case 2]{\includegraphics[width=0.3\linewidth]{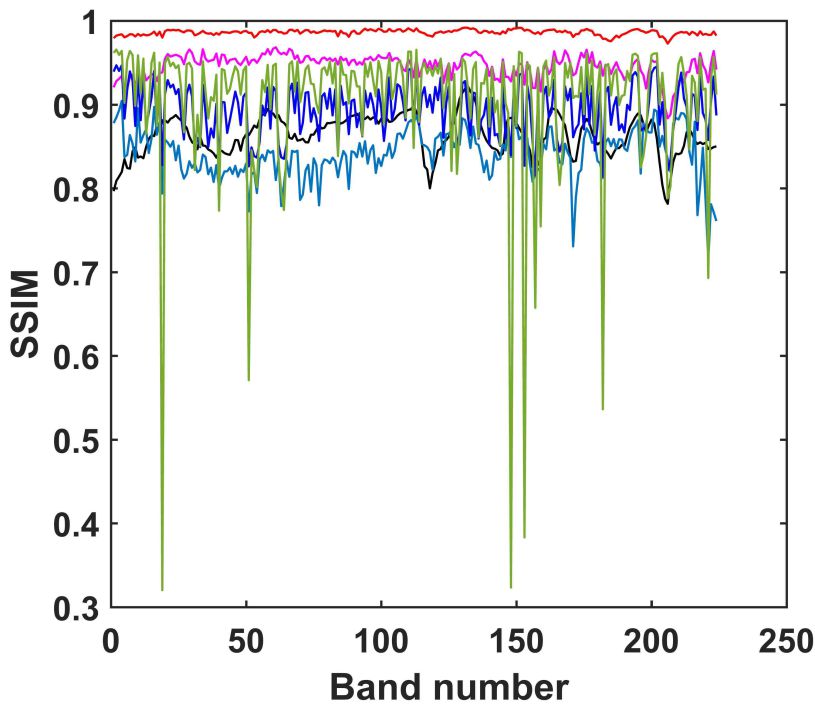}}%
	\hfil
	\subfloat[Noise Case 3]{\includegraphics[width=0.3\linewidth]{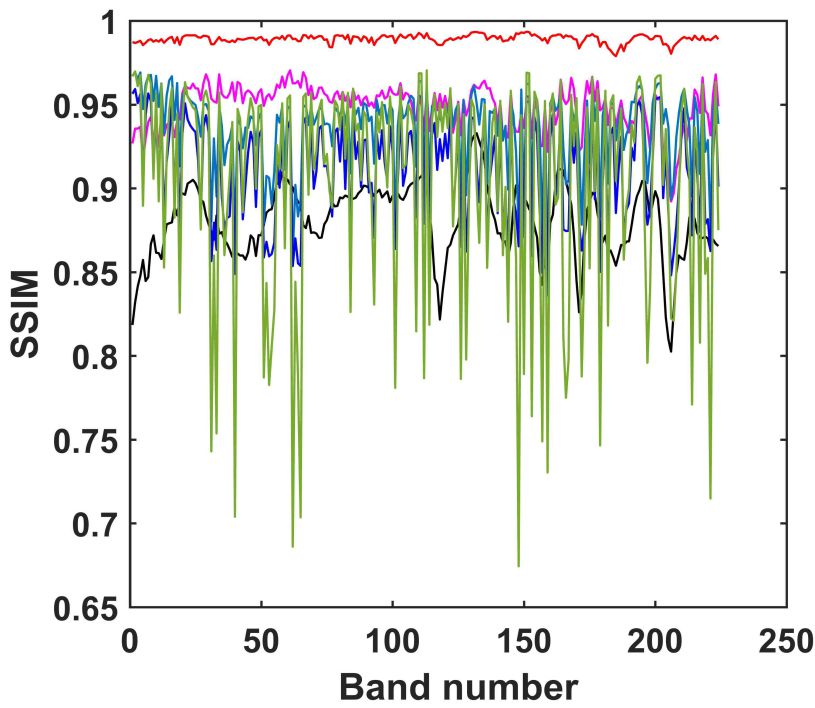}}%
	\hfil
	\subfloat[Noise Case 4]{\includegraphics[width=0.3\linewidth]{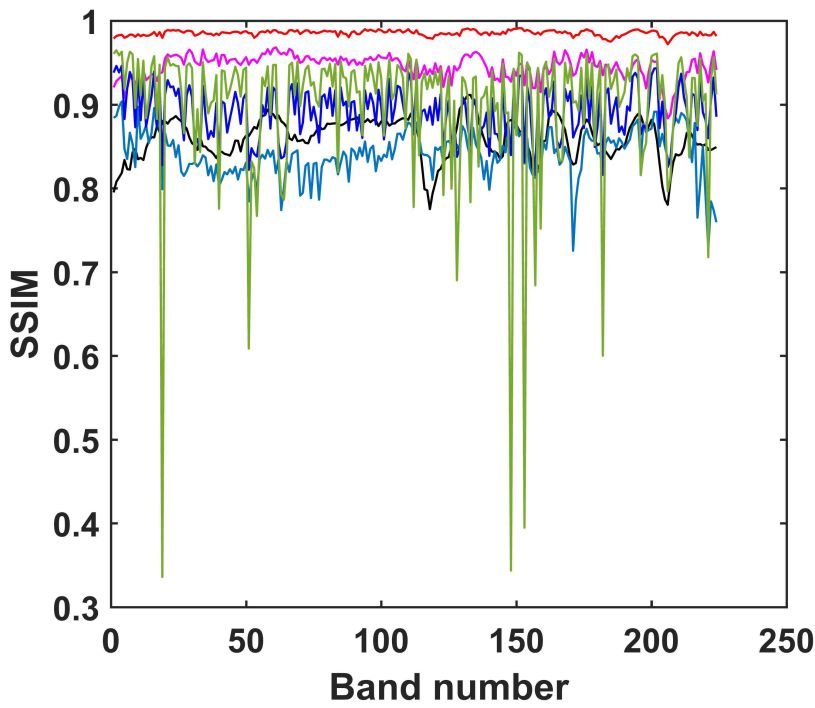}}%
	\hfil
	\subfloat[Noise Case 5]{\includegraphics[width=0.3\linewidth]{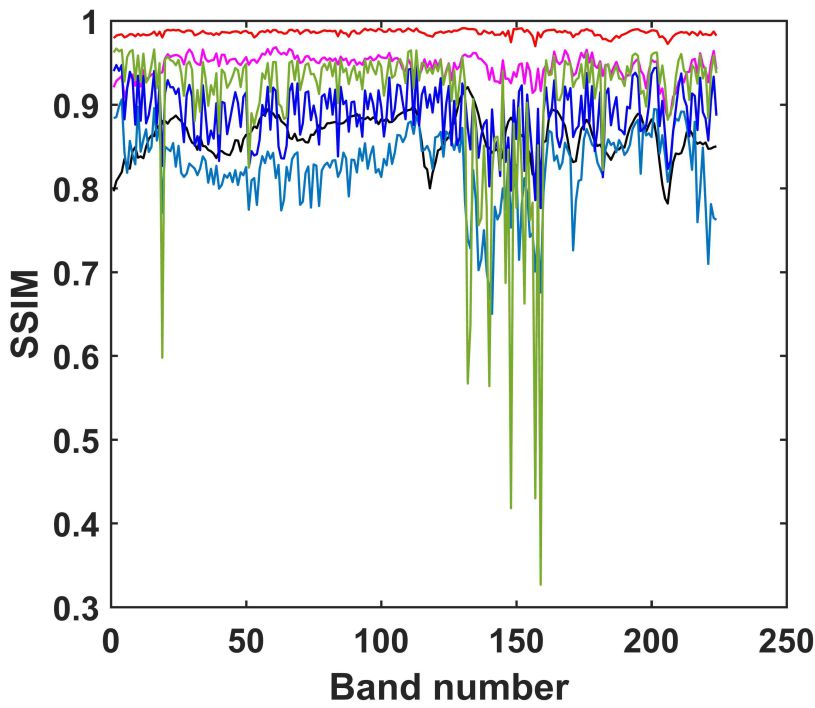}}%
	\hfil
	\subfloat[Noise Case 6]{\includegraphics[width=0.3\linewidth]{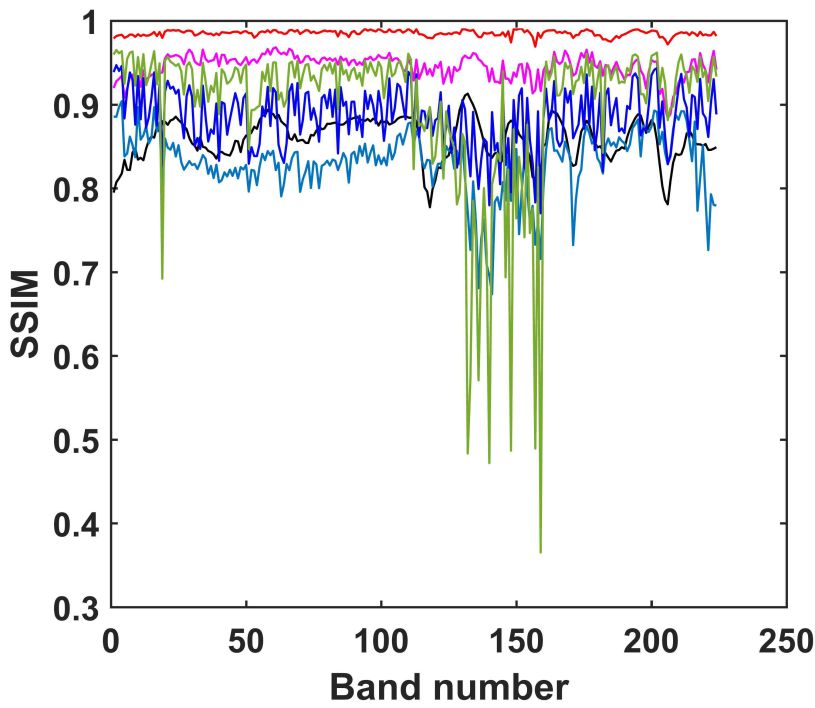}}%
	\caption{Detailed SSIM values of different denoising methods in each band of USGS Indian Pines.}
	\label{fig:ssim_indian}
\end{figure*}	

\subsection{Simulated HSI dataset experiments}

In this subsection, we select two HSI datasets to apply simulated experiments, see Fig.\ref{3dcubefig}.
The first dataset is the Pavia City Centre\footnote{http://www.ehu.es/ccwintco/index.php/Hyperspectral\_Remote\_Sensing\\ \_Scenes}, which is filmed by the reflection optical system imaging spectrometer (ROSIS-03).
The size of Pavia City Centre is 1096$\times$1096, with a total of 102 bands.
Because some of the bands in the Pavia City Centre dataset are heavily polluted by noise, they can not be used as a reference for denoising results.
Therefore, this part of the heavily polluted data has been removed.
Due to space limitations, we select data with a spatial size of 200$\times$200 and a total of 80 bands for simulated experiments in this subsection.
The second dataset is the USGS Indian Pines
dataset\footnote{https://engineering.purdue.edu/~biehl/MultiSpec/hyperspectral.html}, and the spectral signatures
are extracted from the USGS digital spectral library. The size of the USGS Indian Pines data was $145 \times 145 \times 224$.

To simulate noisy HSI data, we add several types of noise to the original HSI data, which can be divided into the following six cases:

Case 1: In this case, the same intensity noise
is added to all the bands. Specifically, the variance of Gaussian white noise is set to 0.1, while the percentage of impulse noise is set to 0.2.

Case 2: In this case, we only add Gaussian white noise with 0.1 variance to the clean HSI,
to verify the removal performance of the proposed model for a single Gaussian noise.

Case 3: In practice, the noise intensity in each band is also different,
and the HSIs are not only contaminated by a single noise.
To simulate this case, we also add Gaussian noise and impulse noise into HSIs.
However, the variance of Gaussian white noise and the percentages of impulse noise in each band are randomly selected from 0 to 0.2.

Case 4: Based on Case 3, some stripes are additionally added from band 44 to band 64 in Pavia City Centre dataset, and from band 111 to band 140 in USGS Indian Pines dataset.
The number of stripes in each band is randomly selected from 20 to 40.

Case 5: Based on Case 3, deadlines are additionally added from band 54 to band 74 in Pavia City Centre dataset, and from band 131 to band 160 in USGS Indian Pines dataset.
The number of deadlines in each band is randomly selected from 3 to 10, and the pixel width of deadlines is randomly generated from 1 to 3.

Case 6: In this case, the Gaussian noise and impulse noise in Case 3, deadlines in Case 4 and stripe noise in Case 5 are simultaneously added to the clean HSIs.

\begin{figure*}[!t]
	\centering
	\subfloat[Original image]{\includegraphics[width=0.22\linewidth]{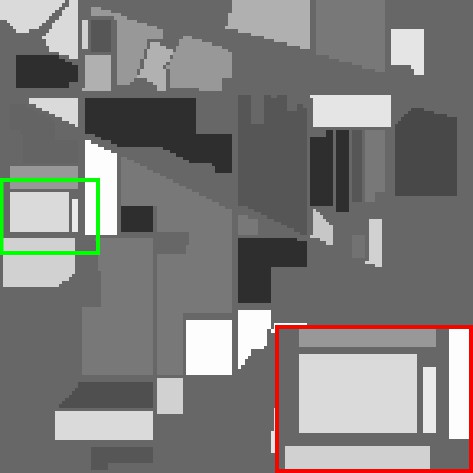}}%
	\hfil
	\subfloat[Noisy image]{\includegraphics[width=0.22\linewidth]{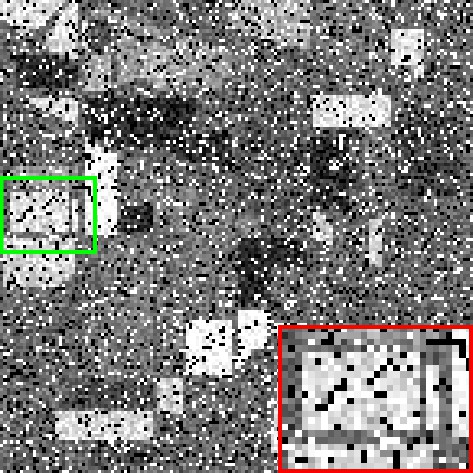}}%
	\hfil
	\subfloat[BM3D]{\includegraphics[width=0.22\linewidth]{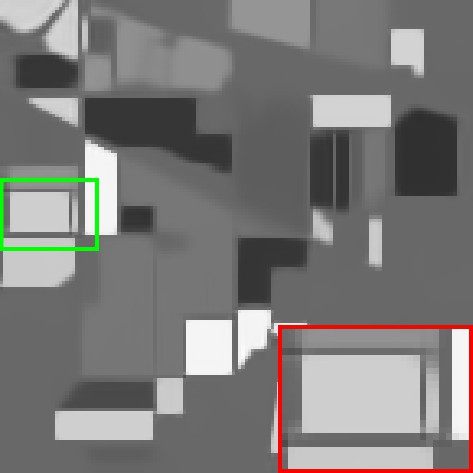}}%
	\hfil
	\subfloat[LRTA]{\includegraphics[width=0.22\linewidth]{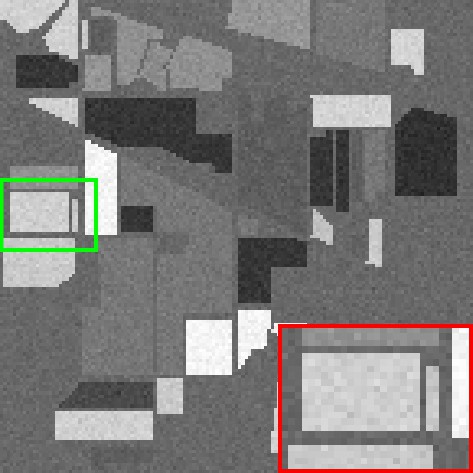}}%
	\hfil
	\subfloat[NAILRMA]{\includegraphics[width=0.22\linewidth]{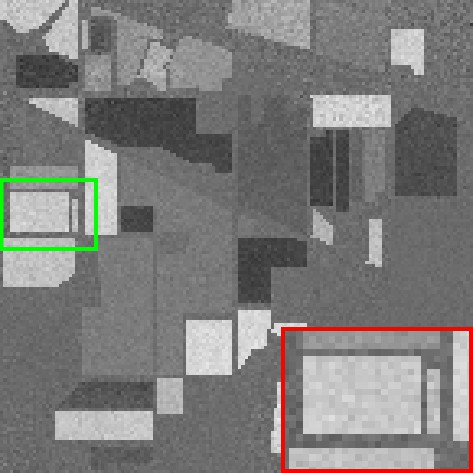}}%
	\hfil
	\subfloat[LRMR]{\includegraphics[width=0.22\linewidth]{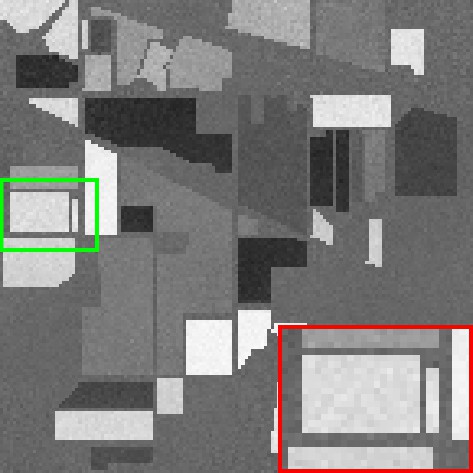}}%
	\hfil
	\subfloat[LLRSSTV]{\includegraphics[width=0.22\linewidth]{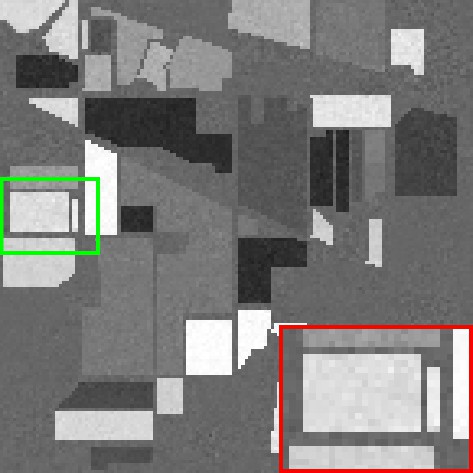}}%
	\hfil
	\subfloat[LLxRGTV]{\includegraphics[width=0.22\linewidth]{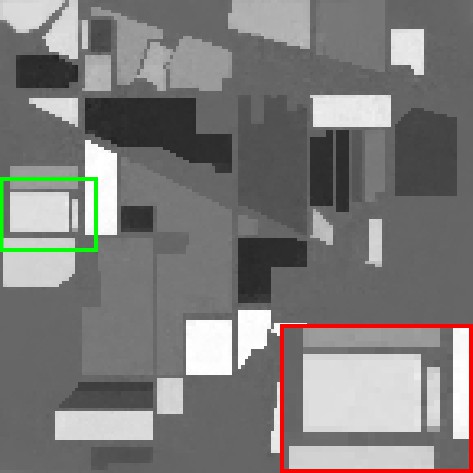}}%
	\caption{Denoised results in the simulated experiments of USGS Indian Pines dataset in Case 1. The PSNR of (c)-(h) are 27.6649 dB, 27.9513 dB, 24.4111 dB, 33.1047 dB, 34.2022 dB and 37.1525 dB, respectively.}
	\label{fig:p02g01_band125_indian}
\end{figure*}

\begin{figure*}[!t]
	\centering
	\subfloat[Original image]{\includegraphics[width=0.22\linewidth]{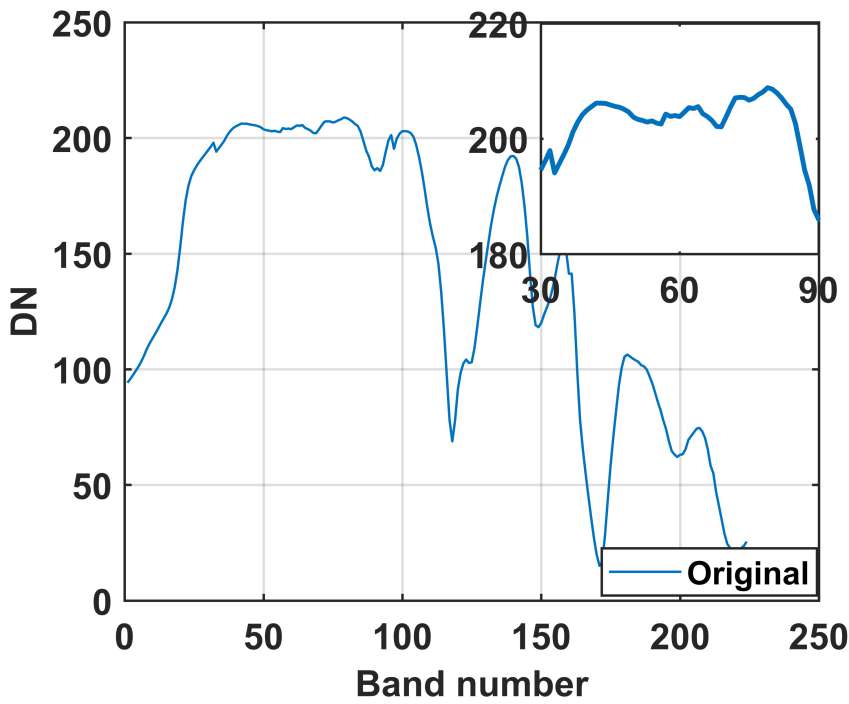}}%
	\hfil
	\subfloat[Noisy image]{\includegraphics[width=0.22\linewidth]{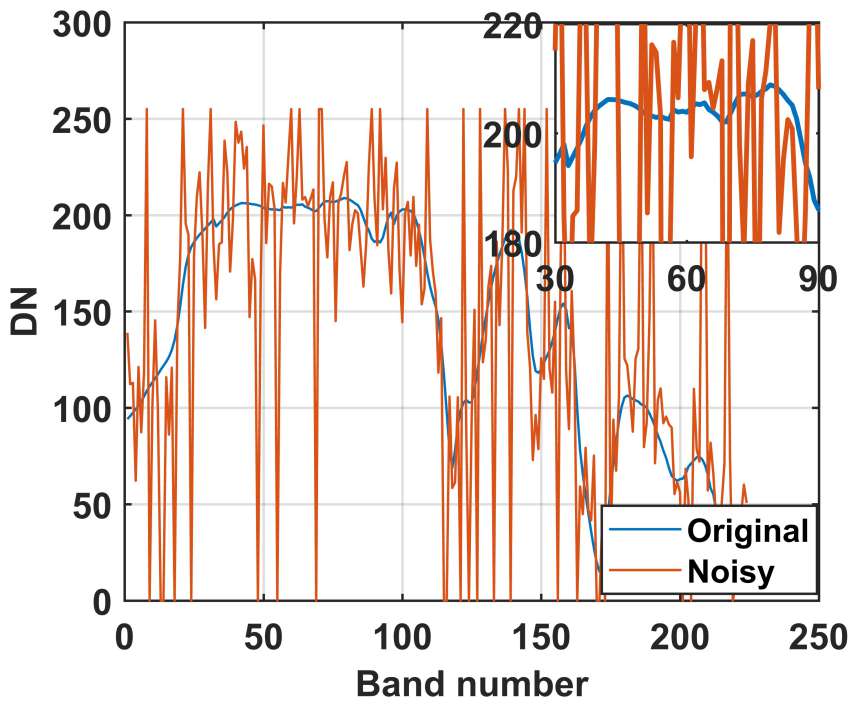}}%
	\hfil
	\subfloat[BM3D]{\includegraphics[width=0.22\linewidth]{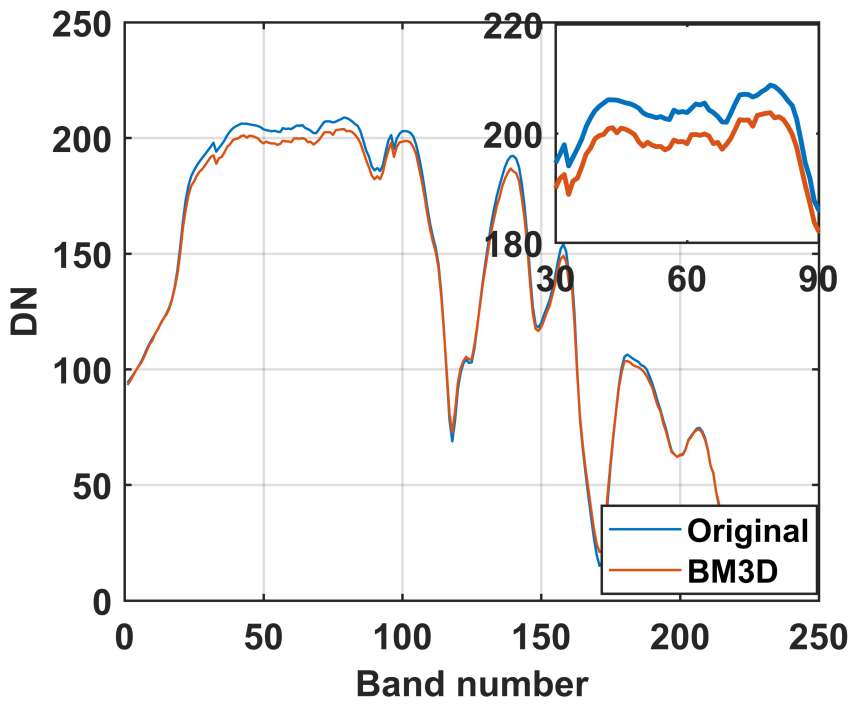}}%
	\hfil
	\subfloat[LRTA]{\includegraphics[width=0.22\linewidth]{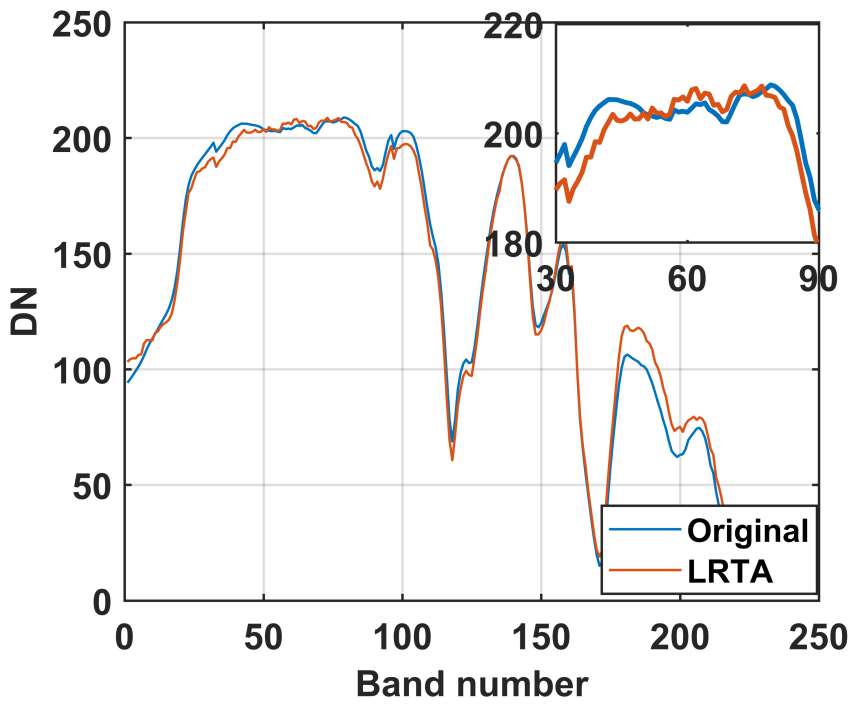}}%
	\hfil
	\subfloat[NAILRMA]{\includegraphics[width=0.22\linewidth]{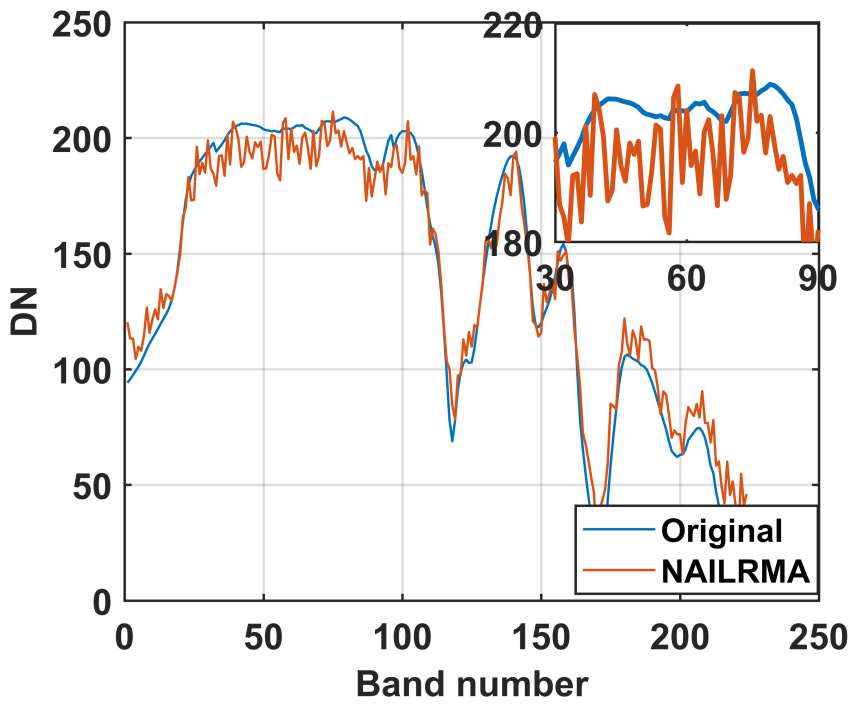}}%
	\hfil
	\subfloat[LRMR]{\includegraphics[width=0.22\linewidth]{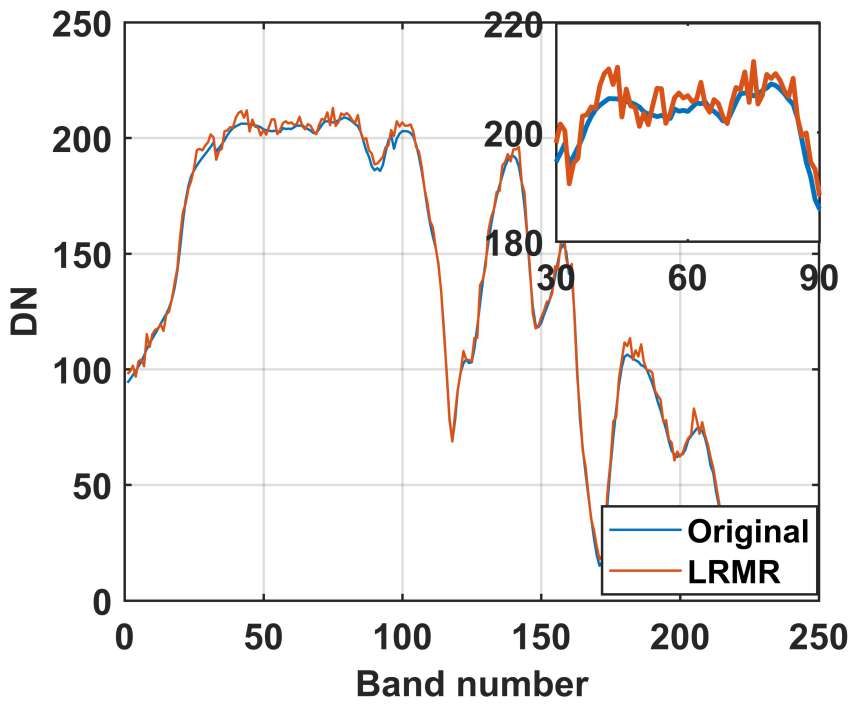}}%
	\hfil
	\subfloat[LLRSSTV]{\includegraphics[width=0.22\linewidth]{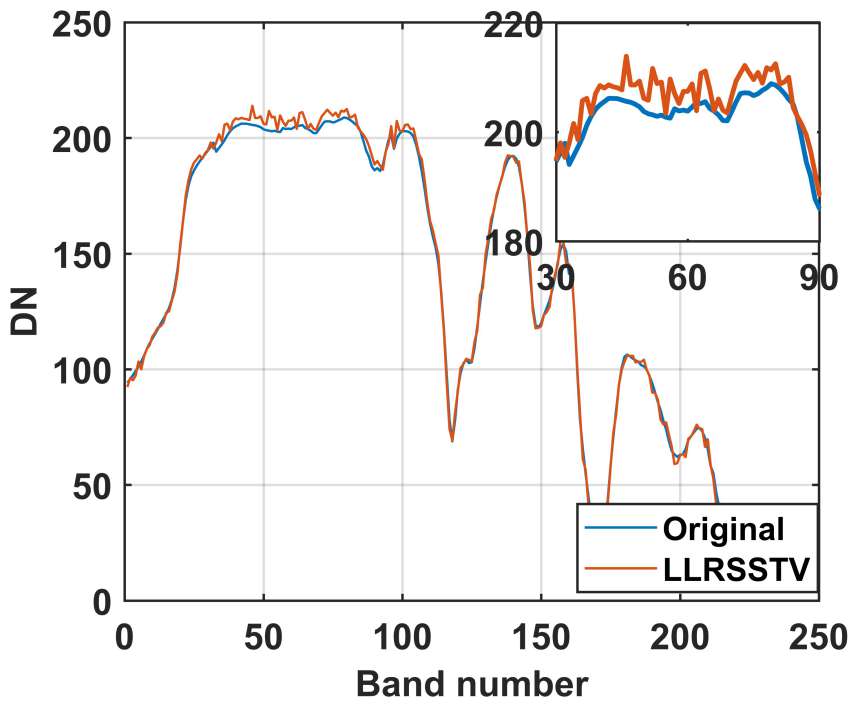}}%
	\hfil
	\subfloat[LLXRGTV]{\includegraphics[width=0.22\linewidth]{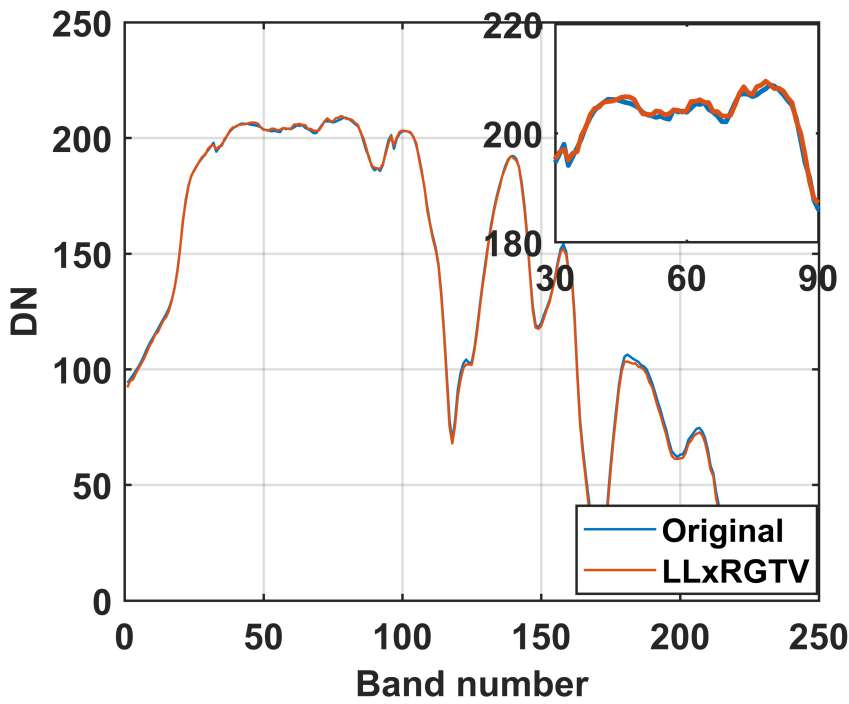}}%
	\caption{Spectrum of pixel (110, 110) in the denoised results of USGS Indian Pines dataset  in noise Case 1.}
	\label{fig:Spectrum_p02g01_indian}
\end{figure*}

\begin{figure*}[!t]
	\centering
	\subfloat[Original image]{\includegraphics[width=0.22\linewidth]{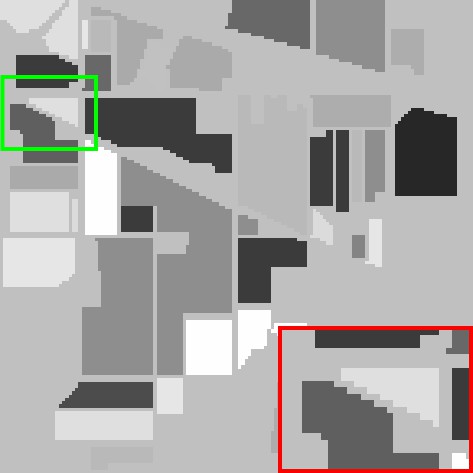}}%
	\hfil
	\subfloat[Noisy image]{\includegraphics[width=0.22\linewidth]{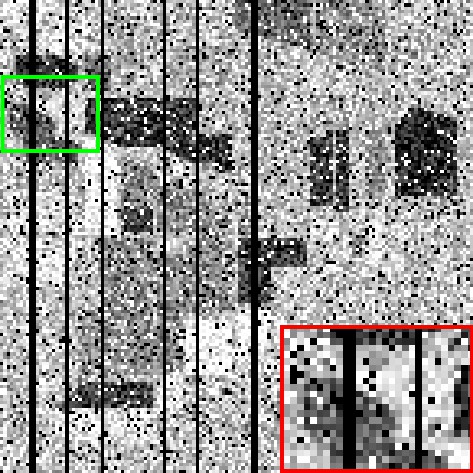}}%
	\hfil
	\subfloat[BM3D]{\includegraphics[width=0.22\linewidth]{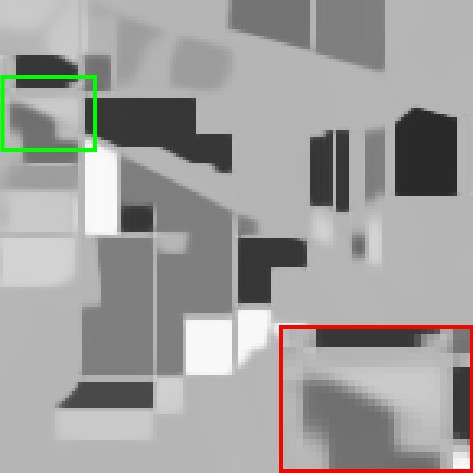}}%
	\hfil
	\subfloat[LRTA]{\includegraphics[width=0.22\linewidth]{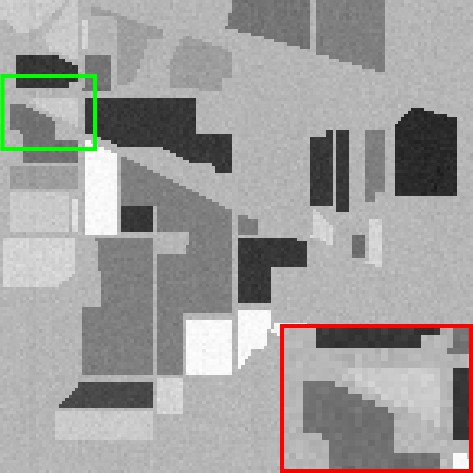}}%
	\hfil
	\subfloat[NAILRMA]{\includegraphics[width=0.22\linewidth]{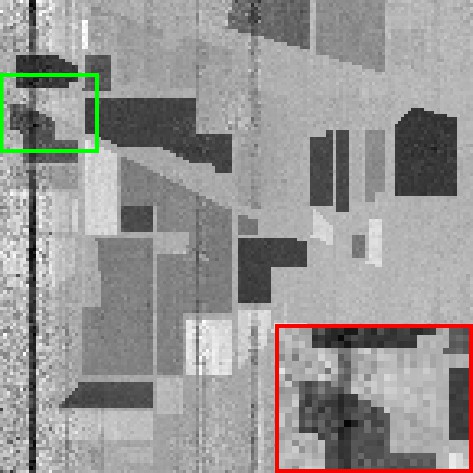}}%
	\hfil
	\subfloat[LRMR]{\includegraphics[width=0.22\linewidth]{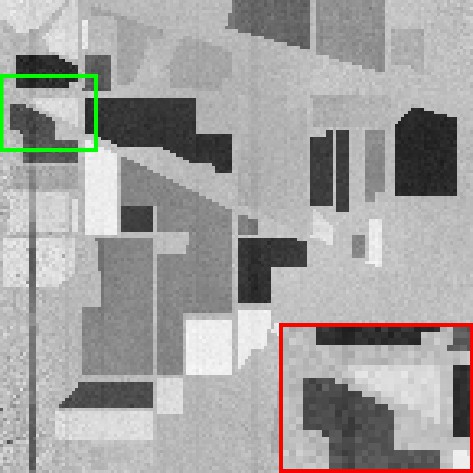}}%
	\hfil
	\subfloat[LLRSSTV]{\includegraphics[width=0.22\linewidth]{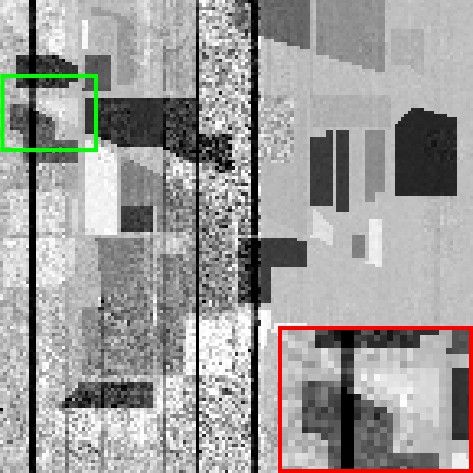}}%
	\hfil
	\subfloat[LLxRGTV]{\includegraphics[width=0.22\linewidth]{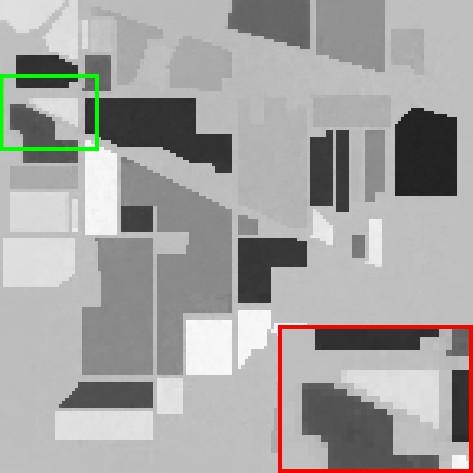}}%
	\caption{Denoised results in the simulated experiments of USGS Indian Pines dataset in Case 5. The PSNR of (c)-(h) are 25.9026 dB, 26.1950 dB, 19.3116 dB, 25.9038 dB, 16.1044 dB and 35.1042 dB, respectively.}
	\label{fig:pgd_band140_indian}
\end{figure*}

\begin{figure*}[!t]
	\centering
	\subfloat[Original image]{\includegraphics[width=0.22\linewidth]{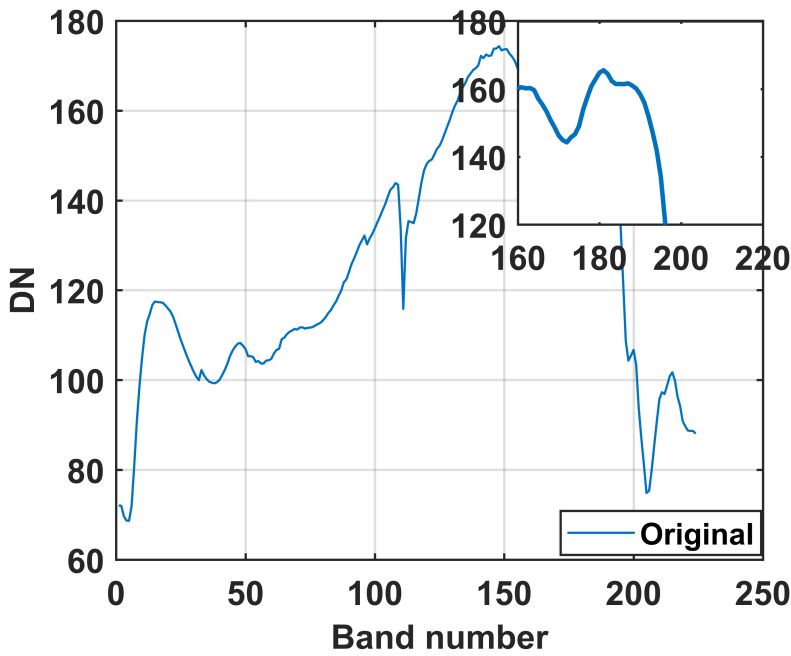}}%
	\hfil
	\subfloat[Noisy image]{\includegraphics[width=0.22\linewidth]{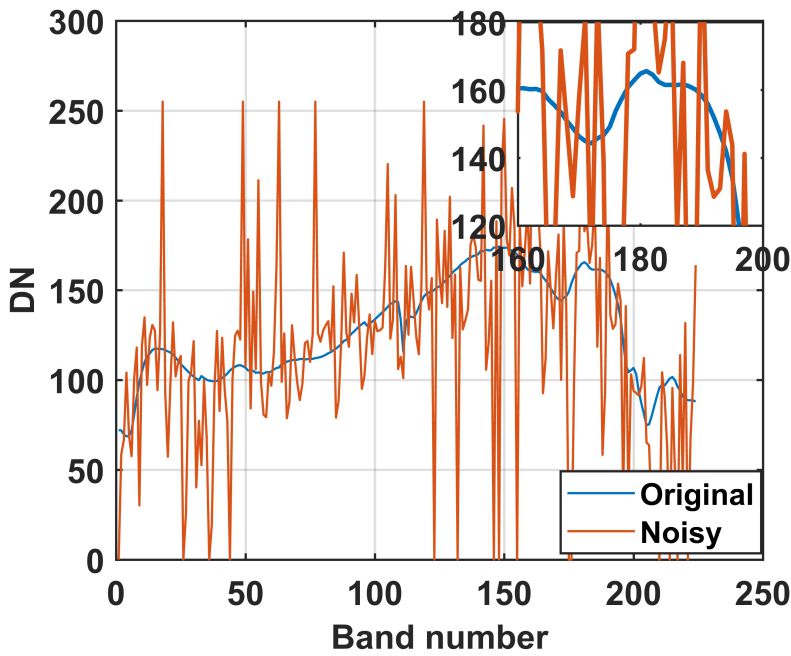}}%
	\hfil
	\subfloat[BM3D]{\includegraphics[width=0.22\linewidth]{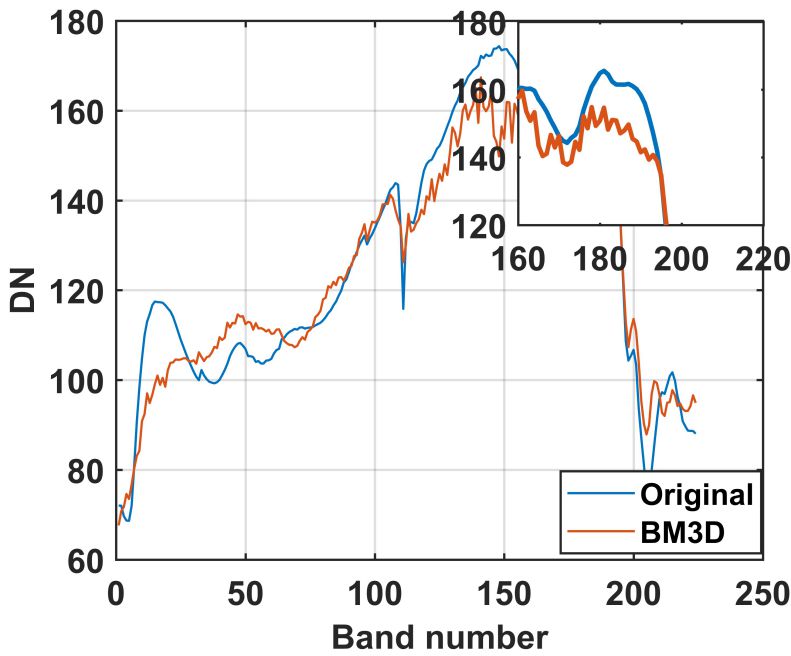}}%
	\hfil
	\subfloat[LRTA]{\includegraphics[width=0.22\linewidth]{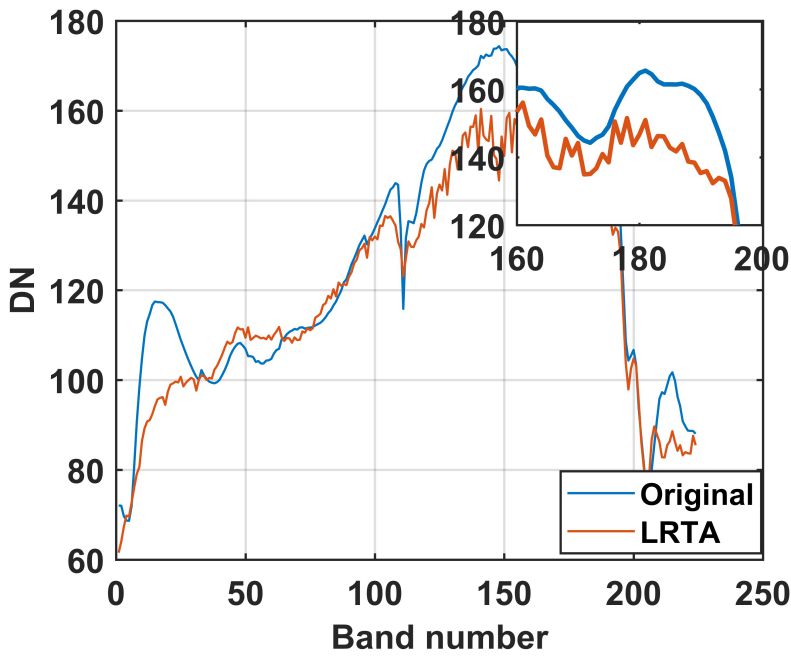}}%
	\hfil
	\subfloat[NAILRMA]{\includegraphics[width=0.22\linewidth]{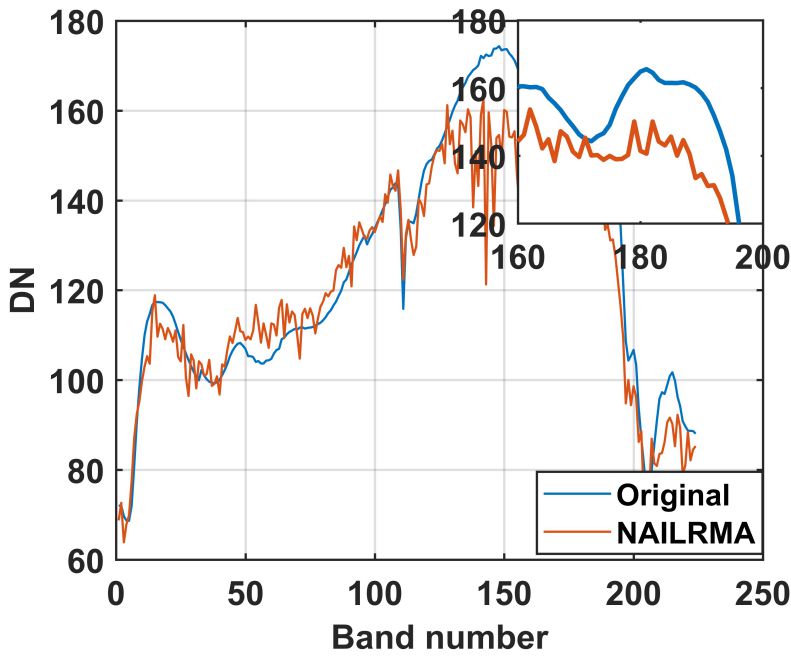}}%
	\hfil
	\subfloat[LRMR]{\includegraphics[width=0.22\linewidth]{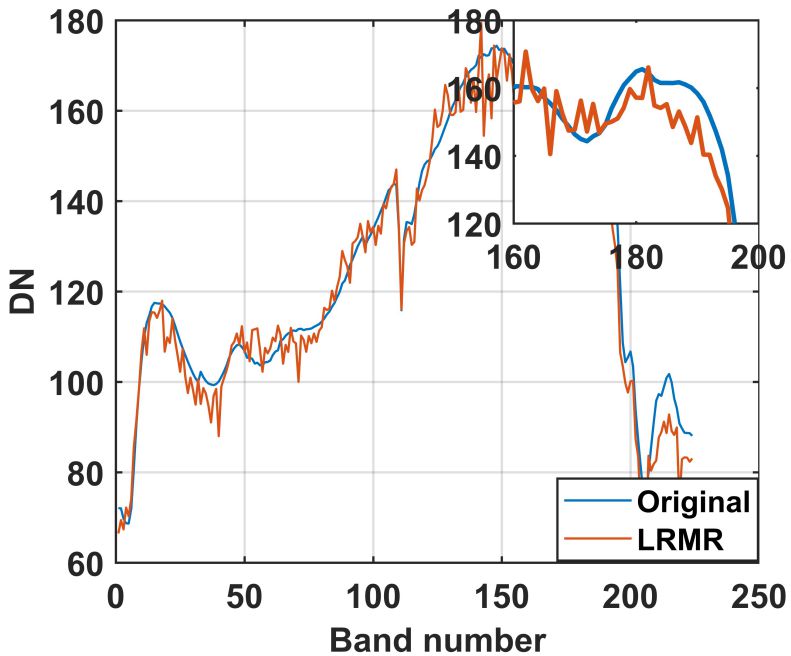}}%
	\hfil
	\subfloat[LLRSSTV]{\includegraphics[width=0.22\linewidth]{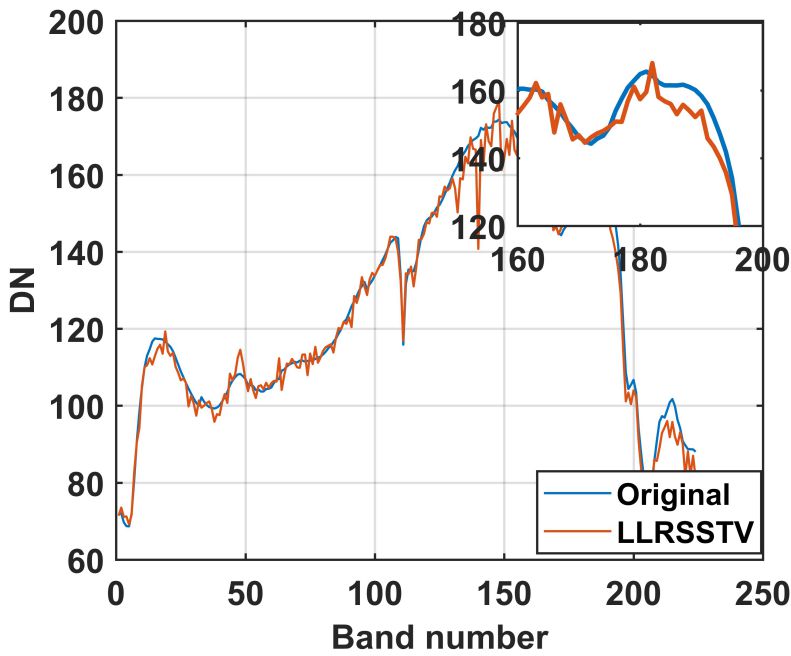}}%
	\hfil
	\subfloat[LLxRGTV]{\includegraphics[width=0.22\linewidth]{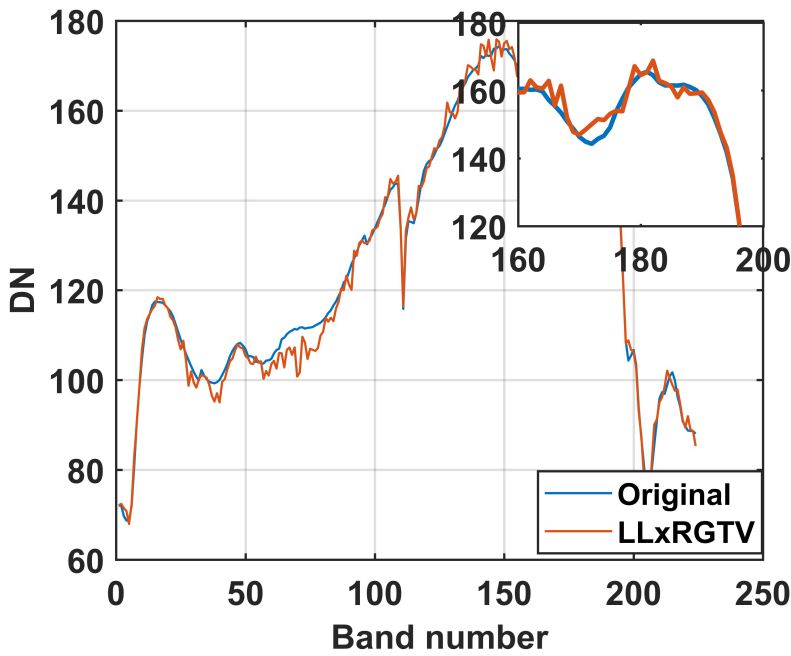}}%
	\caption{Spectrum of pixel (20, 40) in the denoised results of USGS Indian Pines in noise Case 5. }
	\label{fig:Spectrum_pgd_indian}
\end{figure*}

\begin{figure*}[!t]
	\centering
	\subfloat[Original image]{\includegraphics[width=0.22\linewidth]{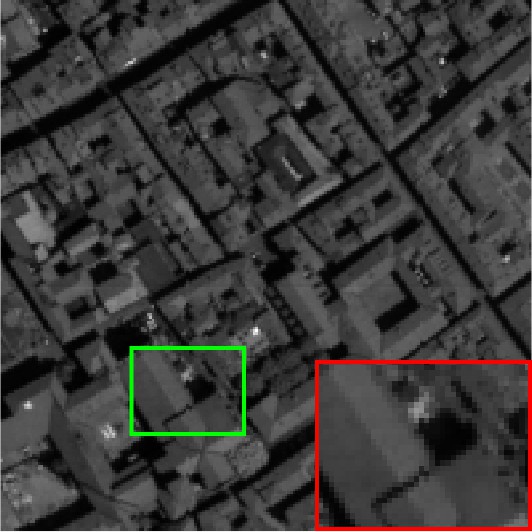}}%
	\hfil
	\subfloat[Noisy image]{\includegraphics[width=0.22\linewidth]{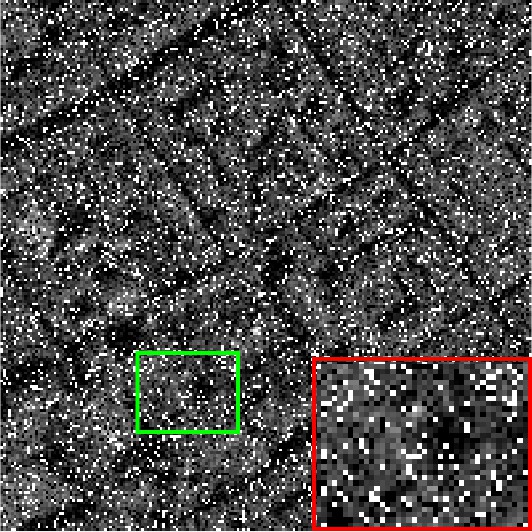}}%
	\hfil
	\subfloat[BM3D]{\includegraphics[width=0.22\linewidth]{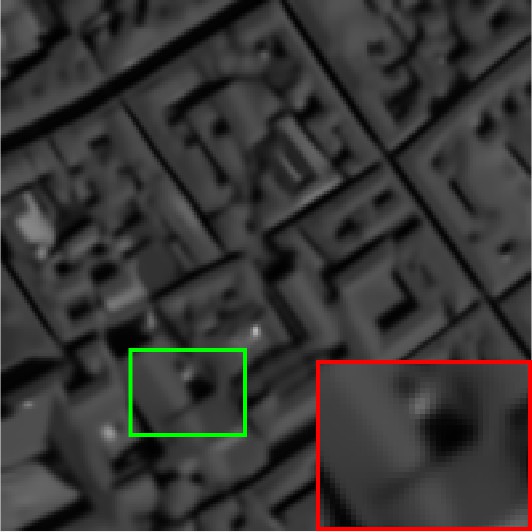}}%
	\hfil
	\subfloat[LRTA]{\includegraphics[width=0.22\linewidth]{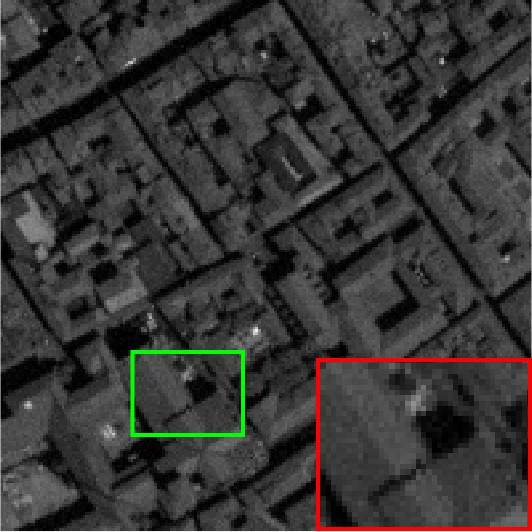}}%
	\hfil
	\subfloat[NAILRMA]{\includegraphics[width=0.22\linewidth]{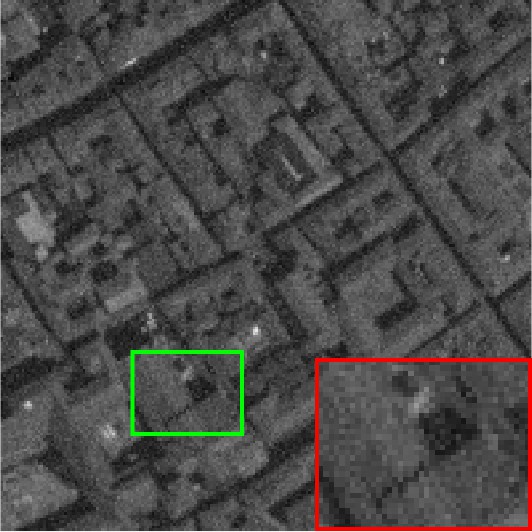}}%
	\hfil
	\subfloat[LRMR]{\includegraphics[width=0.22\linewidth]{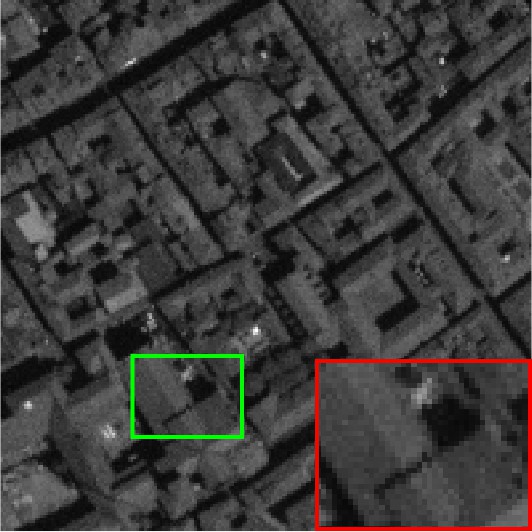}}%
	\hfil
	\subfloat[LLRSSTV]{\includegraphics[width=0.22\linewidth]{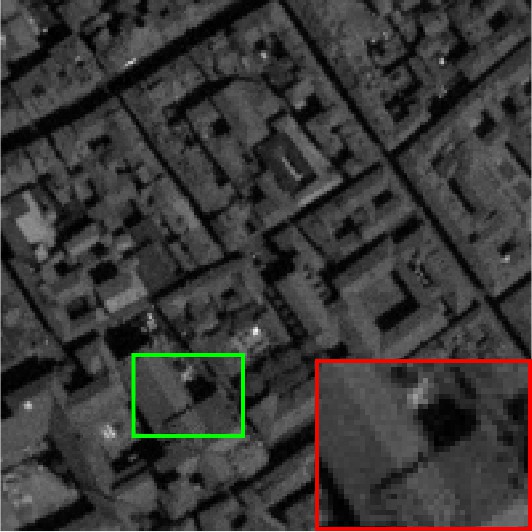}}%
	\hfil
	\subfloat[LLxRGTV]{\includegraphics[width=0.22\linewidth]{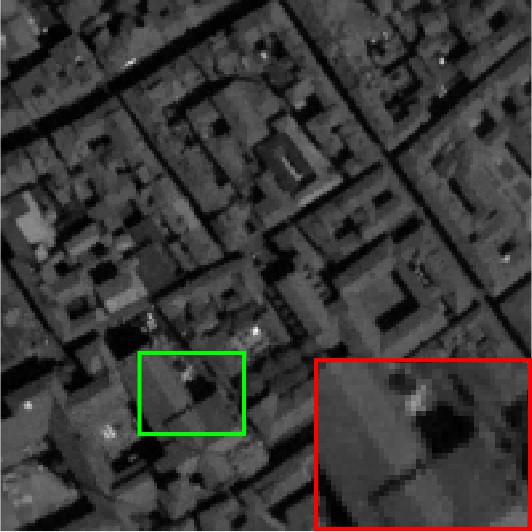}}%
	\caption{Denoised results of Pavia City Centre dataset in noise Case 1.}
	\label{fig:p02g01_band52_pavia}
\end{figure*}

\begin{figure*}[!t]
	\centering
	\subfloat[Original image]{\includegraphics[width=0.22\linewidth]{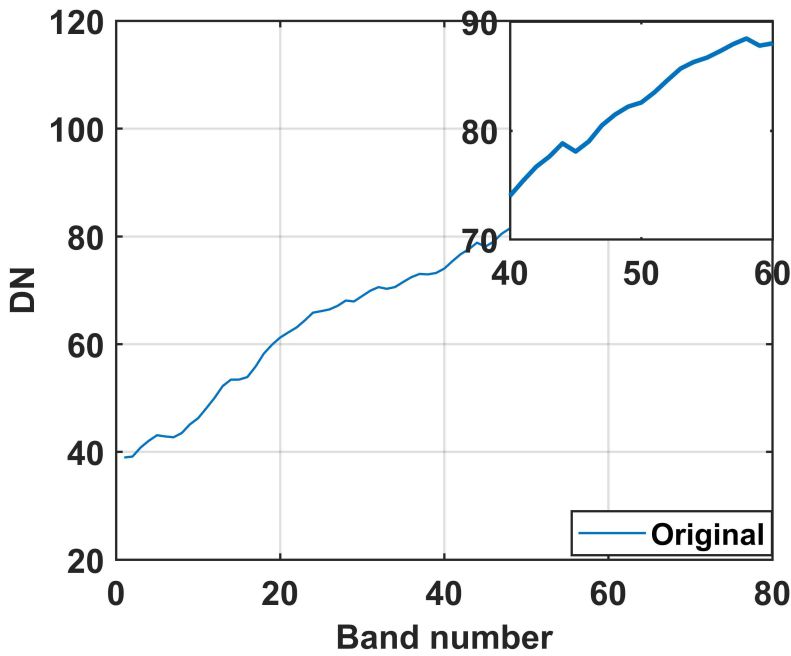}}%
	\hfil
	\subfloat[Noisy image]{\includegraphics[width=0.22\linewidth]{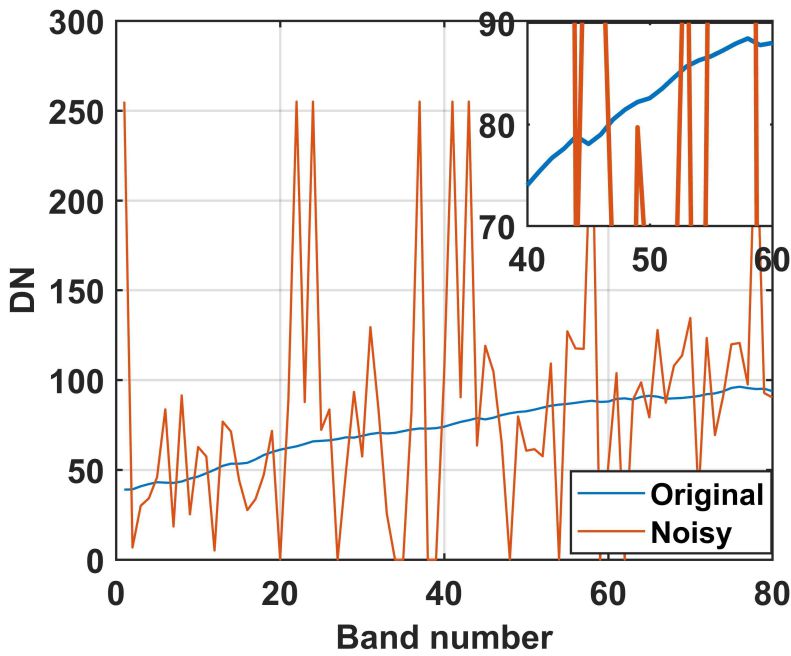}}%
	\hfil
	\subfloat[BM3D]{\includegraphics[width=0.22\linewidth]{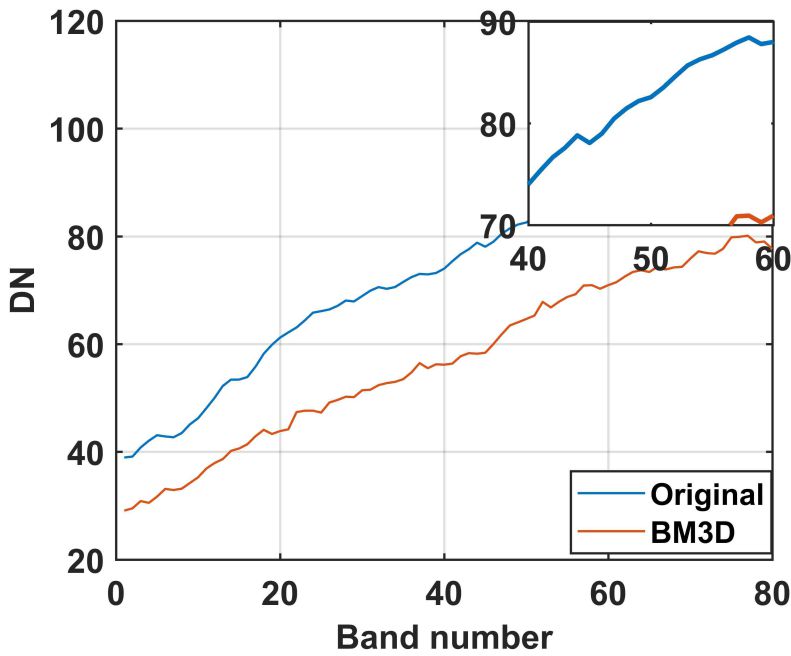}}%
	\hfil
	\subfloat[LRTA]{\includegraphics[width=0.22\linewidth]{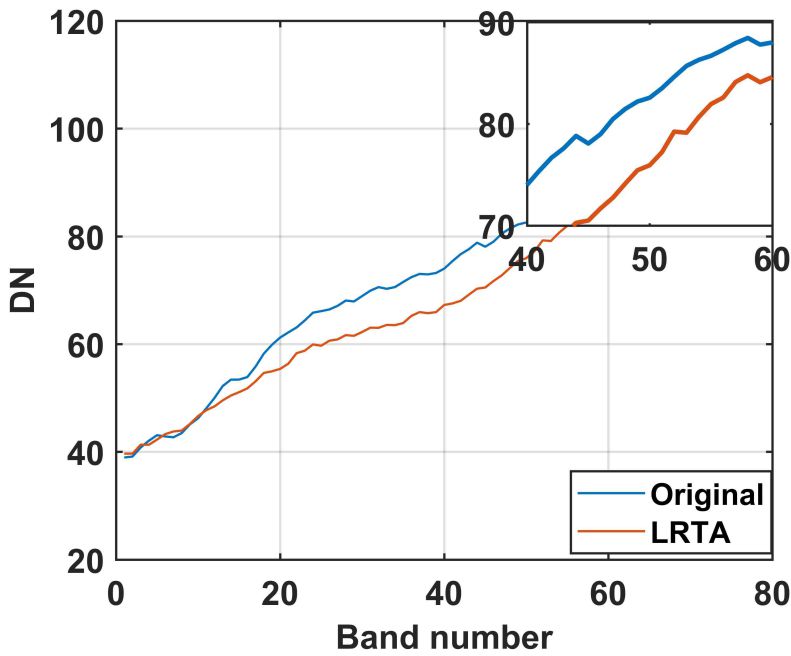}}%
	\hfil
	\subfloat[NAILRMA]{\includegraphics[width=0.22\linewidth]{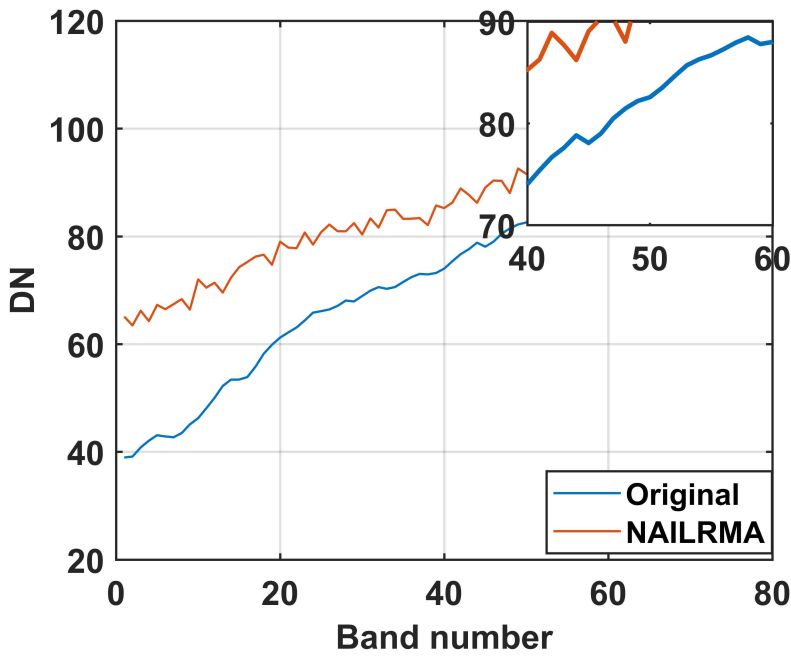}}%
	\hfil
	\subfloat[LRMR]{\includegraphics[width=0.22\linewidth]{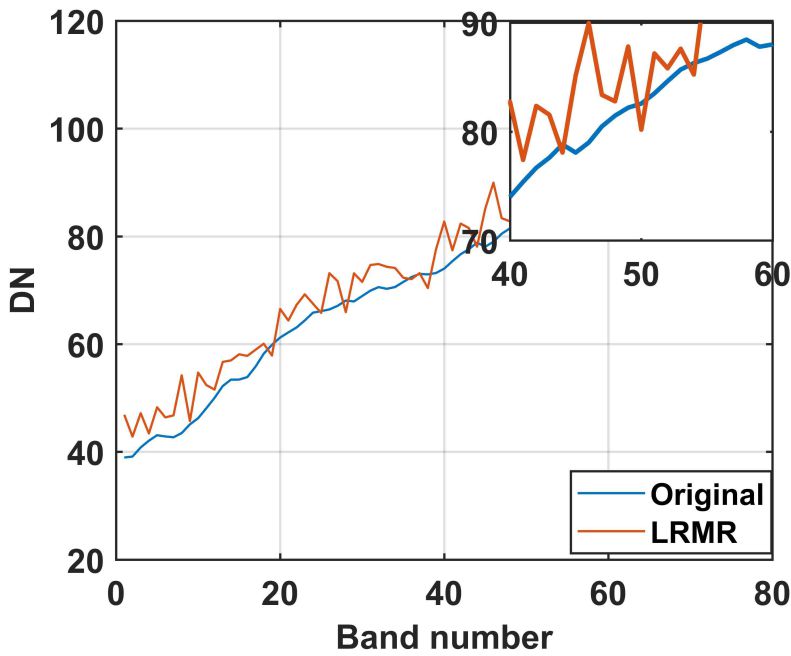}}%
	\hfil
	\subfloat[LLRSSTV]{\includegraphics[width=0.22\linewidth]{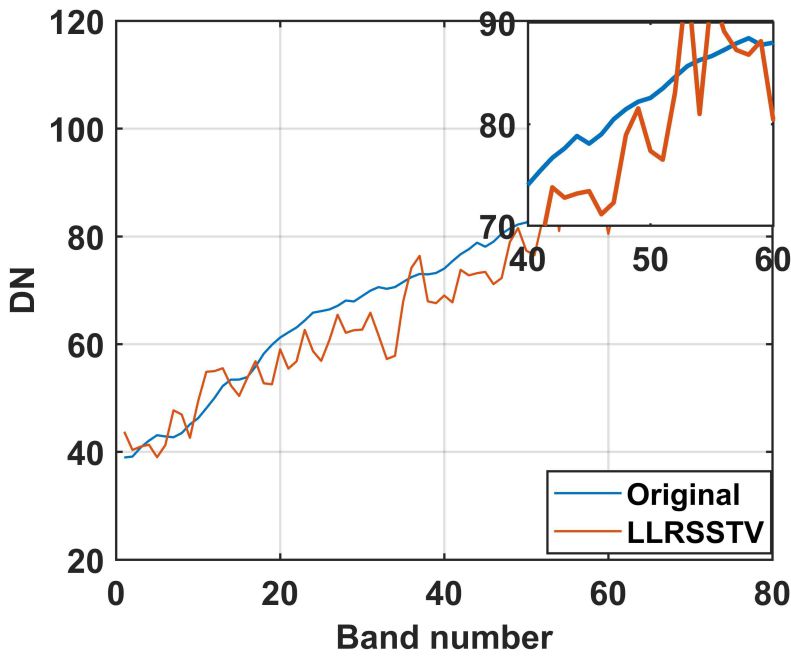}}%
	\hfil
	\subfloat[LLxRGTV]{\includegraphics[width=0.22\linewidth]{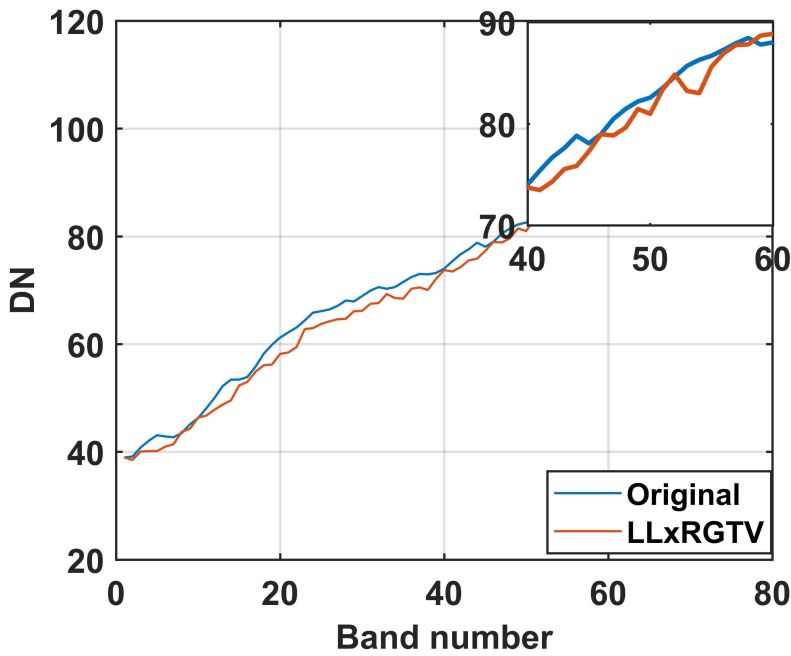}}%
	\caption{Spectrum of pixel (50, 100) in Pavia City Centre dataset of noise Case 1.}
	\label{fig:Spectrum_p02g01_pavia}
\end{figure*}

\begin{figure*}[!t]
	\centering
	\subfloat[Original image]{\includegraphics[width=0.22\linewidth]{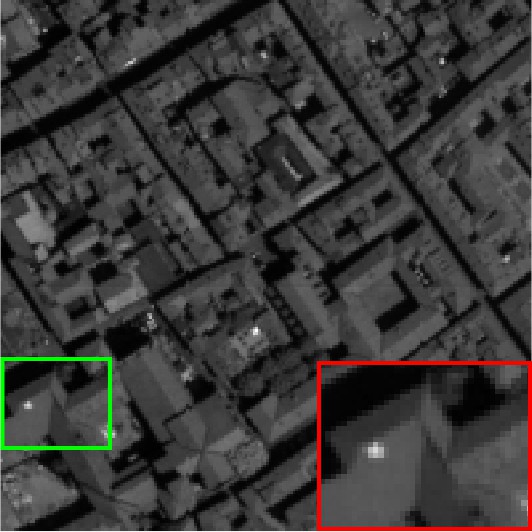}}%
	\hfil
	\subfloat[Noisy image]{\includegraphics[width=0.22\linewidth]{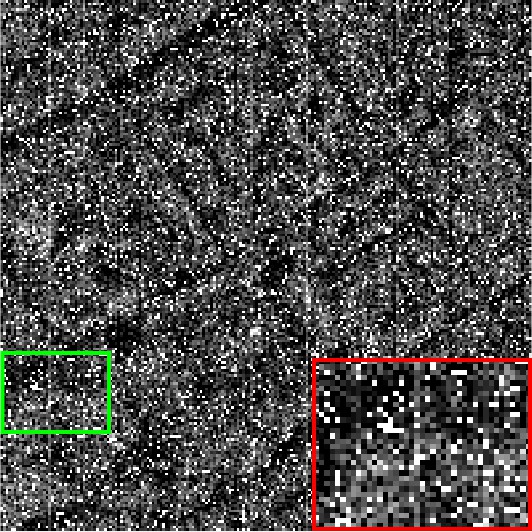}}%
	\hfil
	\subfloat[BM3D]{\includegraphics[width=0.22\linewidth]{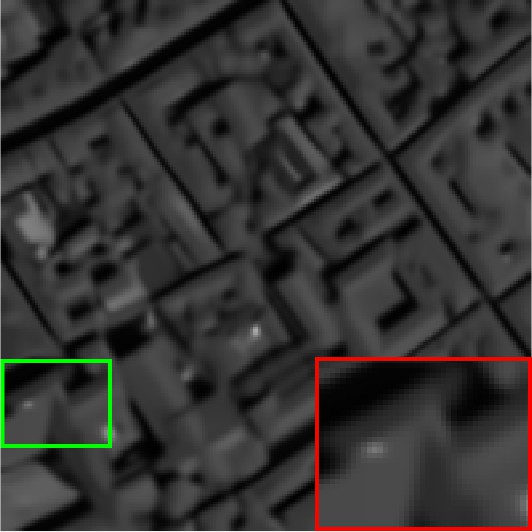}}%
	\hfil
	\subfloat[LRTA]{\includegraphics[width=0.22\linewidth]{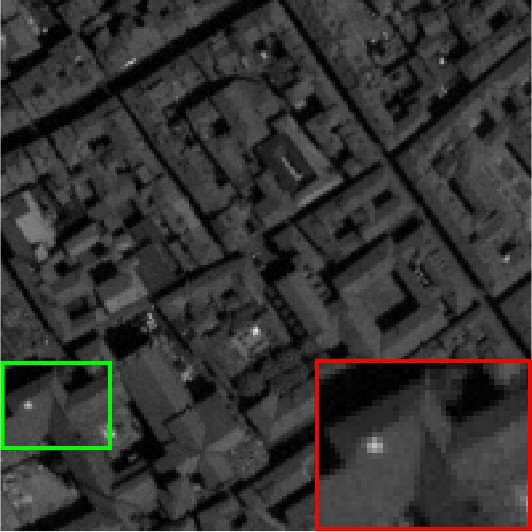}}%
	\hfil
	\subfloat[NAILRMA]{\includegraphics[width=0.22\linewidth]{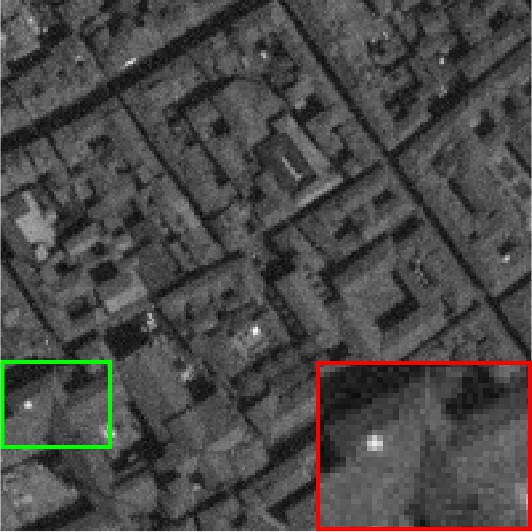}}%
	\hfil
	\subfloat[LRMR]{\includegraphics[width=0.22\linewidth]{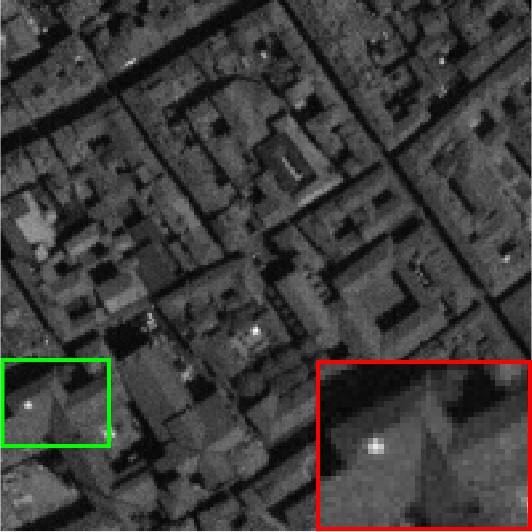}}
	\hfil
	\subfloat[LLRSSTV]{\includegraphics[width=0.22\linewidth]{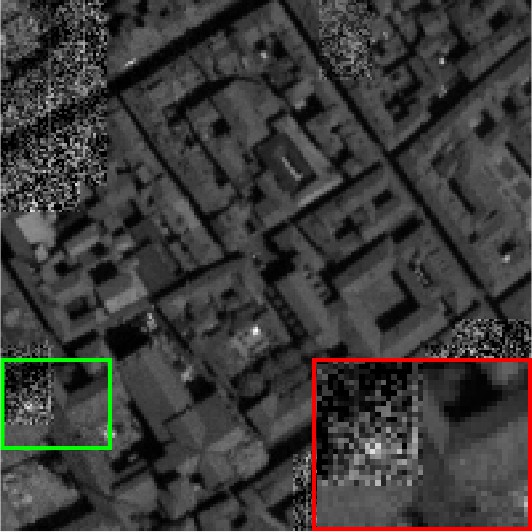}}%
	\hfil
	\subfloat[LLxRGTV]{\includegraphics[width=0.22\linewidth]{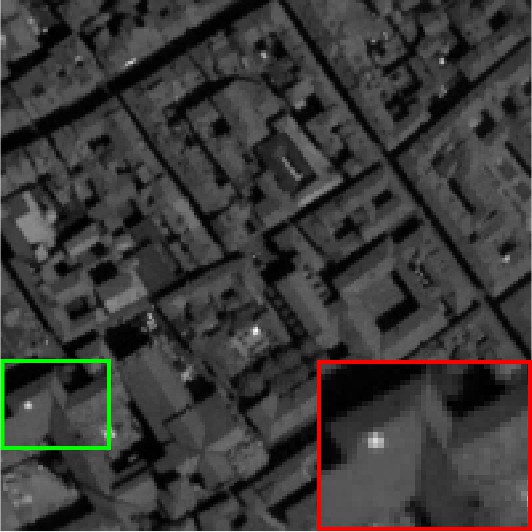}}%
	\caption{Denoised results of Pavia City Centre dataset in noise Case 4.}
	\label{fig:pgs_band52_pavia}
\end{figure*}

\begin{figure*}[!t]
	\centering
	\subfloat[Original image]{\includegraphics[width=0.22\linewidth]{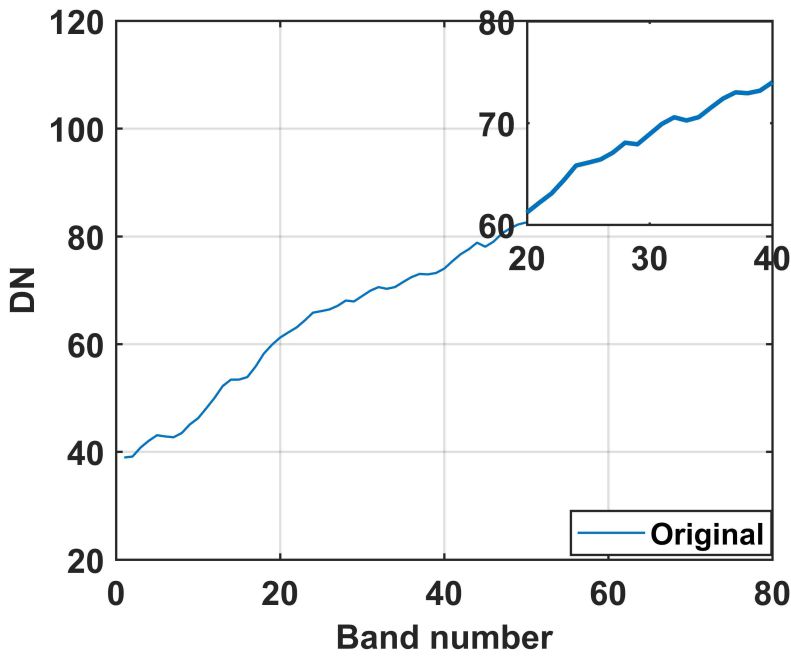}}%
	\hfil
	\subfloat[Noisy image]{\includegraphics[width=0.22\linewidth]{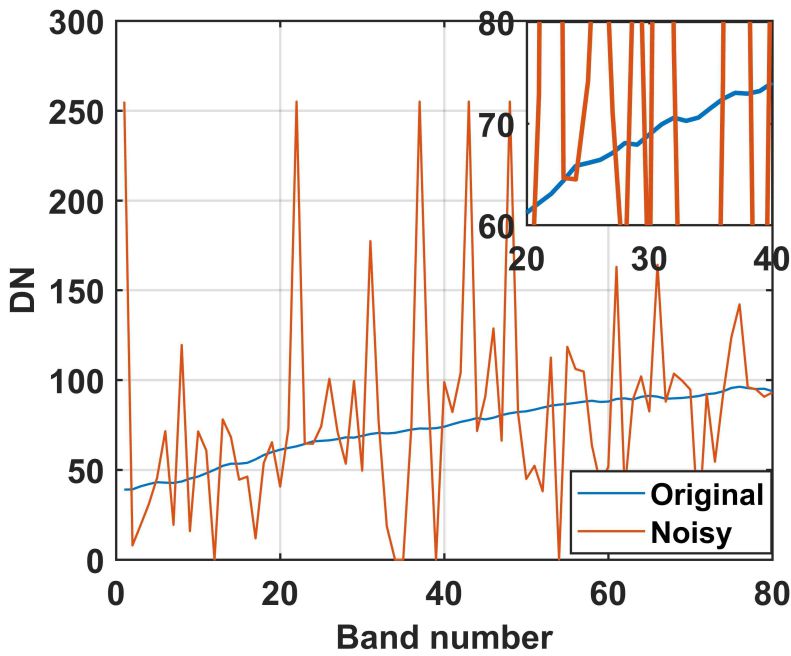}}%
	\hfil
	\subfloat[BM3D]{\includegraphics[width=0.22\linewidth]{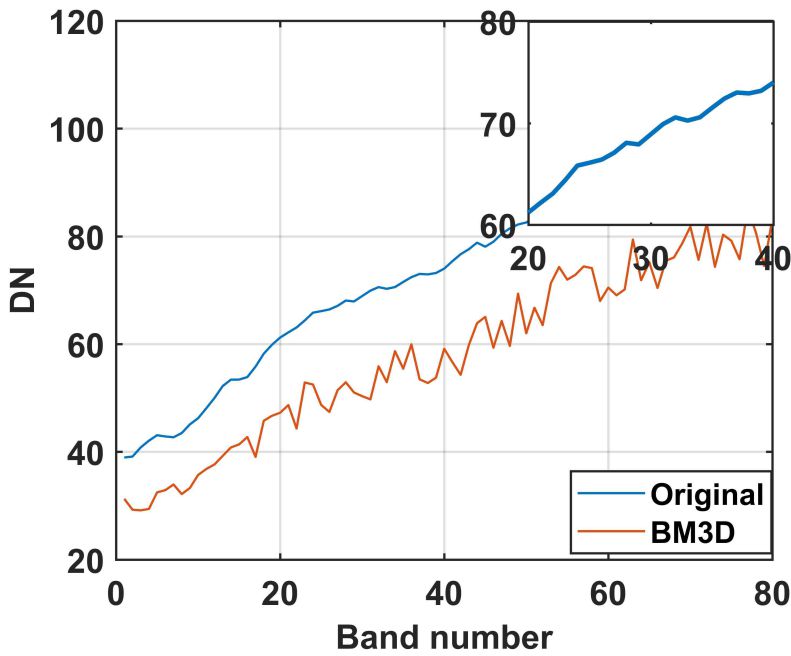}}%
	\hfil
	\subfloat[LRTA]{\includegraphics[width=0.22\linewidth]{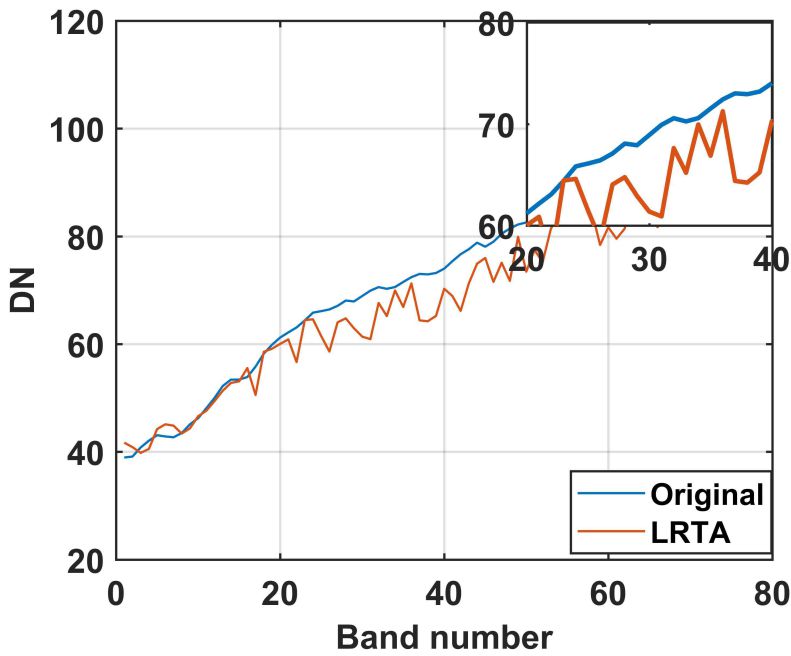}}%
	\hfil
	\subfloat[NAILRMA]{\includegraphics[width=0.22\linewidth]{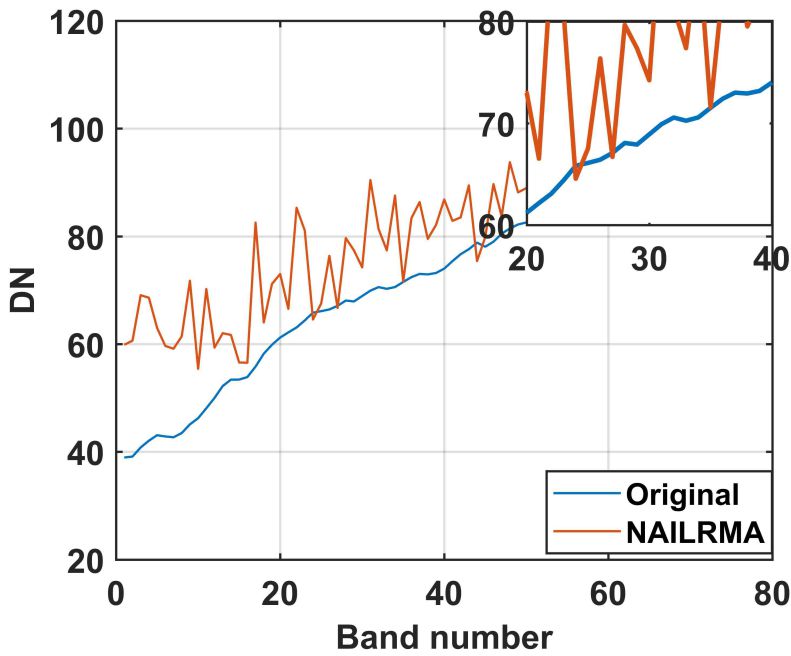}}%
	\hfil
	\subfloat[LRMR]{\includegraphics[width=0.22\linewidth]{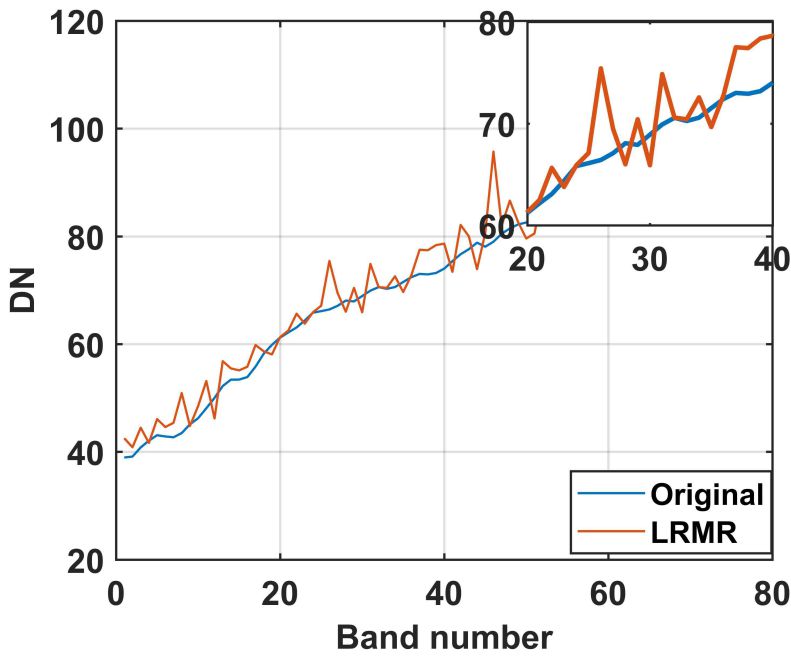}}%
	\hfil
	\subfloat[LLRSSTV]{\includegraphics[width=0.22\linewidth]{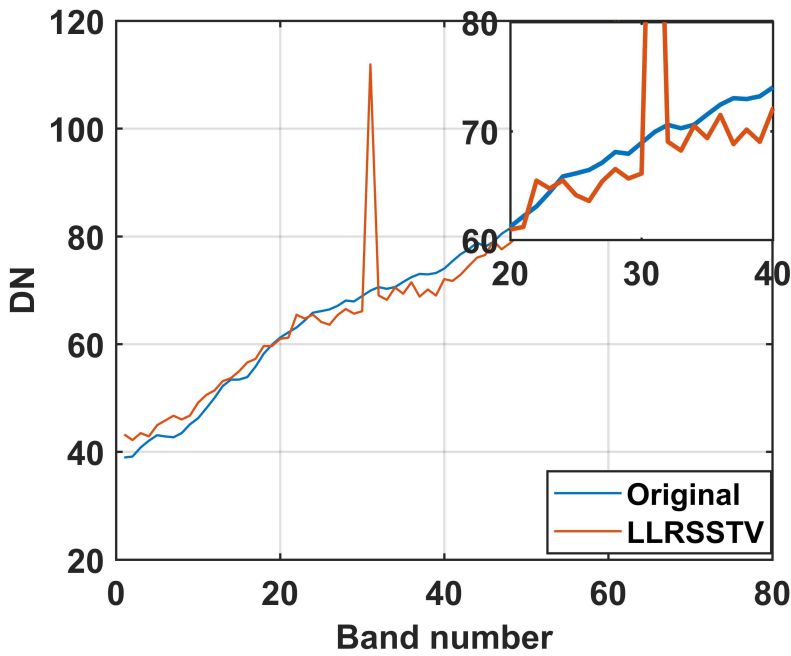}}%
	\hfil
	\subfloat[LLxRGTV]{\includegraphics[width=0.22\linewidth]{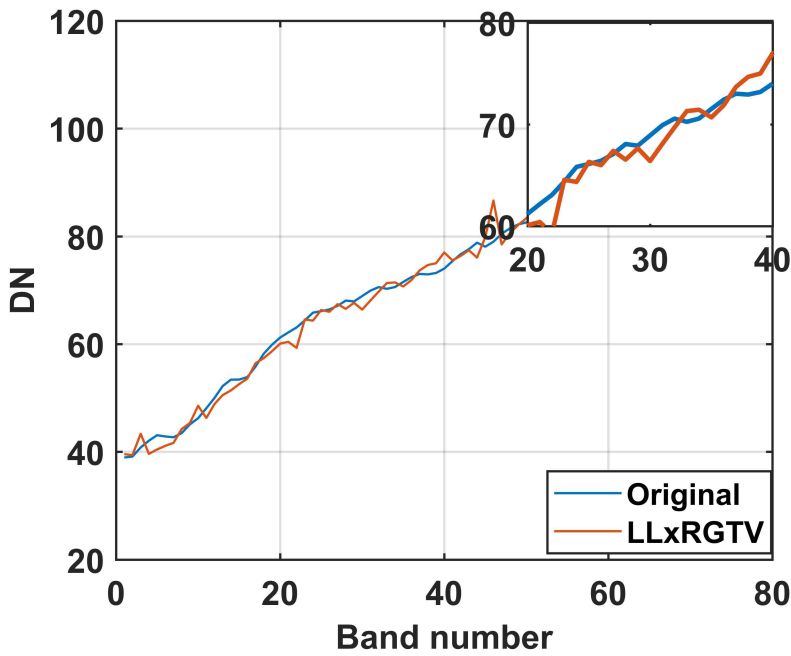}}%
	\caption{Spectrum of pixel (50, 100) in the Pavia City Centre dataset of noise Case 4.}
	\label{fig:Spectrum_pgs_pavia}
\end{figure*}

In this part, based on the simulation results under the above six noise cases, the proposed model is evaluated from three aspects, i.e., visual evaluation, spectral feature analysis and quantitative index evaluation.
For visual evaluation, in noise Case 1, we show the 125-th band of USGS Indian Pines and 52-th band of Pavia City Center in Fig. \ref{fig:p02g01_band125_indian} and Fig. \ref{fig:p02g01_band52_pavia}, respectively.
With the noise Case 4, Fig. \ref{fig:pgs_band52_pavia} shows the 52-th band of Pavia City Center.
With the noise Case 5, the 140-th band of USGS Indian Pines is shown in Fig. \ref{fig:pgd_band140_indian}.
Compared to other models, it can be seen that the result of our model is closest to the original reference image.
In addition, in order to further compare the performance of the models,
we show the spectral characteristics between the clean HSIs and the restored HSIs in Fig. \ref{fig:Spectrum_pgd_indian}, Fig. \ref{fig:Spectrum_p02g01_indian}, Fig. \ref{fig:Spectrum_p02g01_pavia} and Fig. \ref{fig:Spectrum_pgs_pavia}.
It is also clear that the spectral characteristics in results of our model are also closest to ones in the true image.
For quantitative comparison, Table \ref{tab:indian} and Table \ref{tab:pavia} list the PQIs of all the compared models in the six noise cases.
The best results for each PQI are marked in bold.
It is clear from Table 1 that in all noise cases our model achieves the best results compared to other methods.
It is worth noting that the proposed model is about 3.5 dB better in MPSNR compared to the suboptimal method.

\begin{table*}[!t]
	\caption{Quantitative evaluation of different methods of Pavia City Centre in all noise cases}
	\centering
	\label{tab:pavia}
	\begin{tabular}{llllllllllllr}
		\hline \hline
		Noise Case & Level & Evaluation index & BM3D & LRTA & NAILRMA & LRMR & LLRSSTV & LLxRGTV  \\
		\hline \hline
		\multirow{5}{*}{Case 1} & & MPSNR & 26.561 & 29.404 & 21.721 & 31.173 & \underline{32.581} & \textbf{33.212}\\
		& G=0.1 & MSSIM & 0.773 & 0.905 & 0.701 & 0.903 & \underline{0.921} & \textbf{0.937}\\
		& P=0.2 & MFSIM & 0.846 & 0.947 & 0.873 & 0.945 & \underline{0.959} & \textbf{0.964}\\
		& & ERGAS & 172.041 & 126.62 & 305.818 & 102.671 & \underline{87.53} & \textbf{81.856}\\
		& & MSAD & \underline{6.55} & 7.329 & 9.196 & 6.927 & 6.814 & \textbf{6.006}\\
		\hline
		\multirow{5}{*}{Case 2} & & MPSNR & 26.89 & 31.4 & 25.365 & 32.317 & \underline{34.424} & \textbf{36.27}\\
		& impulse & MSSIM & 0.779 & \underline{0.947} & 0.809 & 0.922 & 0.943 & \textbf{0.965}\\
		& +Gaussian & MFSIM & 0.85 & 0.968 & 0.909 & 0.955 &\underline{ 0.968} & \textbf{0.98}\\
		& & ERGAS & 167.352 & 107.701 & 212.022 & \underline{91.916} & 103.399 & \textbf{59.941}\\
		& & MSAD & 6.891 & \underline{6.585} & 9.347 & 6.744 & 9.594 & \textbf{4.965}\\
		\hline
		\multirow{5}{*}{Case 3} & & MPSNR & 27.204 & 32.471 & \textbf{37.978} & 33.72 & 35.604 & \underline{37.268}\\
		& & MSSIM & 0.784 & 0.955 & \textbf{0.975} & 0.941 & 0.951 & \underline{0.969}\\
		& Gaussian & MFSIM & 0.854 & 0.972 & \textbf{0.983} & 0.961 & 0.972 & \underline{0.982}\\
		& & ERGAS & 161.019 & 96.268 & \textbf{50.731} & 79.992 & 77.685 & \underline{53.635}\\
		& & MSAD & 6.601 & 6.185 & \textbf{4.561} & 7.584 & 8.04 & \underline{4.788}\\
		\hline
		\multirow{5}{*}{Case 4} & & MPSNR & 26.858 & 31.285 & 25.344 & 32.034 & \underline{34.015} & \textbf{35.992}\\
		& impulse & MSSIM & 0.778 & \underline{0.946} & 0.808 & 0.92 & 0.941 & \textbf{0.964}\\
		& +Gaussian & MFSIM & 0.85 & 0.967 & 0.908 & 0.953 & \underline{0.967} & \textbf{0.979}\\
		& +Stripes & ERGAS & 167.937 & 108.713 & 212.04 & \underline{94.221} & 104.946 & \textbf{61.264}\\
		& & MSAD & 6.901 & \underline{6.593} & 9.354 & 6.917 & 9.876 & \textbf{5.037}\\
		\hline
		\multirow{5}{*}{Case 5} & & MPSNR & 26.748 & 31.078 & 25.448 & 32.103 & \underline{33.991} & \textbf{35.957}\\
		& Gaussian & MSSIM &  0.777 & \underline{0.945} & 0.807 & 0.921 & 0.941 & \textbf{0.964}\\
		& +impulse & MFSIM & 0.849 & 0.967 & 0.907 & 0.953 & \underline{0.967} & \textbf{0.98}\\
		& +deadline & ERGAS & 170.534 & 112.466 & 210.545 & \underline{93.807} & 105.49 & \textbf{61.789}\\
		& & MSAD & 7.04 & \underline{6.774} & 9.636 & 6.943 & 9.795 & \textbf{5.117}\\
		\hline
		\multirow{5}{*}{Case 6} & & MPSNR & 26.712 & 30.966 & 25.359 & 31.824 & \underline{33.875} & \textbf{35.711}\\
		& Gaussian & MSSIM & 0.776 & \underline{0.944} & 0.805 & 0.918 & 0.939 & \textbf{0.963}\\
		& +impulse & MFSIM & 0.848 & 0.966 & 0.905 & 0.952 & \underline{0.966} & \textbf{0.979}\\
		& +deadline & ERGAS & 171.29 & 113.668 & 211.803 & \underline{96.265} & 104.097 & \textbf{63.048}\\
		& +stripe & MSAD & 7.074 & \underline{6.809} & 9.681 & 7.14 & 9.797 & \textbf{5.204}\\
		\hline \hline
	\end{tabular}
\end{table*}

\begin{figure*}[!t]
	\centering
	\includegraphics[width=0.5\linewidth]{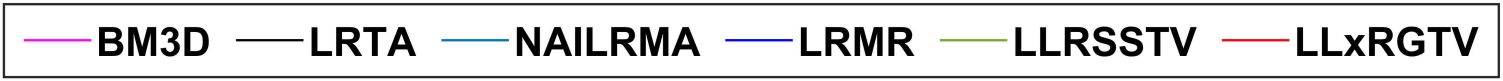}\\
	\subfloat[Noise Case 1]{\includegraphics[width=0.3\linewidth]{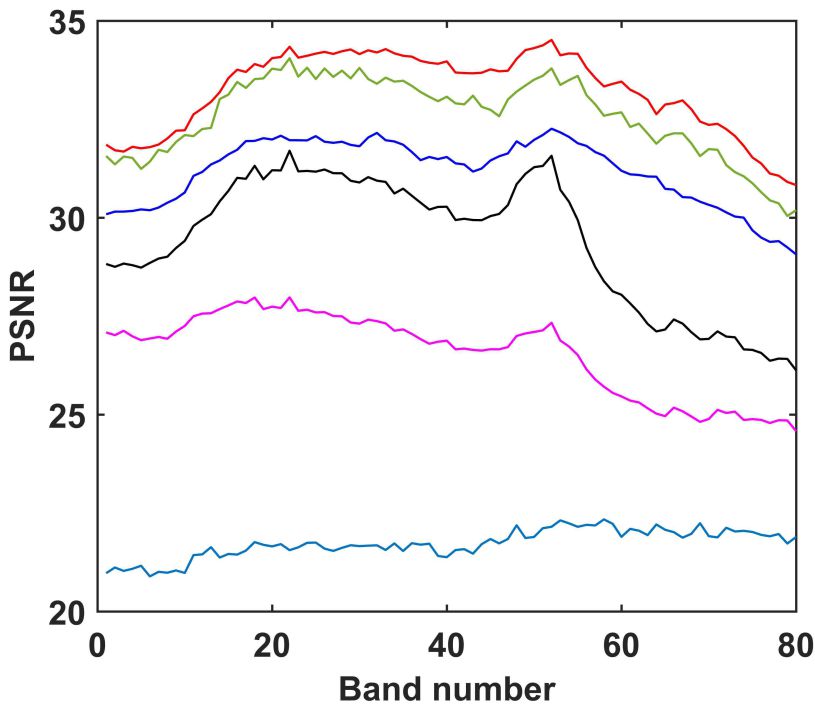}}%
	\hfil
	\subfloat[Noise Case 2]{\includegraphics[width=0.3\linewidth]{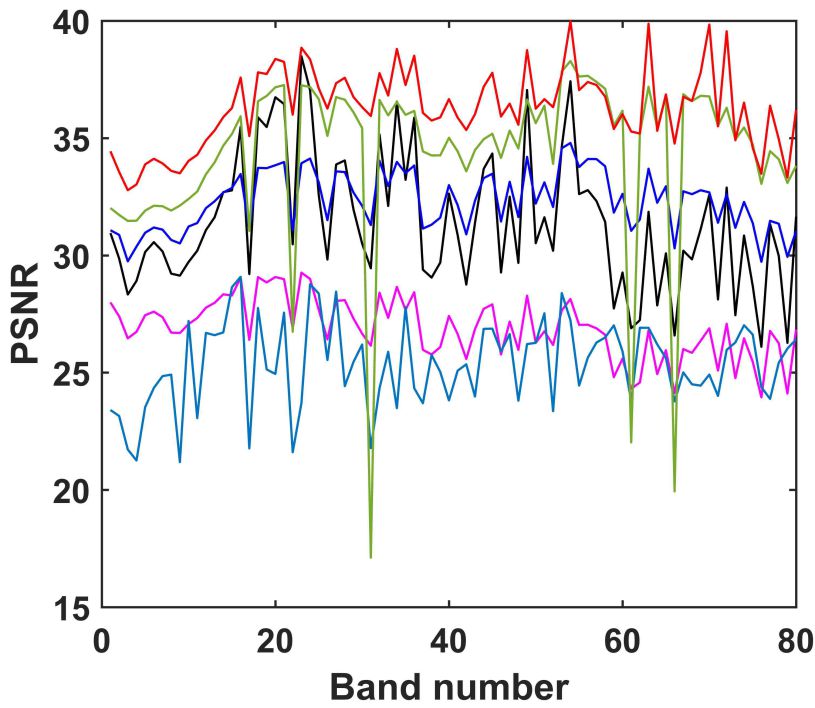}}%
	\hfil
	\subfloat[Noise Case 3]{\includegraphics[width=0.3\linewidth]{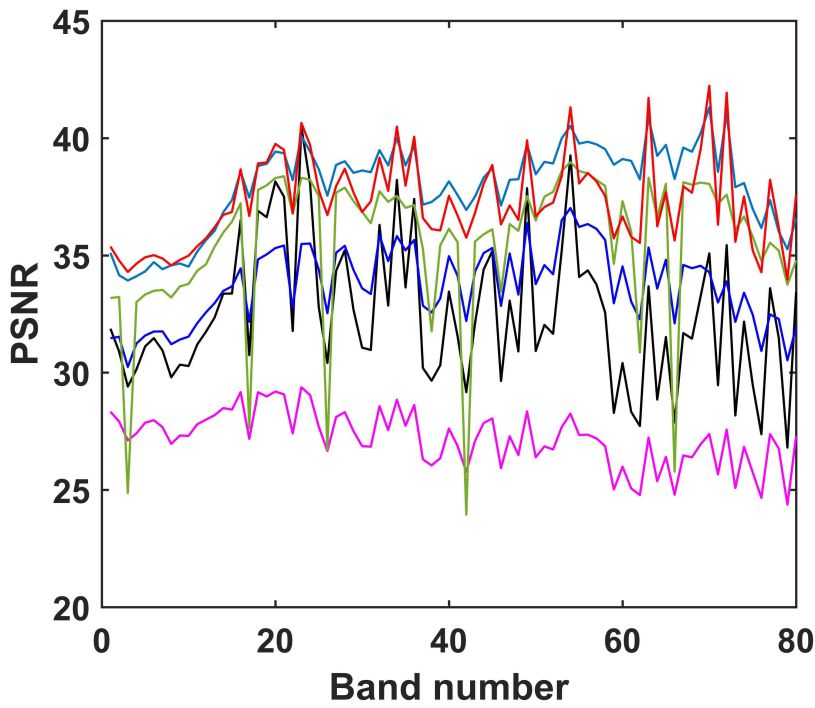}}%
	\hfil
	\subfloat[Noise Case 4]{\includegraphics[width=0.3\linewidth]{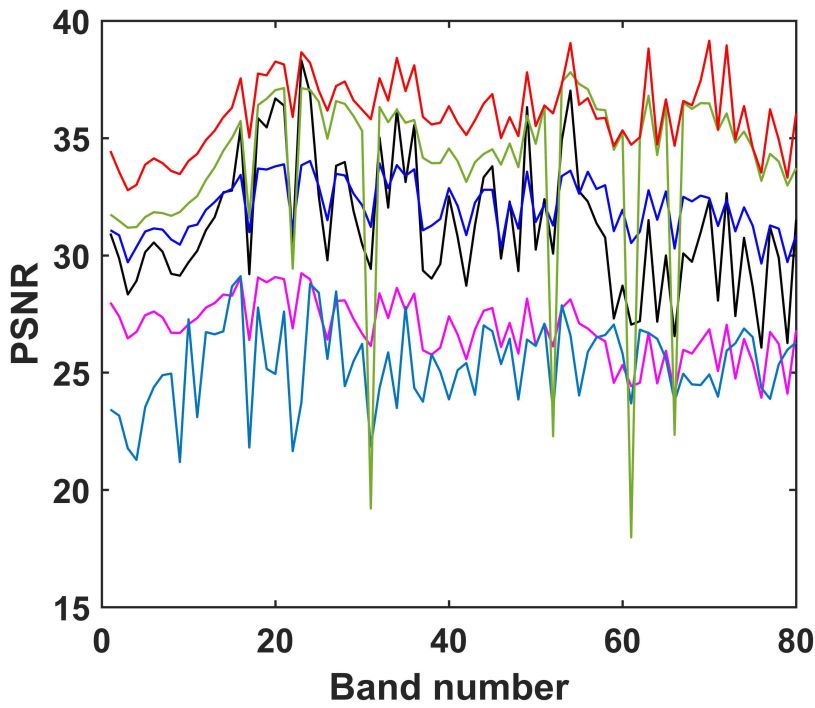}}%
	\hfil
	\subfloat[Noise Case 5]{\includegraphics[width=0.3\linewidth]{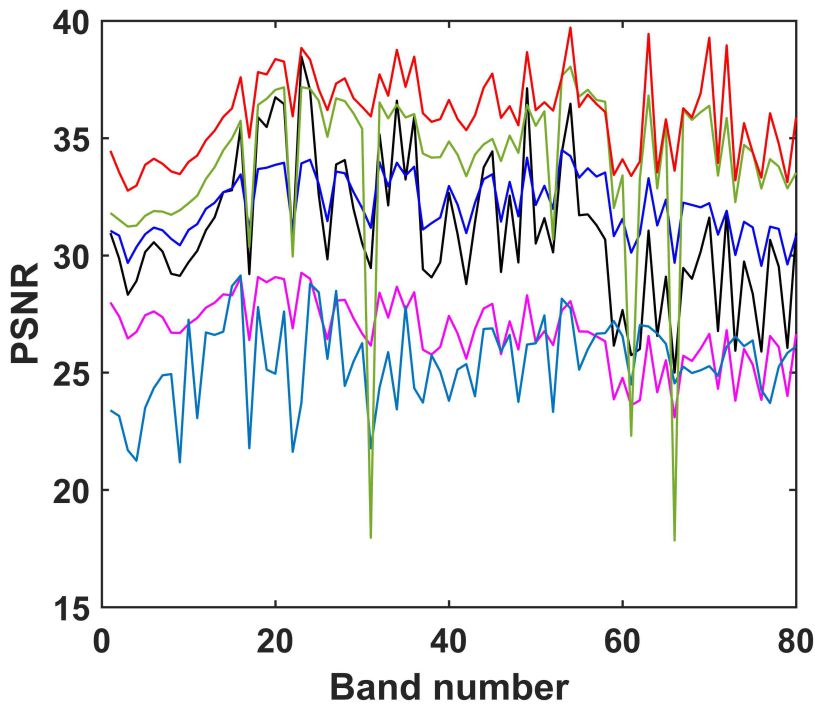}}%
	\hfil
	\subfloat[Noise Case 6]{\includegraphics[width=0.3\linewidth]{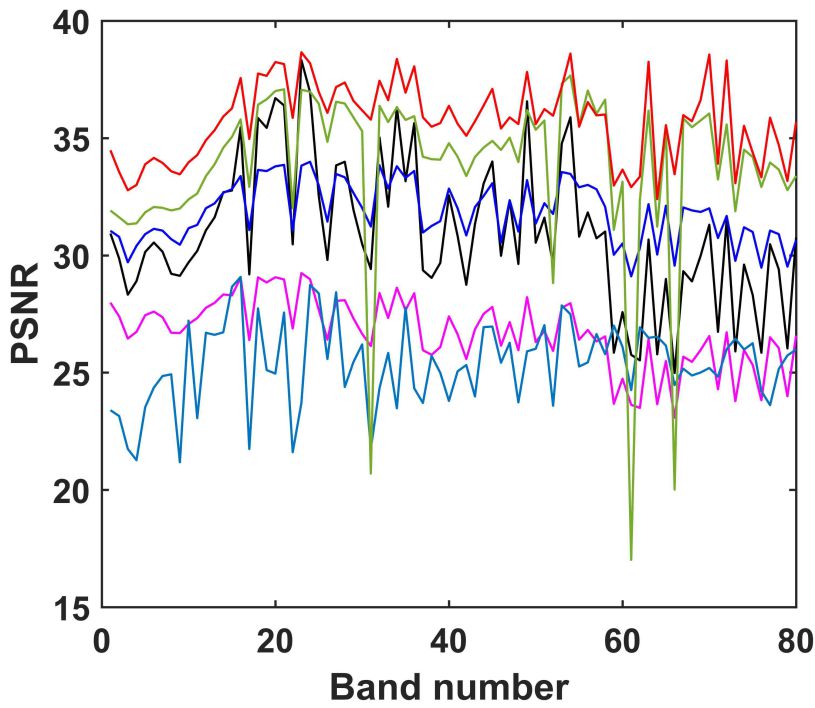}}%
	\caption{Detailed quantitative evaluation PSNR of Pavia City Centre dataset of different methods for each band.}
	\label{fig:psnr_pavia}
\end{figure*}

\begin{figure*}[!t]
	\centering
	\includegraphics[width=0.5\linewidth]{tuli_pavia.jpg}\\
	\subfloat[Noise Case 1]{\includegraphics[width=0.3\linewidth]{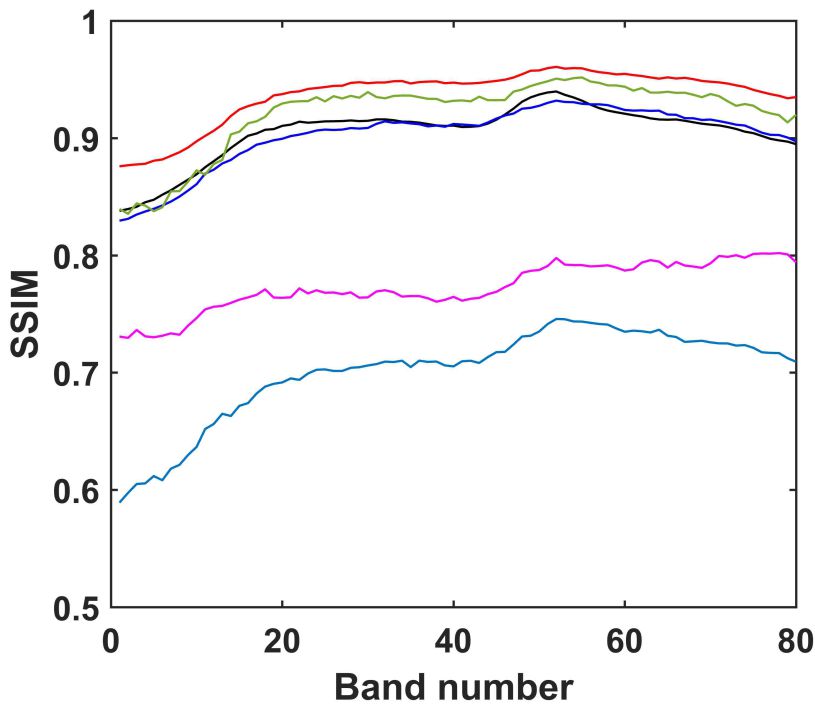}}%
	\hfil
	\subfloat[Noise Case 2]{\includegraphics[width=0.3\linewidth]{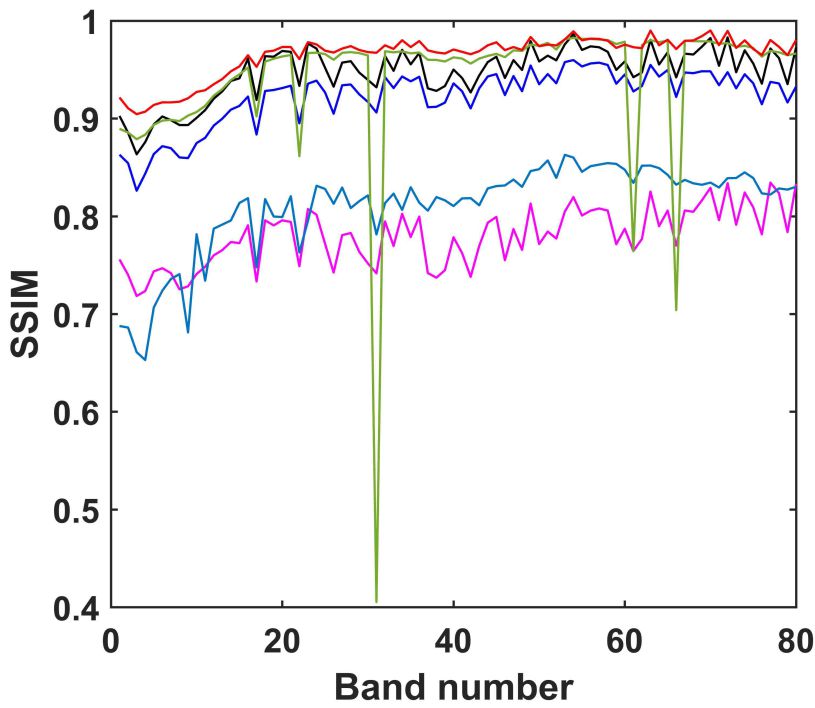}}%
	\hfil
	\subfloat[Noise Case 3]{\includegraphics[width=0.3\linewidth]{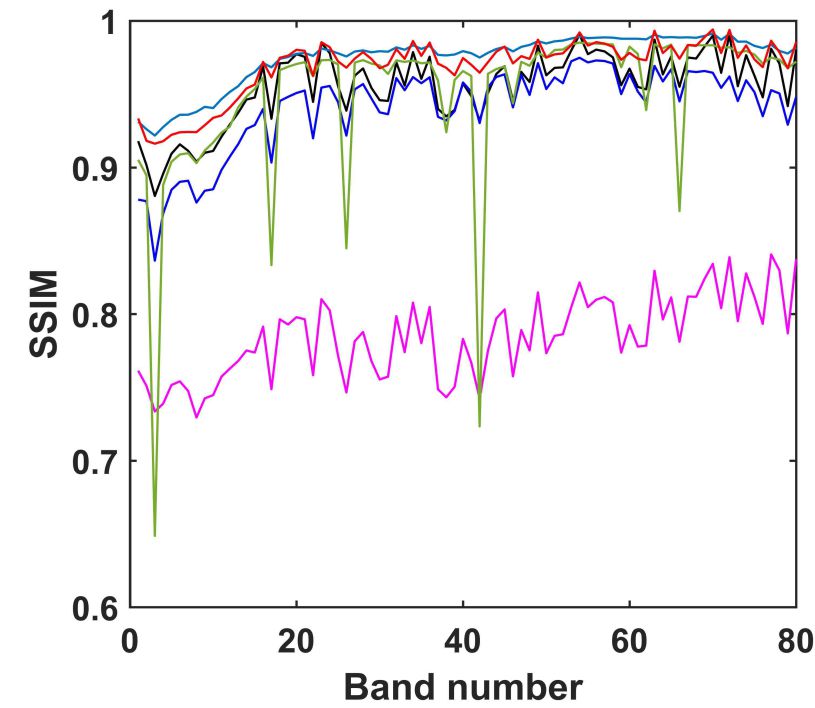}}%
	\hfil
	\subfloat[Noise Case: 4]{\includegraphics[width=0.3\linewidth]{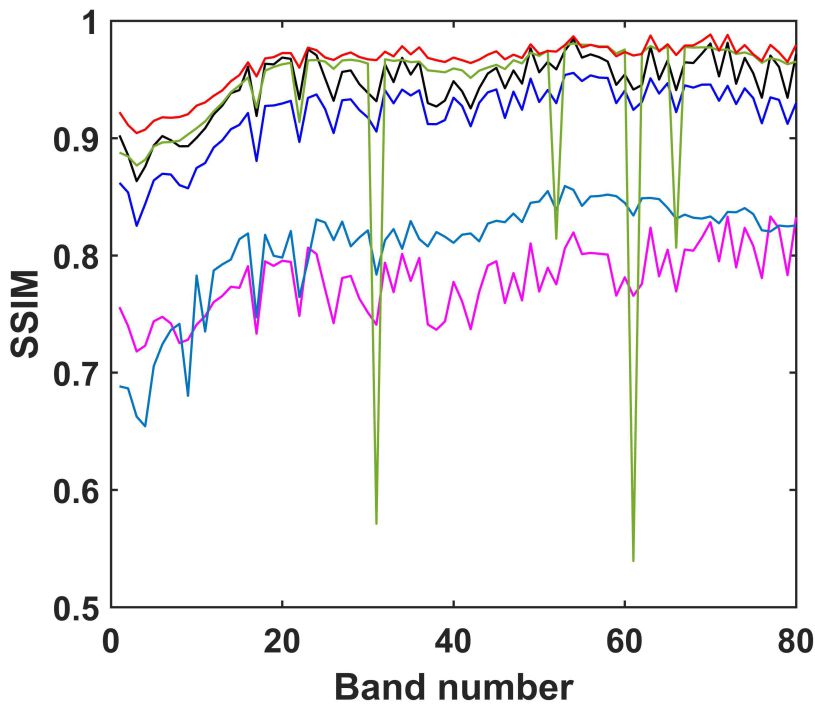}}%
	\hfil
	\subfloat[Noise Case: 5]{\includegraphics[width=0.3\linewidth]{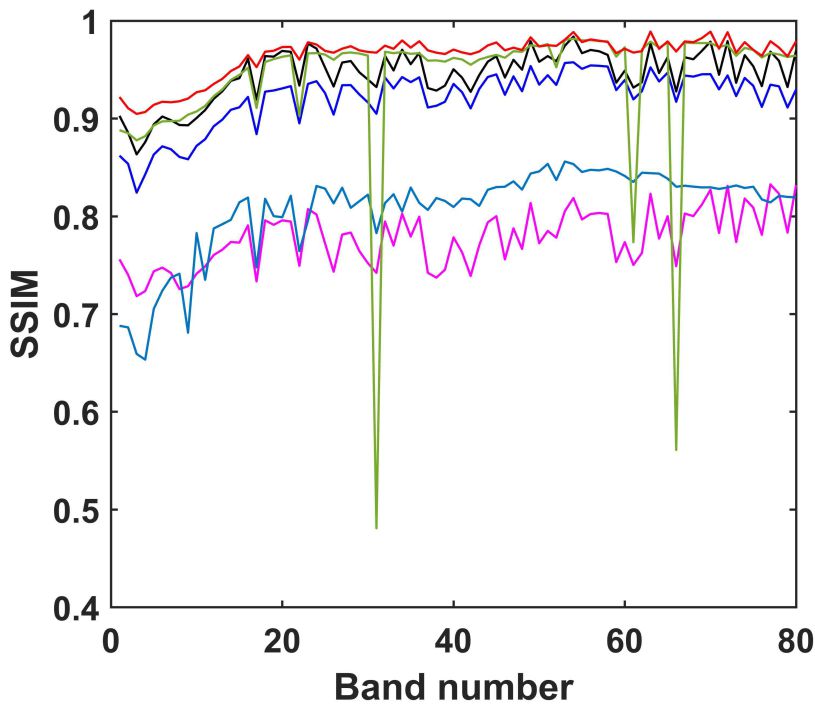}}%
	\hfil
	\subfloat[Noise Case: 6]{\includegraphics[width=0.3\linewidth]{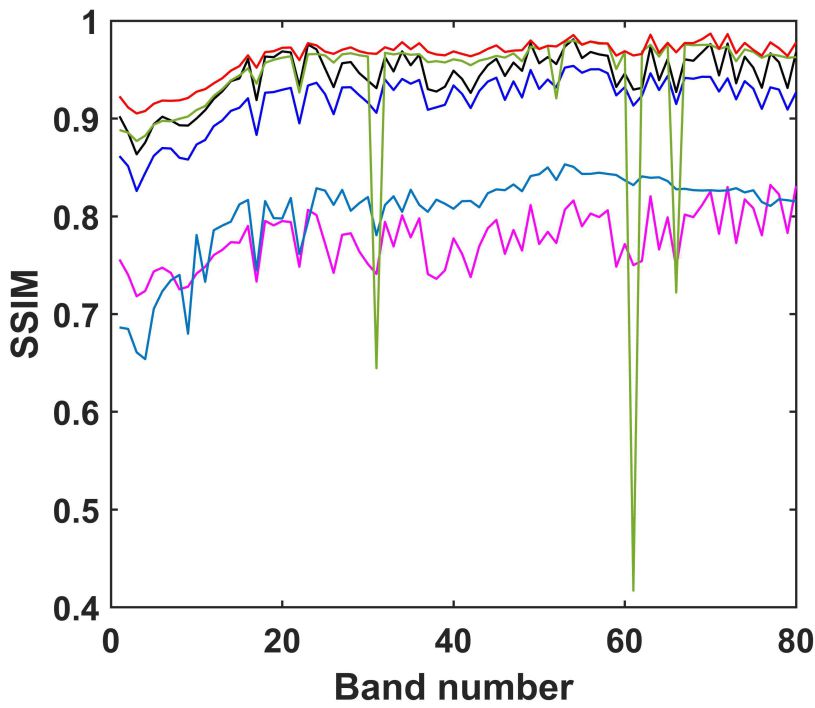}}%
	\caption{Detailed quantitative evaluation SSIM of Pavia City Centre dataset of different methods for each band.}
	\label{fig:ssim_pavia}
\end{figure*}

\begin{figure}[t]
	\centering
	\includegraphics[width=0.4\linewidth]{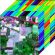}		
	\includegraphics[width=0.4\linewidth]{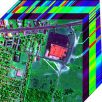}	
	\caption{Dataset used in the real data experiments.
		Left: AVIRIS Indian Pines  dataset (R: 1, G: 103, B: 220).
		Right: HYDICE Urban dataset (R: 20, G: 90, B: 180). }
	\label{3dcubeofurban}
\end{figure}

\subsection{Real HSI dataset experiments}

In this subsection, two HSI datasets with real noise are adopted in our experiments.
They are AVIRIS Indian Piness dataset and  HYDICE Urban dataset in Fig. \ref{3dcubeofurban}.

\subsubsection{AVIRIS Indian Pines dataset}
\label{section:AVIRIS_indian_analysis}

The AVIRIS Indian Piness dataset \footnote{https://engineering.purdue.edu/~biehl/MultiSpec/hyperspectral.html} is collected by the Airborne Visible Infrared Imaging Spectrometer (AVIRIS).
The size of the space is $145\times145$, with a total of 220 bands.
Some bands of the dataset are seriously polluted by impulse noise and Gaussian white noise, while some bands preserve high image quality.

\begin{figure*}[!t]
	\centering
	\subfloat[Original band]{\includegraphics[width=0.15\linewidth]{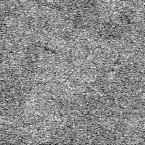}}%
	\hfil
	\subfloat[LRTA]{\includegraphics[width=0.15\linewidth]{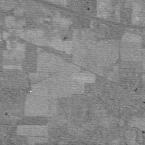}}%
	\hfil
	\subfloat[NAILRMA]{\includegraphics[width=0.15\linewidth]{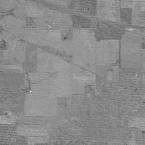}}%
	\hfil
	\subfloat[LRMR]{\includegraphics[width=0.15\linewidth]{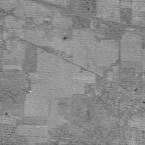}}%
	\hfil
	\subfloat[LLRSSTV]{\includegraphics[width=0.15\linewidth]{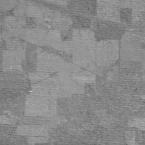}}%
	\hfil
	\subfloat[LLxRGTV]{\includegraphics[width=0.15\linewidth]{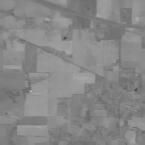}}%
	\caption{Denoised results in the real experiments of AVIRIS Indian Pines.}
	\label{fig:real_indian_b108}
\end{figure*}

\begin{figure*}[!t]
	\centering
	\subfloat[Original band]{\includegraphics[width=0.15\linewidth]{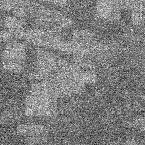}}%
	\hfil
	\subfloat[LRTA]{\includegraphics[width=0.15\linewidth]{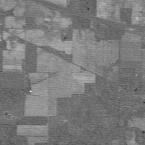}}%
	\hfil
	\subfloat[NAILRMA]{\includegraphics[width=0.15\linewidth]{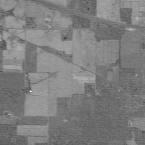}}%
	\hfil
	\subfloat[LRMR]{\includegraphics[width=0.15\linewidth]{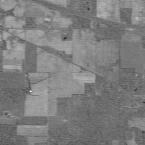}}%
	\hfil
	\subfloat[LLRSSTV]{\includegraphics[width=0.15\linewidth]{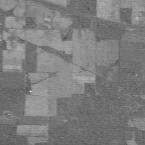}}%
	\hfil
	\subfloat[LLxRGTV]{\includegraphics[width=0.15\linewidth]{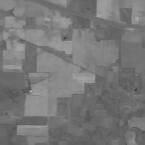}}%
	\caption{Denoised results in the real experiments of AVIRIS Indian Pines.}
	\label{fig:real_indian_b220}
\end{figure*}

The performance of the proposed model is evaluated from two aspects, i.e., visual evaluation and vertical mean profiles analysis.
For visual evaluation, Fig. \ref{fig:real_indian_b108} and Fig. \ref{fig:real_indian_b220}  show the comparison of the 108-th and 220-th band of AVIRIS Indian Pines before and after denoising.
It can be seen that the original image is heavily polluted by noise and its ground features are basically unrecognizable.
Although all the comparison methods have restored the main information of the features,
the restored images still have some residual noise locally and its local textures are still not recognized.
The proposed model not only restores the main information of the image, but also removes local noise.
Therefore, the restored image has clearer texture information.

\begin{figure*}[!t]
	\centering
	\subfloat[Original band]{\includegraphics[width=0.15\linewidth]{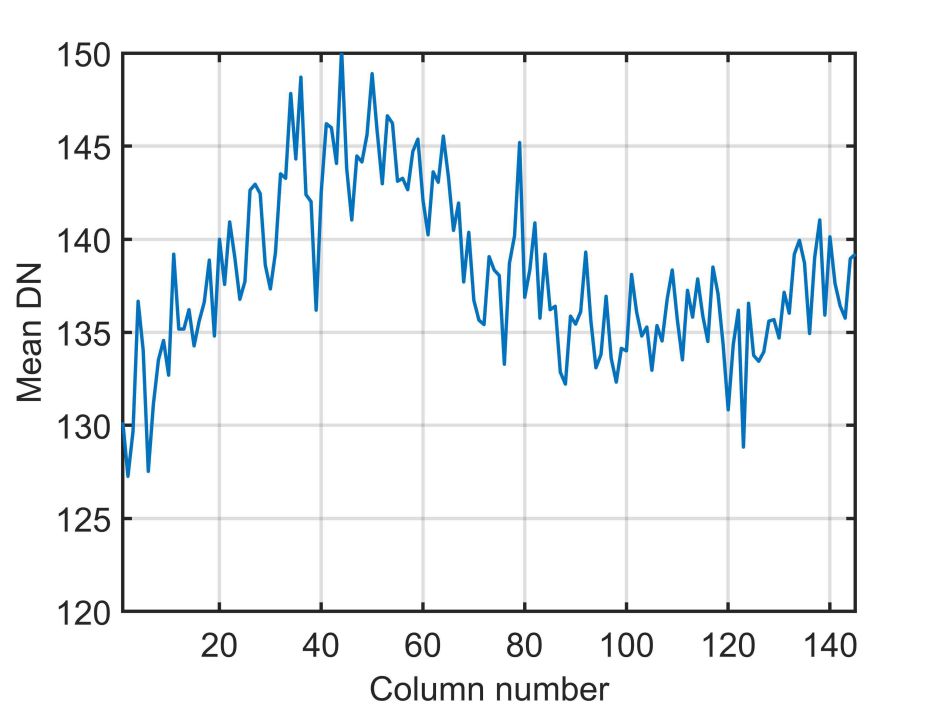}}%
	\hfil
	\subfloat[LRTA]{\includegraphics[width=0.15\linewidth]{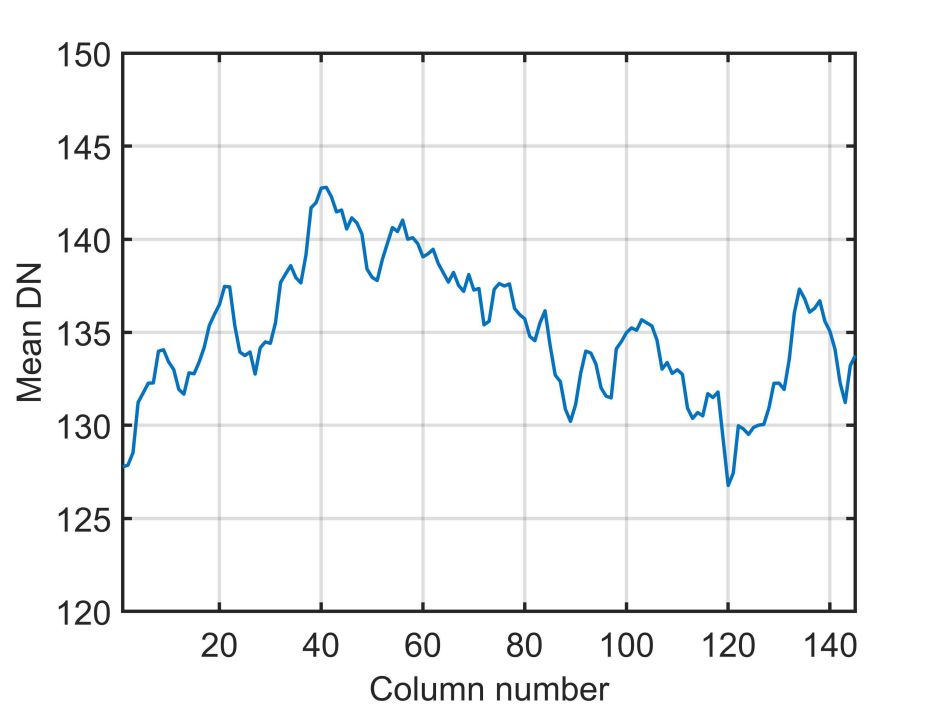}}%
	\hfil
	\subfloat[NAILRMA]{\includegraphics[width=0.15\linewidth]{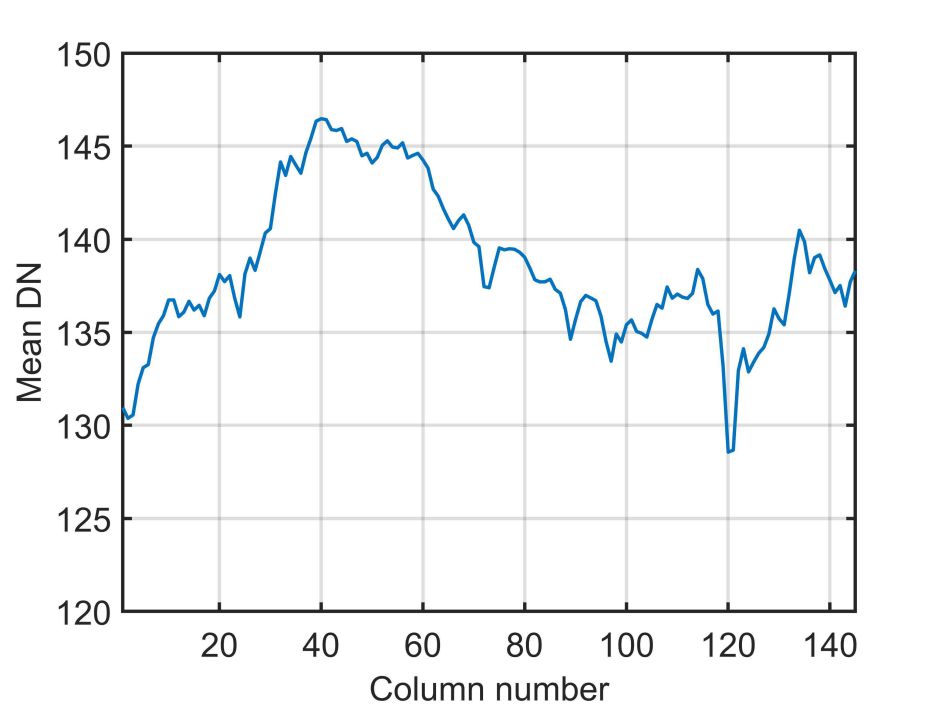}}%
	\hfil
	\subfloat[LRMR]{\includegraphics[width=0.15\linewidth]{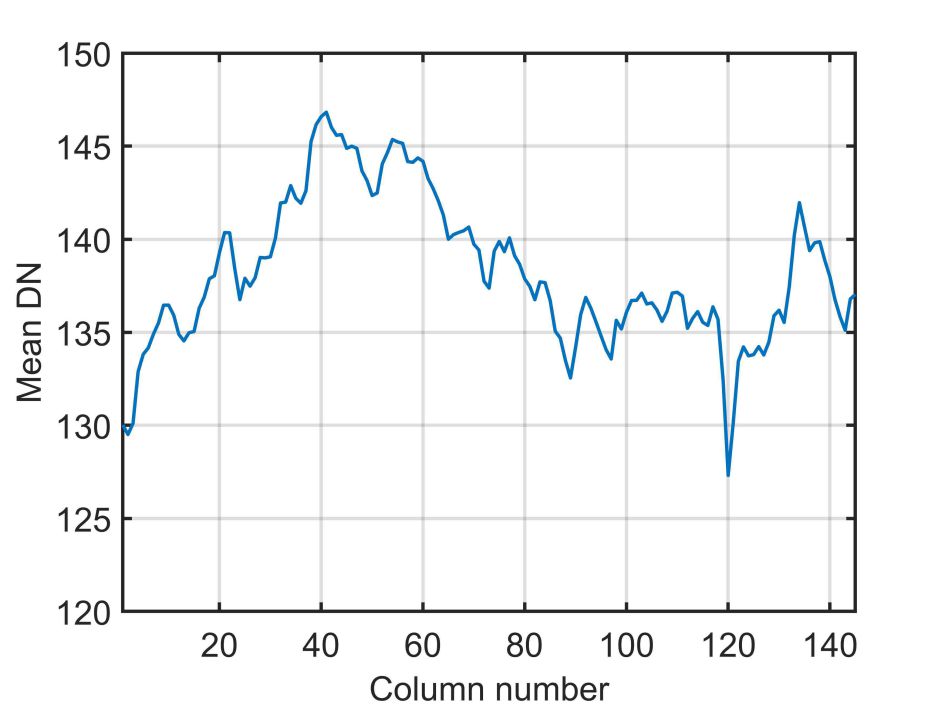}}%
	\hfil
	\subfloat[LLRSSTV]{\includegraphics[width=0.15\linewidth]{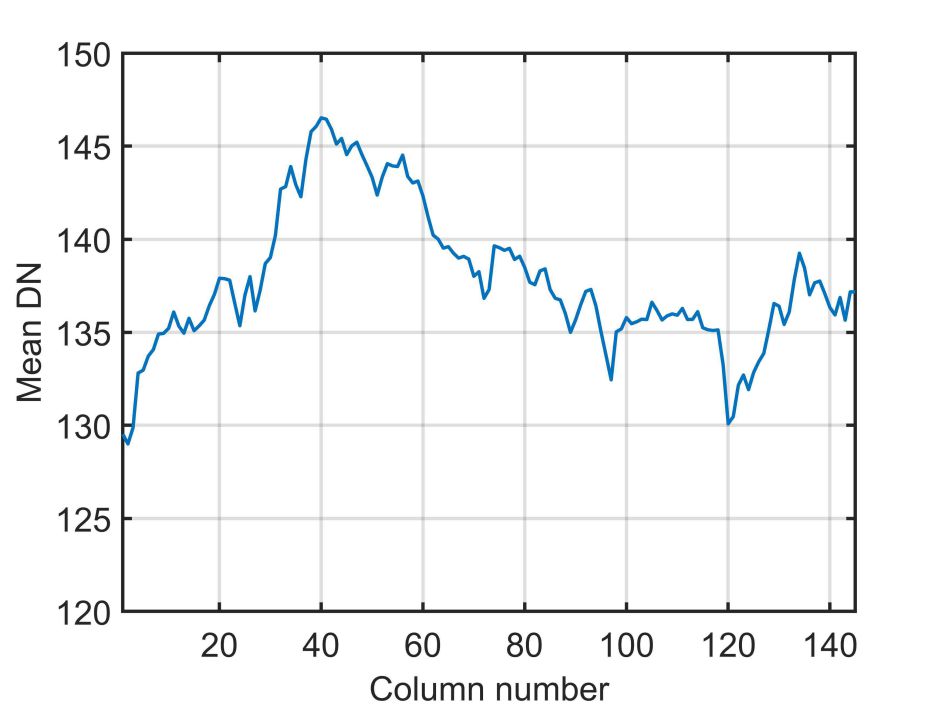}}%
	\hfil
	\subfloat[LLxRGTV]{\includegraphics[width=0.15\linewidth]{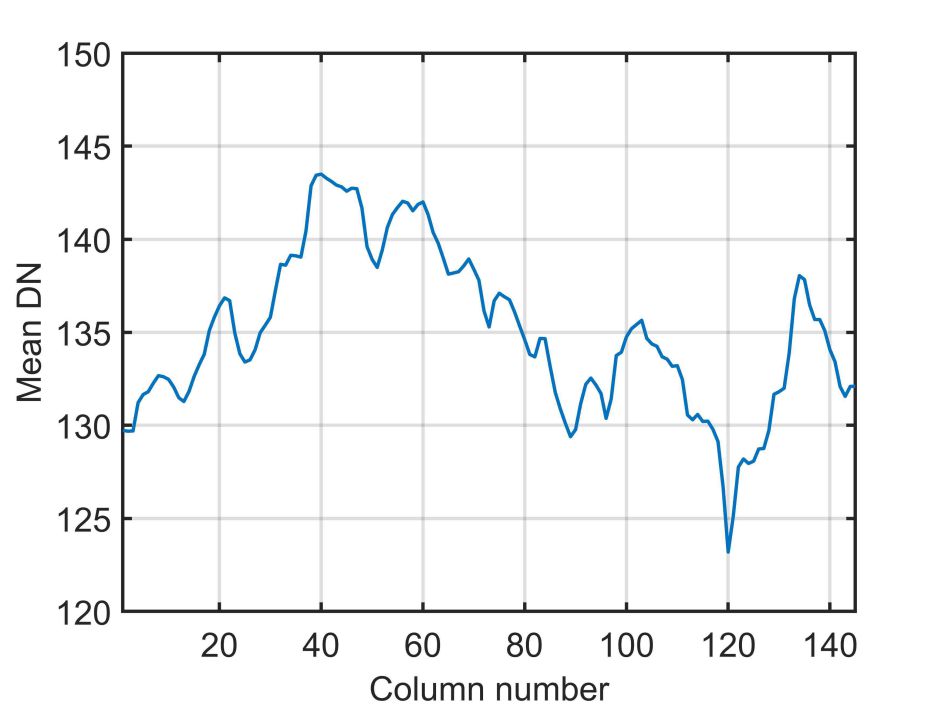}}%
	\caption{Vertical mean profiles of band 108 in the AVIRIS Indian Pines dataset experiment.}
	\label{fig:DN_real_indian_b108}
\end{figure*}

\begin{figure*}[!t]
	\centering
	\subfloat[Original band]{\includegraphics[width=0.15\linewidth]{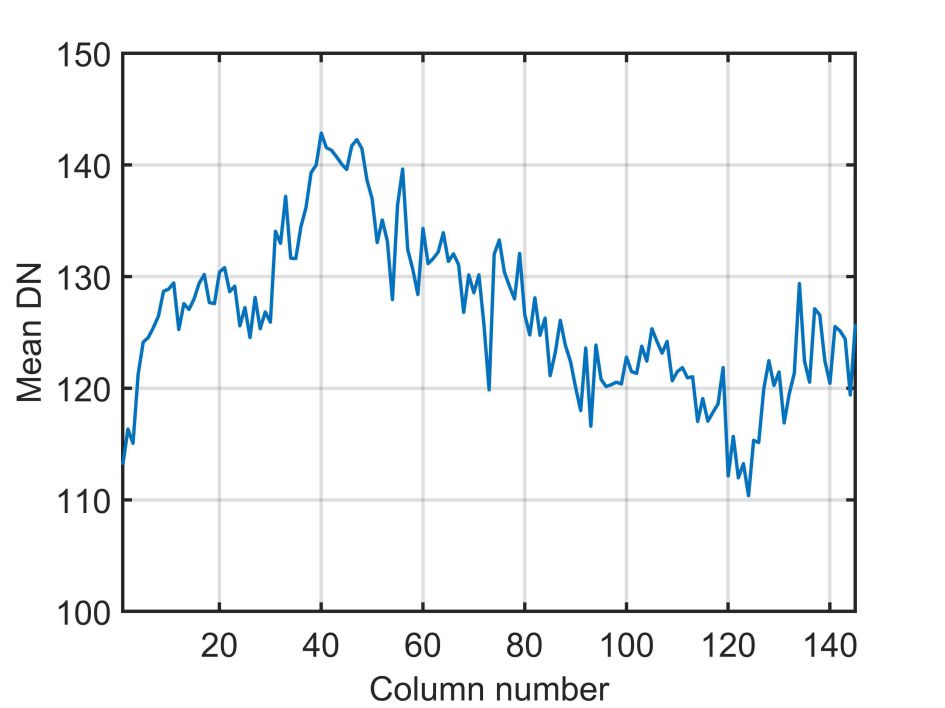}}%
	\hfil
	\subfloat[LRTA]{\includegraphics[width=0.15\linewidth]{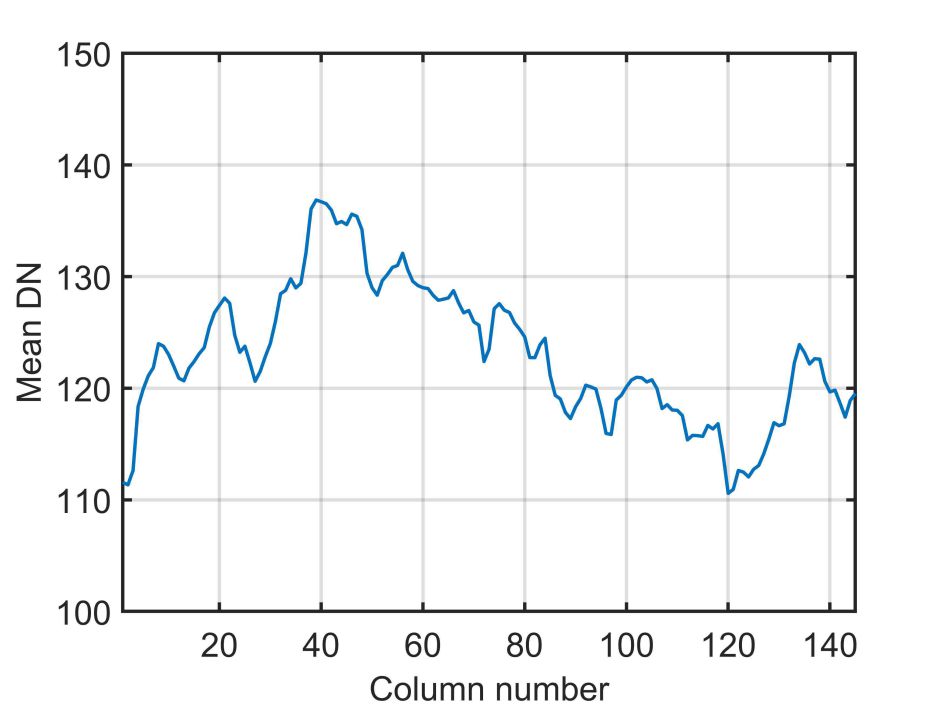}}%
	\hfil
	\subfloat[NAILRMA]{\includegraphics[width=0.15\linewidth]{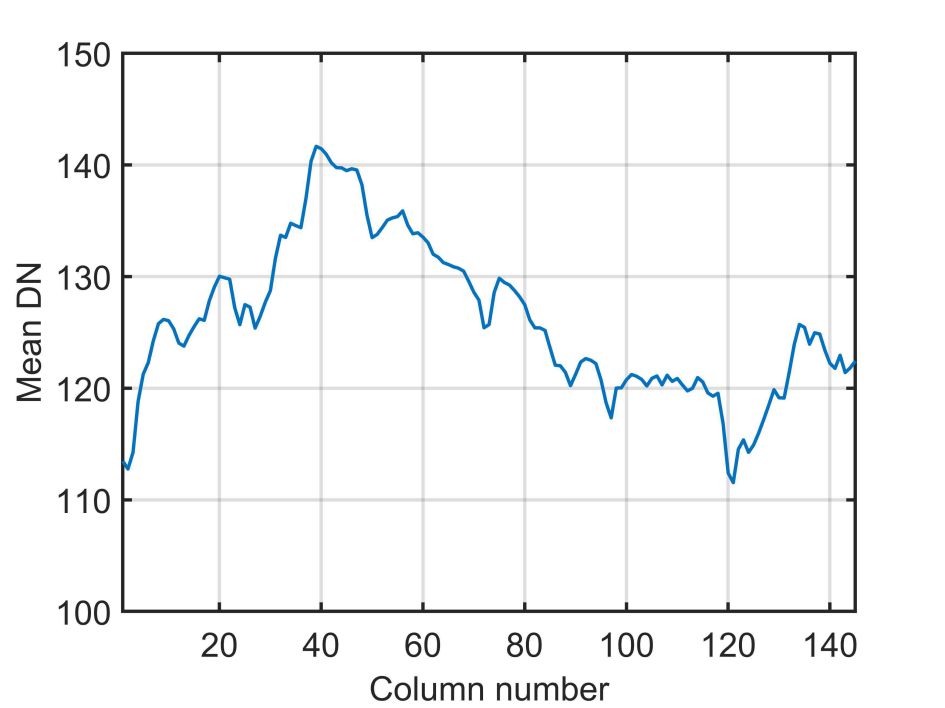}}%
	\hfil
	\subfloat[LRMR]{\includegraphics[width=0.15\linewidth]{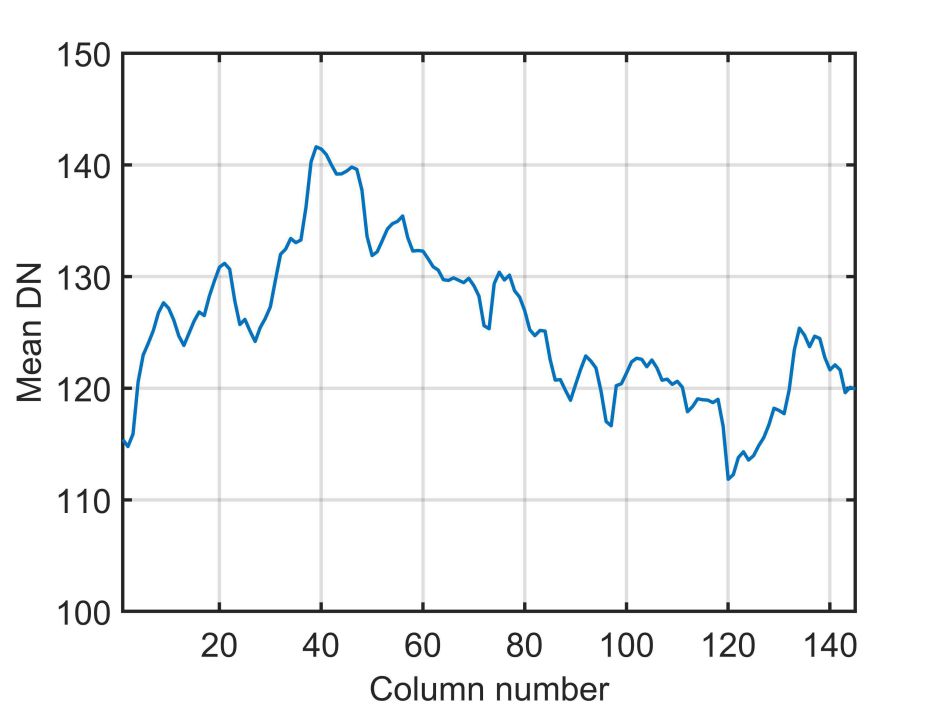}}%
	\hfil
	\subfloat[LLRSSTV]{\includegraphics[width=0.15\linewidth]{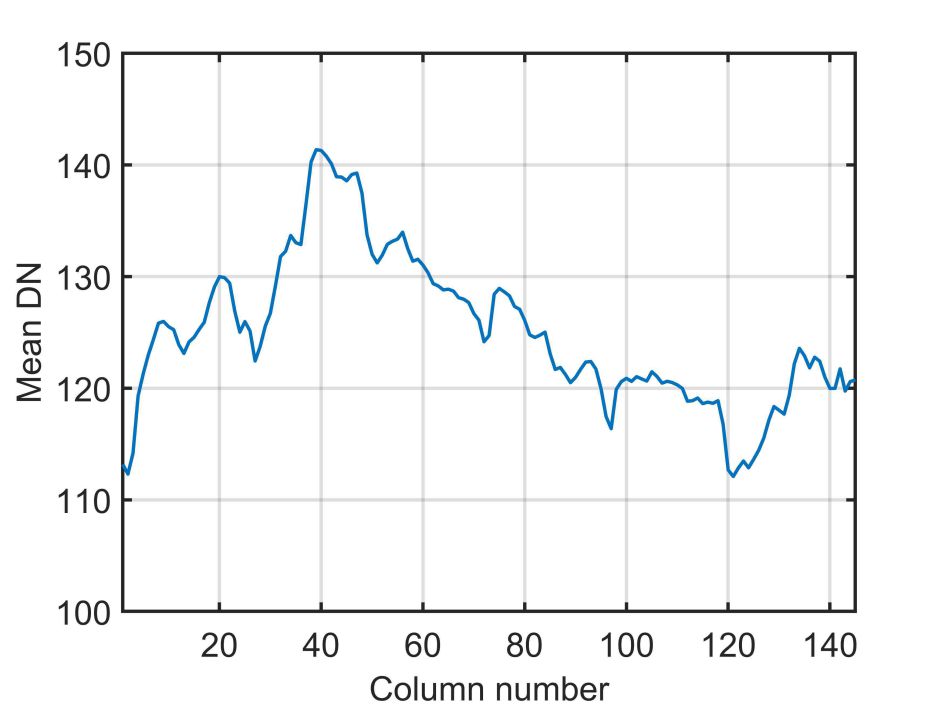}}%
	\hfil
	\subfloat[LLxRGTV]{\includegraphics[width=0.15\linewidth]{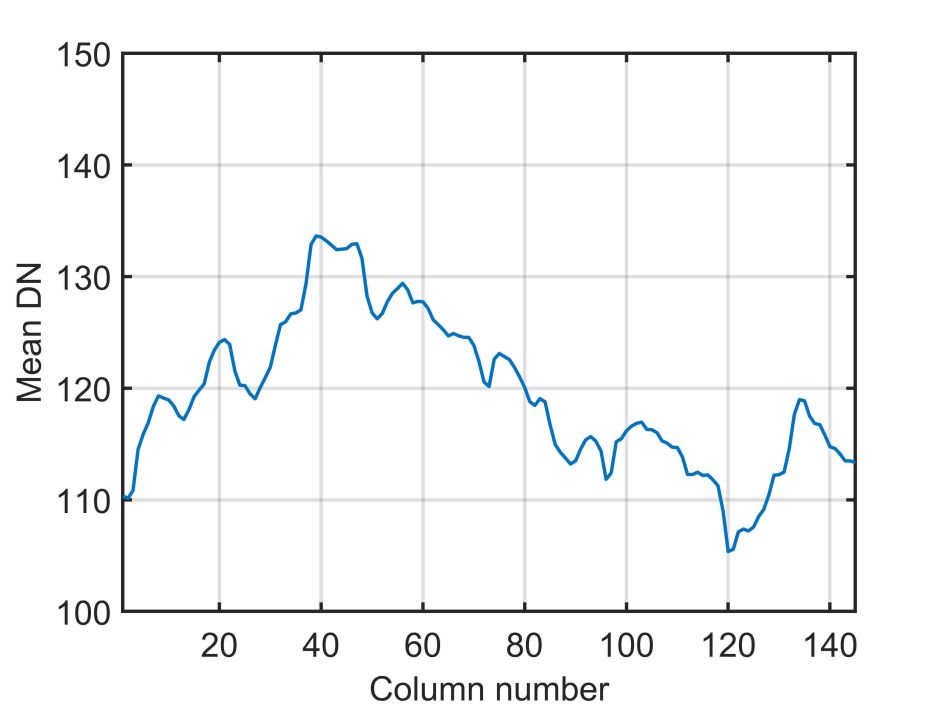}}%
	\caption{Vertical mean profiles of band 220 in AVIRIS Indian Pines dataset experiment.}
	\label{fig:DN_real_indian_b220}
\end{figure*}

For vertical mean profiles analysis,
Fig. \ref{fig:DN_real_indian_b108} and Fig. \ref{fig:DN_real_indian_b220}  show the vertical mean profiles of the 108-th and 220-th band.
The smaller the fluctuation of the mean profiles is, the higher the image quality is.
Due to the existence of mixed noise, the mean profile curves of the noisy image appear to fluctuate rapidly.
After denoising, one can see that the mean profile curves of our method are the most stable and its fluctuation is the smallest.
This fact is also consistent with the visual results shown in Fig. \ref{fig:real_indian_b108} and Fig. \ref{fig:real_indian_b220}.

\subsubsection{HYDICE Urban dataset}

\begin{figure*}[!t]
	\centering
	\subfloat[Original band]{\includegraphics[width=0.15\linewidth]{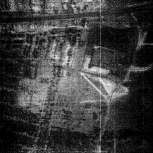}}%
	\hfil
	\subfloat[LRTA]{\includegraphics[width=0.15\linewidth]{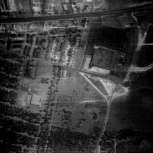}}%
	\hfil
	\subfloat[NAILRMA]{\includegraphics[width=0.15\linewidth]{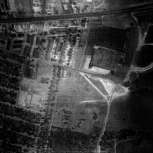}}%
	\hfil
	\subfloat[LRMR]{\includegraphics[width=0.15\linewidth]{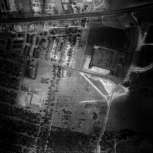}}%
	\hfil
	\subfloat[LLRSSTV]{\includegraphics[width=0.15\linewidth]{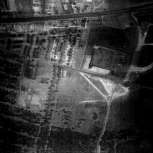}}%
	\hfil
	\subfloat[LLxRGTV]{\includegraphics[width=0.15\linewidth]{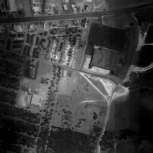}}%
	\caption{Denoised results in the real experiments of HYDICE Urban dataset. }
	\label{fig:real_urban_b139}
\end{figure*}

\begin{figure*}[!t]
	\centering
	\subfloat[Original band]{\includegraphics[width=0.15\linewidth]{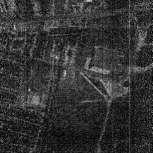}}%
	\hfil
	\subfloat[LRTA]{\includegraphics[width=0.15\linewidth]{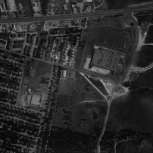}}%
	\hfil
	\subfloat[NAILRMA]{\includegraphics[width=0.15\linewidth]{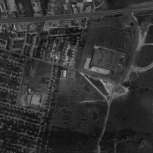}}%
	\hfil
	\subfloat[LRMR]{\includegraphics[width=0.15\linewidth]{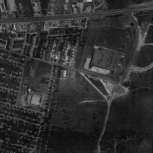}}%
	\hfil
	\subfloat[LLRSSTV]{\includegraphics[width=0.15\linewidth]{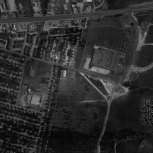}}%
	\hfil
	\subfloat[LLxRGTV]{\includegraphics[width=0.15\linewidth]{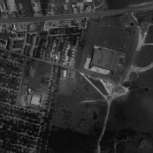}}%
	\caption{Denoised results in the real experiments of HYDICE Urban dataset.}
	\label{fig:real_urban_b207}
\end{figure*}

The original size of HYDICE Urban dataset \footnote{http://www.tec.army.mil/hypercube/} is $207\times207\times210$.
There are 189 bands left after the water absorption bands are excluded.
The right image in Fig. \ref{3dcubeofurban} is a three-dimensional representation of the HSI with false colors synthesized by three bands: 20th, 90th and 180th band.
It can be seen that it is mainly polluted by the atmosphere, water absorption, stripes and other unknown noise.
Similar to the section \ref{section:AVIRIS_indian_analysis}, in this part we also evaluate the proposed model from two aspects: visual evaluation and vertical mean profiles analysis.
As shown in Fig. \ref{fig:real_urban_b139} and Fig. \ref{fig:real_urban_b207}, all the test methods can remove the mixed noises, to some extent.
However, the competitive methods result in local noise residual problem, which leads to the loss of local texture information.
The proposed model can both effectively remove the mixed noises and preserve the local details of the HSI.

\begin{figure*}[!t]
	\centering
	\subfloat[Original band]{\includegraphics[width=0.15\linewidth]{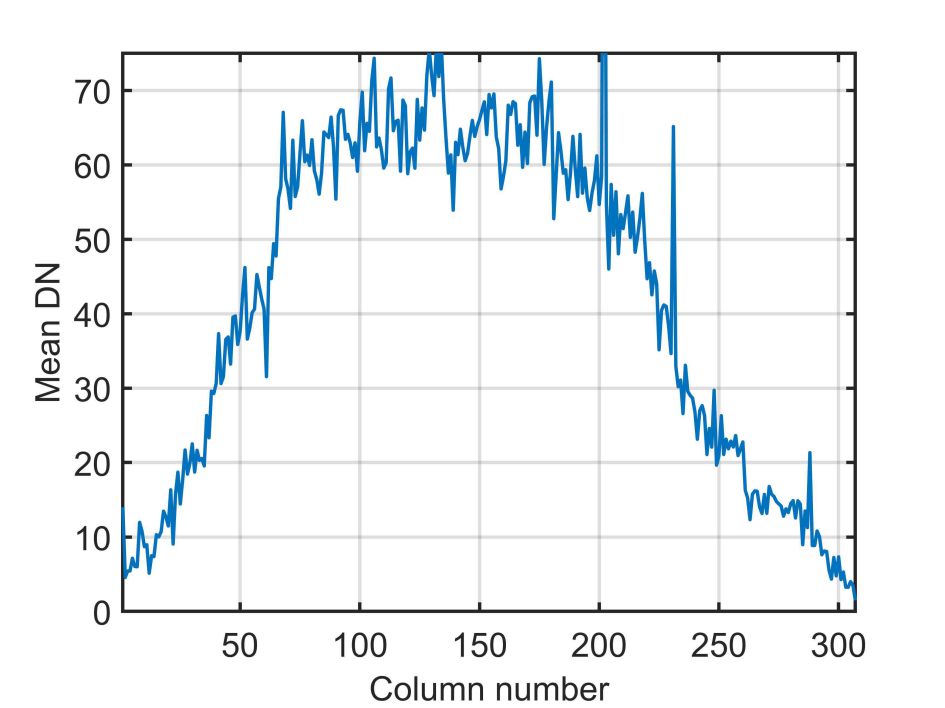}}%
	\hfil
	\subfloat[LRTA]{\includegraphics[width=0.15\linewidth]{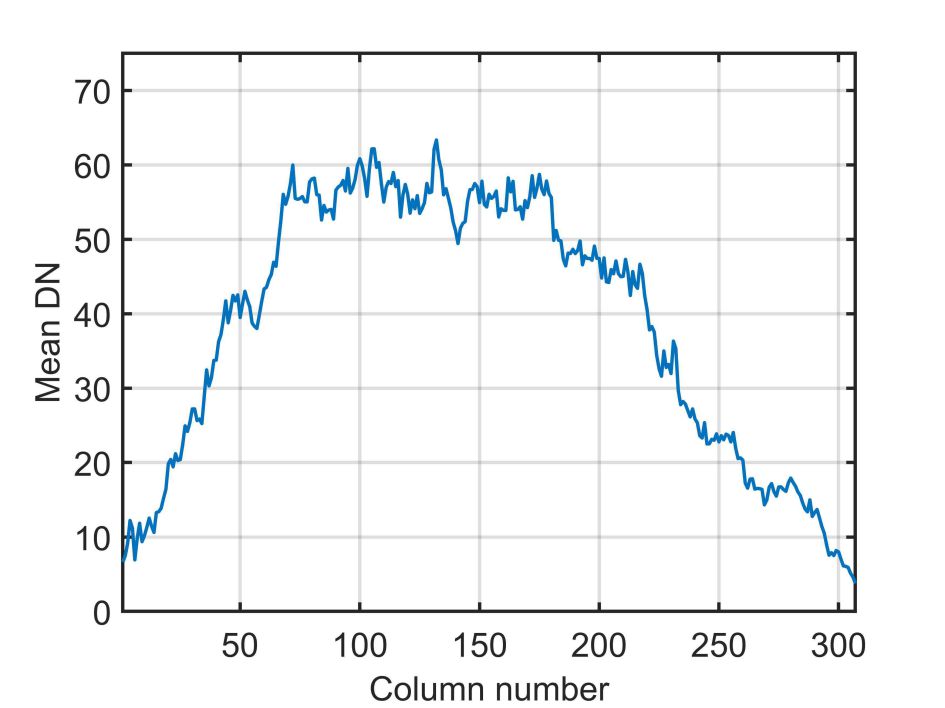}}%
	\hfil
	\subfloat[NAILRMA]{\includegraphics[width=0.15\linewidth]{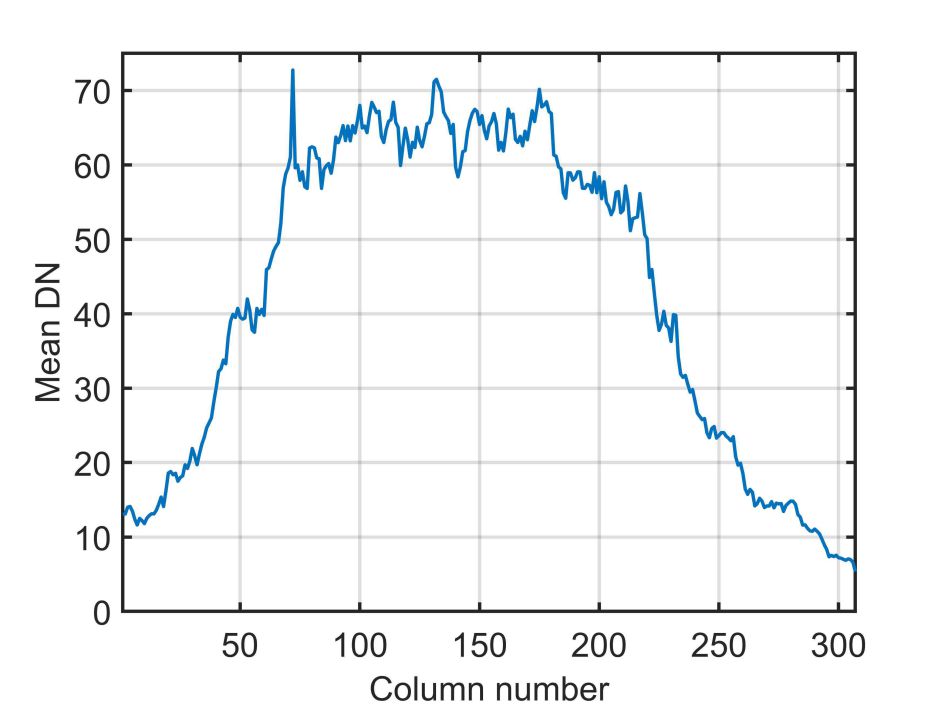}}%
	\hfil
	\subfloat[LRMR]{\includegraphics[width=0.15\linewidth]{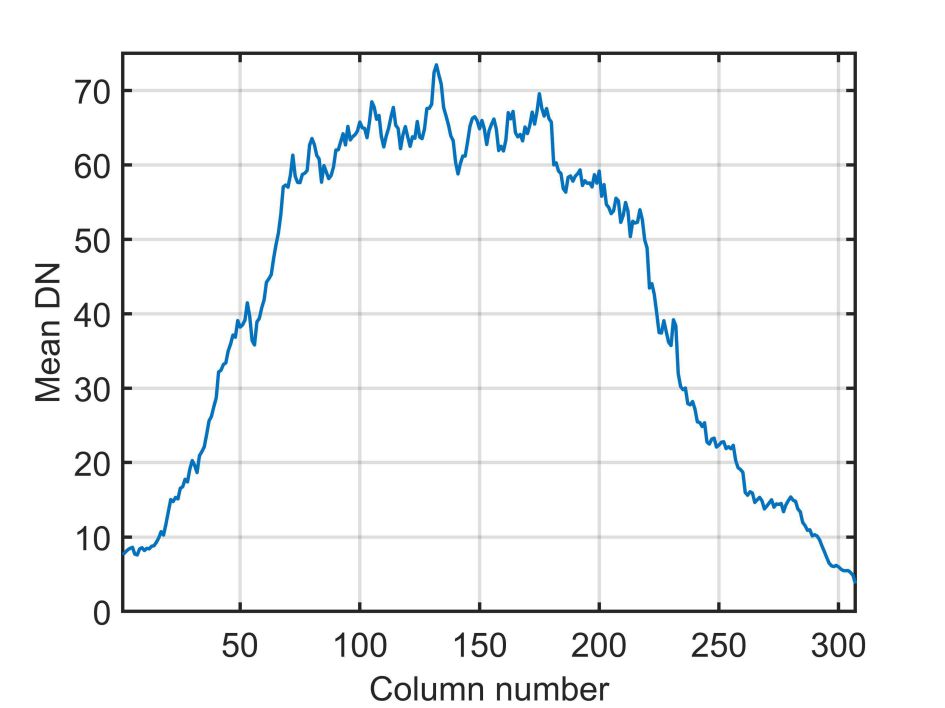}}%
	\hfil
	\subfloat[LLRSSTV]{\includegraphics[width=0.15\linewidth]{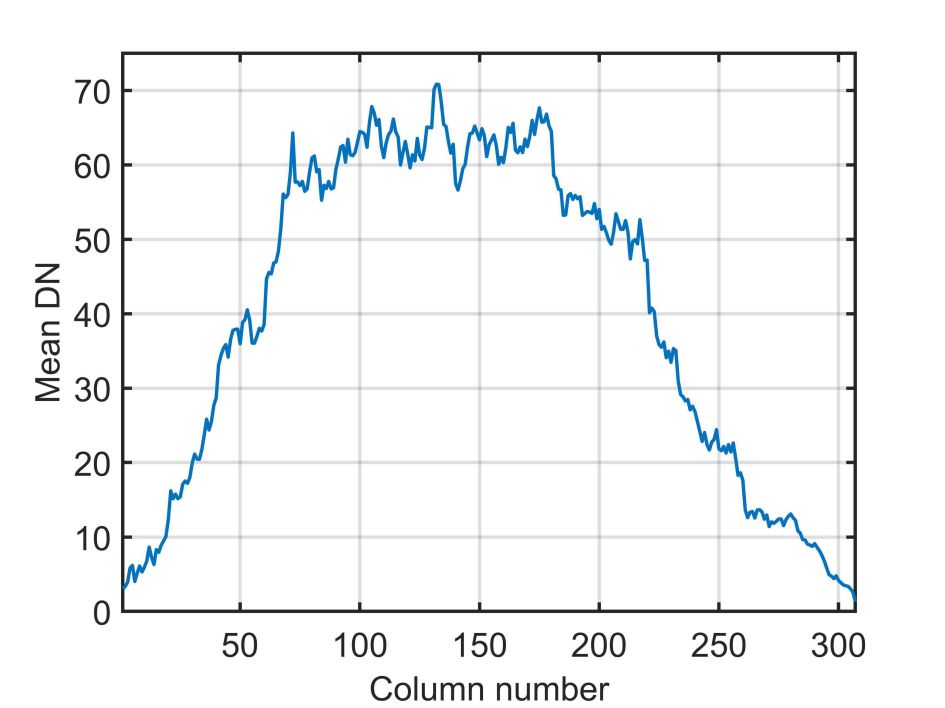}}%
	\hfil
	\subfloat[LLxRGTV]{\includegraphics[width=0.15\linewidth]{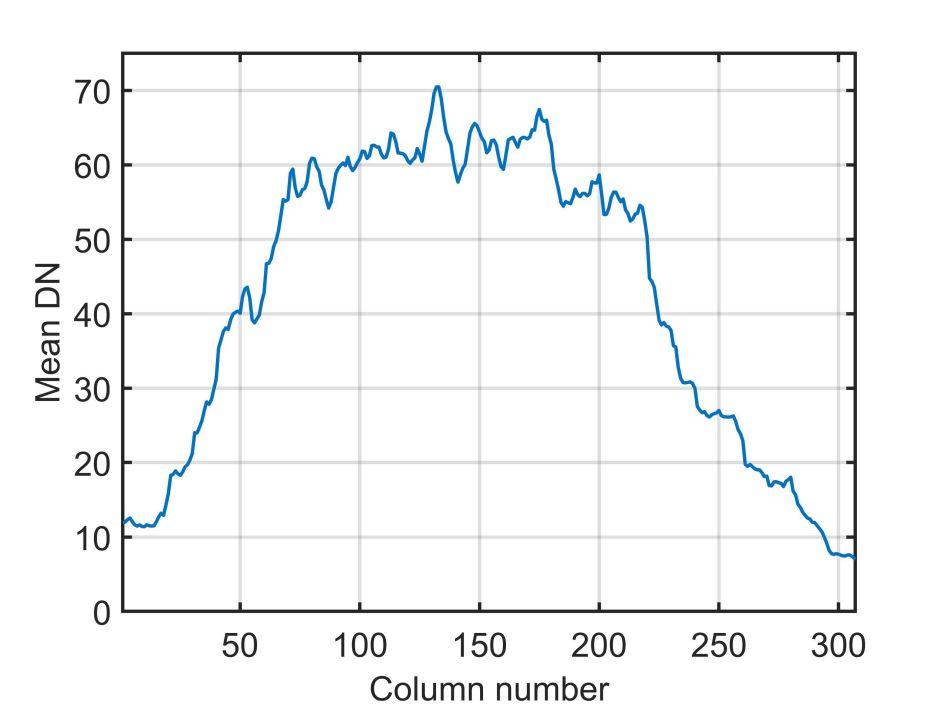}}%
	\caption{Vertical mean profiles of band 139 in the HYDICE Urban dataset experiment.}
	\label{fig:DN_real_urban_b139}
\end{figure*}

\begin{figure*}[!t]
	\centering
	\subfloat[Original band]{\includegraphics[width=0.15\linewidth]{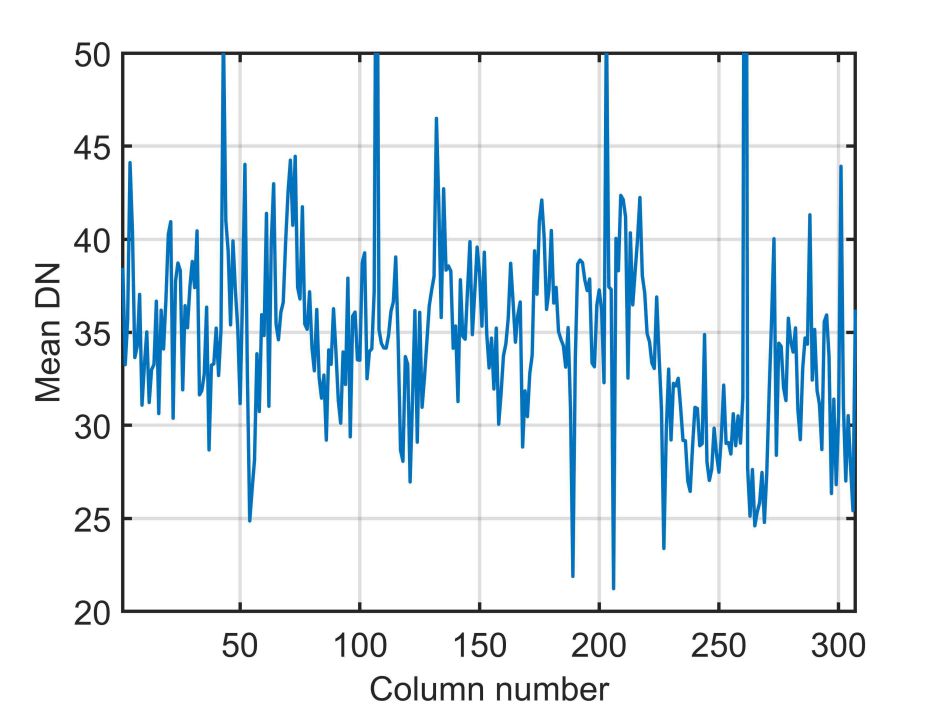}}%
	\hfil
	\subfloat[LRTA]{\includegraphics[width=0.15\linewidth]{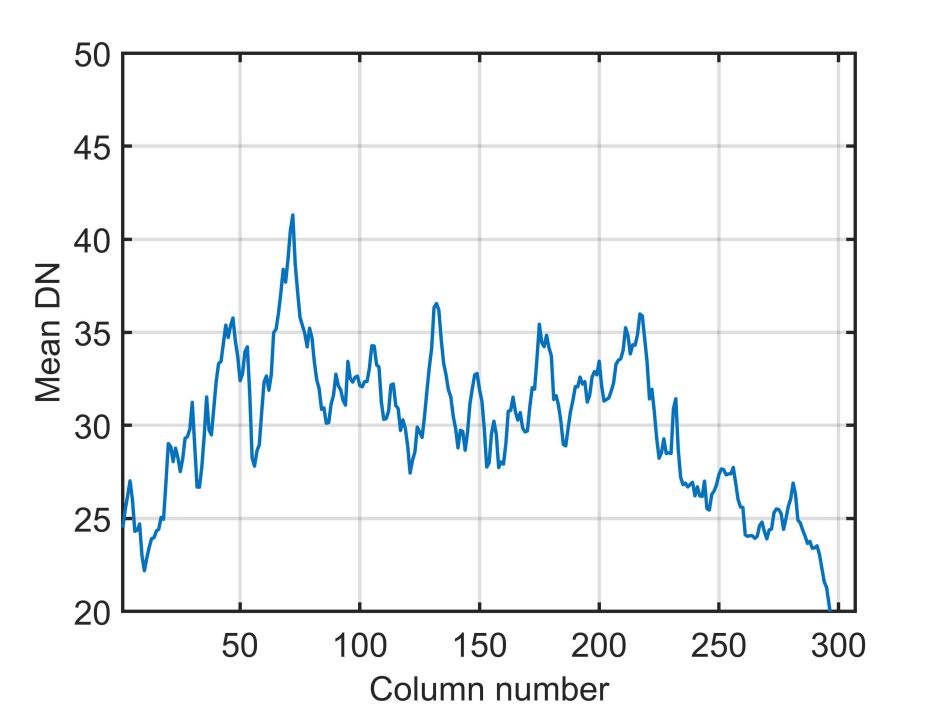}}%
	\hfil
	\subfloat[NAILRMA]{\includegraphics[width=0.15\linewidth]{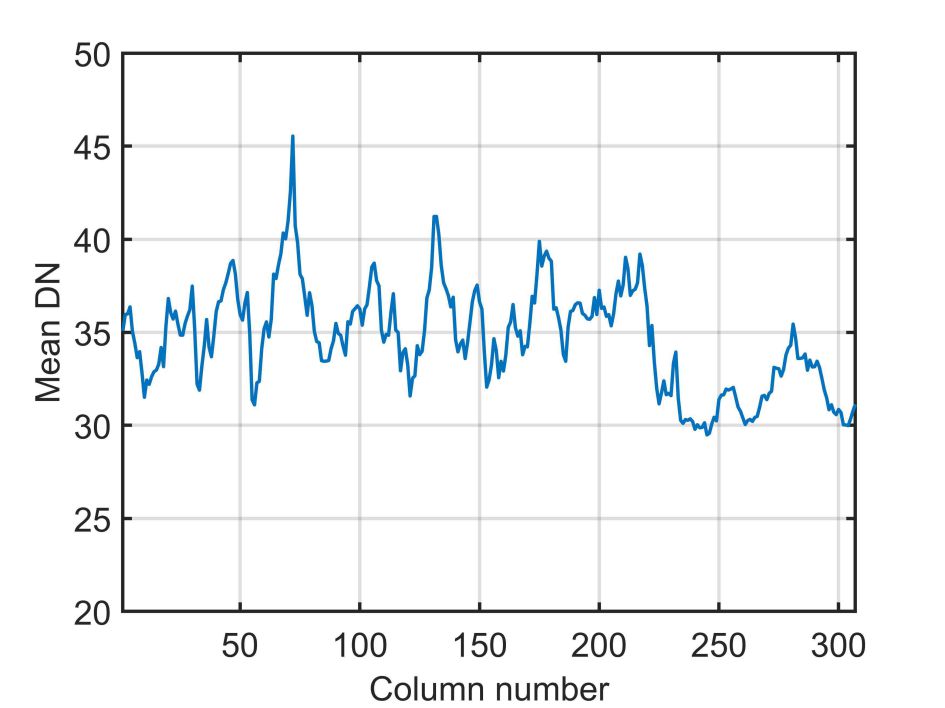}}%
	\hfil
	\subfloat[LRMR]{\includegraphics[width=0.15\linewidth]{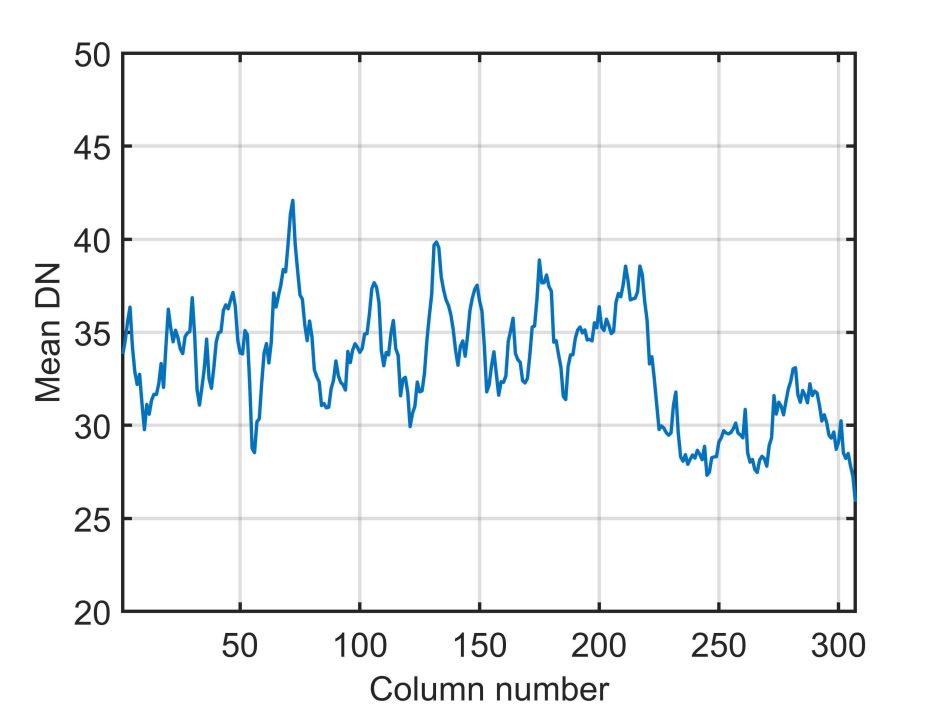}}%
	\hfil
	\subfloat[LLRSSTV]{\includegraphics[width=0.15\linewidth]{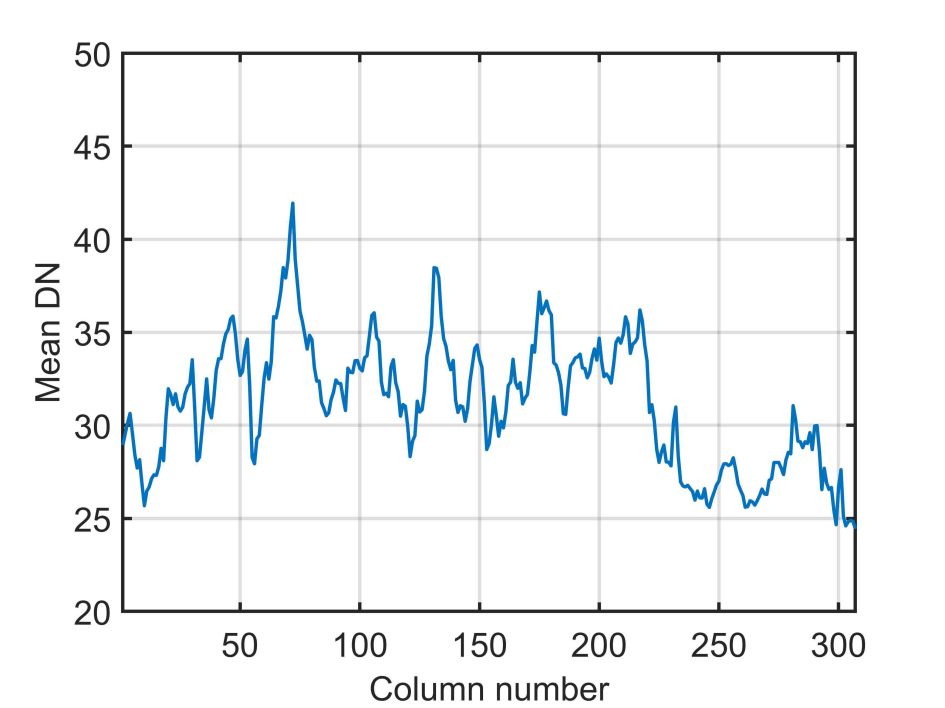}}%
	\hfil
	\subfloat[LLxRGTV]{\includegraphics[width=0.15\linewidth]{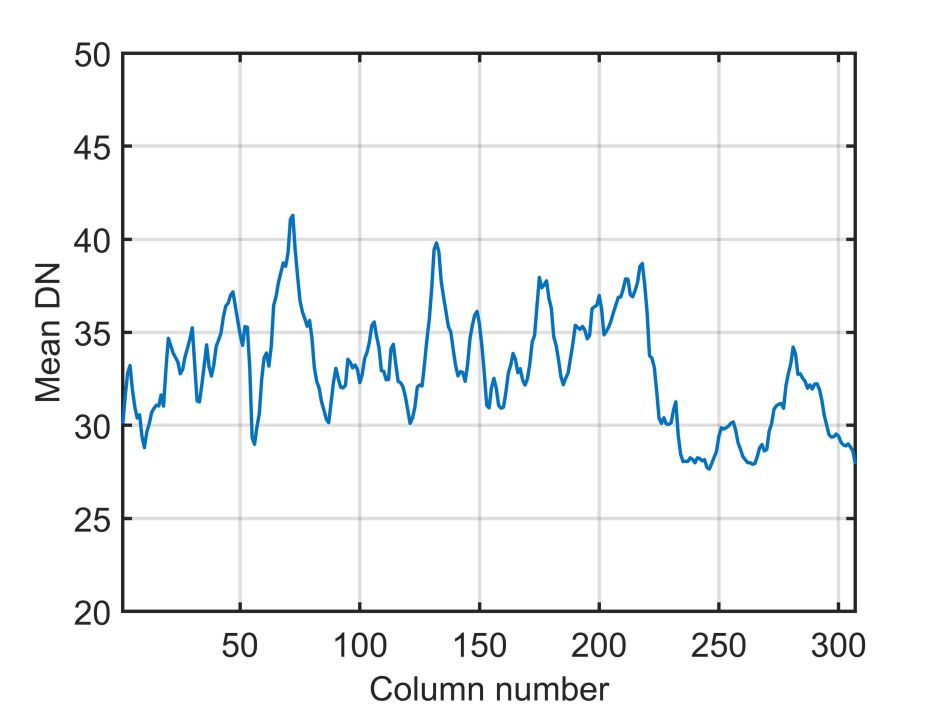}}%
	\caption{Vertical mean profiles of band 207 in HYDICE Urban dataset experiment.}
	\label{fig:DN_real_urban_b207}
\end{figure*}

\subsection{Discussion}
\label{discussion}

\subsubsection{Sensitivity analysis of parameters}

\begin{figure}[t]
	\centering
	\includegraphics[width=0.45\linewidth]{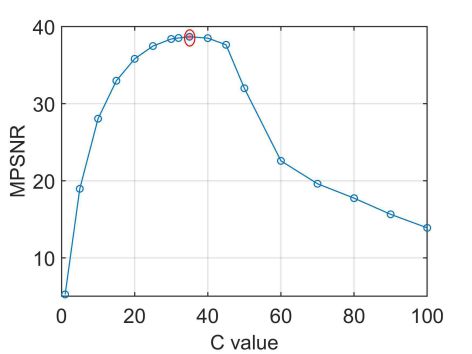}
	\includegraphics[width=0.45\linewidth]{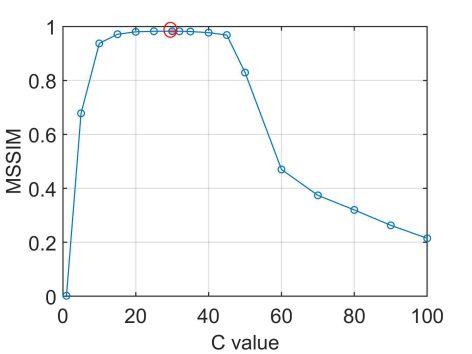}
	\caption{Sensitivity analysis of the $C$ value. (a) Change
		in the MPSNR value, (b) Change in the MSSIM value.}
	\label{fig:Sen_C}
\end{figure}

In the proposed model, there are several parameters that need to be carefully identified.
Specifically, as stated in the RPCA model \cite{RPCA},
the parameter $\lambda =  1/ \sqrt{\max(m, n)p}$ for the sparsity regularization term is good enough to guarantees the exact recovery.
Owing to bring the new TV penalty, our model is different from RPCA.
We thus set $\lambda = C/ \sqrt{\max(m, n)p}$ as the initial value, and adjust the parameter $C$ according to the specific experimental datasets.
Under the condition of USGS Indian Pines dataset, Fig. \ref{fig:Sen_C} reports the MPSNR and MSSIM values of our model according to the different parameters $\lambda$.

\begin{figure}
	\centering
	\includegraphics[width=0.45\linewidth]{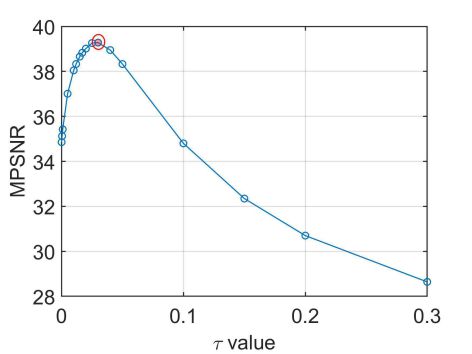}
	\includegraphics[width=0.45\linewidth]{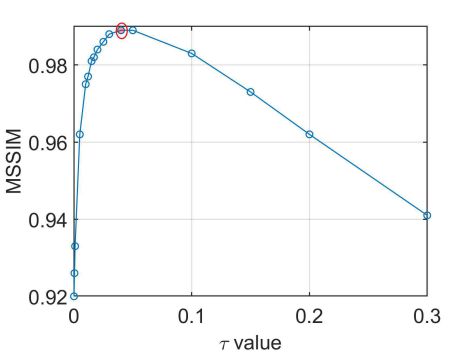}
	\caption{Sensitivity analysis of the $\tau$ value. (a) Change
		in the MPSNR value, (b) Change in the MSSIM value.}
	\label{fig:Sen_tau}
\end{figure}

In the proposed model, $\tau$ also is an important parameter used to balance the influence of TV term and the rest of the regular terms.
Fig. \ref{fig:Sen_tau} shows the MPSNR and MSSIM of our model as $\tau$ varied in the set:
$\{$ 0.0001, 0.0005, 0.001, 0.005, 0.01, 0.012, 0.015,	0.017,	0.02,	0.025,	0.03,	0.04,	0.05,	0.1,	0.15,	0.2,	0.3 $\}$.
Based on the above analysis, in the simulated and real data experiments, we suggest the use of $C=35, \tau= 0.03$.
Finally, we manually set $\gamma$ in $L_{\gamma}$ norm to 0.3.

\subsubsection{Empirical convergence}

\begin{figure}[t]
	\centering
	\subfloat[MPSNR]{\includegraphics[width=0.45\linewidth]{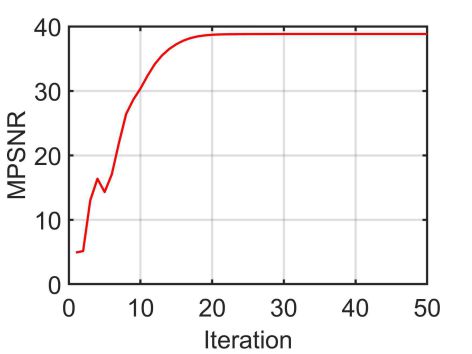}}%
	\hfil
	\subfloat[MSSIM]{\includegraphics[width=0.45\linewidth]{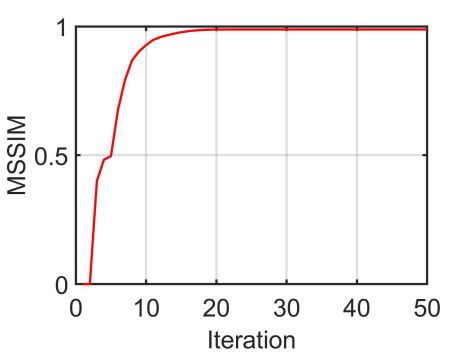}}%
	\caption{MPSNR (a) and MSSIM (b) versus iteration of the proposed model.}
	\label{fig:iteration_psnr_ssim}
\end{figure}

\begin{figure}[t]
	\centering
	\subfloat[Error 1]{\includegraphics[width=0.3\linewidth]{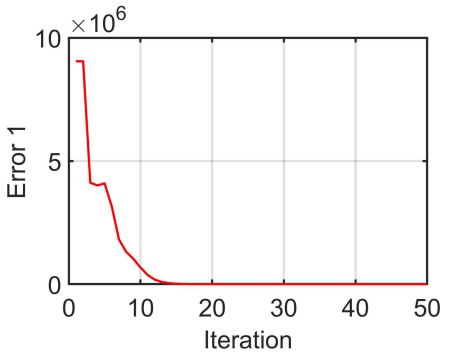}}%
	\hfil
	\subfloat[Error 2]{\includegraphics[width=0.3\linewidth]{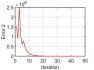}}%
	\hfil
	\subfloat[Error 3]{\includegraphics[width=0.3\linewidth]{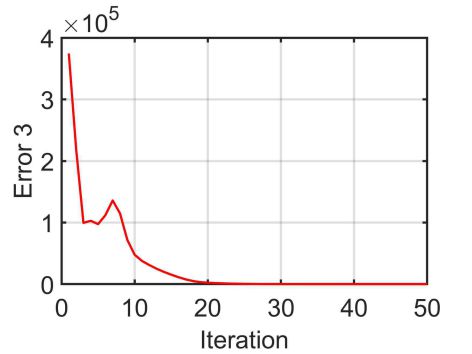}}%
	\caption{Error versus iteration of the proposed model. Error 1= $\|\mathcal{O}_{i,j} - \mathcal{L}_{i,j} - \mathcal{S}_{i,j} - \mathcal{N}_{i,j} \|_F^2$, Error 2= $ \|\mathcal{L}_{i,j} - \mathcal{J}_{i,j} \|_F^2$, Error 3= $ \|\mathcal{J} - \mathcal{X} \|_F^2$.}
	\label{fig:iteration_error}
\end{figure}

To study the empirical convergence of the proposed model, we show the iteration errors, MPSNR and MSSIM of the proposed model with respect to the iteration numbers in
Fig. \ref{fig:iteration_error} and Fig. \ref{fig:iteration_psnr_ssim}.
It can be seen that when the number of iterations reaches a certain threshold, the proposed model is converged.

\section{Conclusion}
\label{conclusion}

In this paper, we propose a new HSI denoising model based on spatial-spectral total variation and nonconvex low-rank tensor approximation.
Instead of using the traditional matrix nuclear norm to explore the overall low-rank prior of the HSIs,
we directly model the tensor and propose a nonconvex approximation to represent the local patch based low-rank structure.
Furthermore, to preserve the global smoothing structure, we introduce the spatial-spectral total variation regularization into the nonconvex local low-rank model,
and propose our denoising model.
An ADMM-based algorithm is designed to efficiently solve the proposed model.
The proposed model has been evaluated on four public HSI datasets,
which show that our model can effectively remove the mixed noise while maintaining the texture information of the HSIs.

%


\bibliographystyle{cas-model2-names}

\bibliography{mybibfile}


%
%

\end{document}